**Title: The PUR-1 Cyber-Physical Digital Twin**

**Authors**

V. Theos[1], J. Lau[1], K. Gkouliaras[1], Z. Dahm[1], K. Vasili[1], N. Fillgrove[1], W. Richards[1], T. Miller[1], B. Jowers[1], and S. Chatzidakis[1,*]

[1]School of Nuclear Engineering, Purdue University, West Lafayette, IN 47907

*schatzid@purdue.edu

**Abstract**

Digital twin technologies have the potential to improve operational flexibility and responsiveness capabilities of nuclear systems. To provide decision support, cyber event characterization, state estimation, predictive control, and real-time dynamic processing of operational data, however, an efficient digital twin needs to integrate multiple models (data-driven as well as physics-based) with explainability while at the same time maintain two-way synchronization with the physical facility at a time constant less than its operational cycle. In this work, we present the Purdue University Reactor One Digital Twin (PUR-1 DT), a cyber-physical digital twin with a complete high-fidelity physics-based and AI-driven virtual model stack (neutronics, thermal-hydraulics, point kinetics) which provides closed-loop explainable diagnostics, forecasting, predictive control, and action recommendation back to the reactor via two-way communications and a cyber-physical testbed. We demonstrate real-time synchronized state estimation and short-term forecasting over a full reactor operational cycle and conduct a series of benchmarking experiments to validate accuracy and latency. Our results show good agreement with experimental results and lay the groundwork for further development and experimental demonstration of DT-enabled functionalities in real-world facilities.

# 1 Introduction

Nuclear energy could play a pivotal role in the global transition towards reliable and clean electricity. It offers a reliable baseload source of power that complements renewable generation and supports grid stability. Despite these significant contributions characterizing nuclear power as a robust and flexible energy option, nuclear systems face challenges such as limited operational flexibility, increased operation and maintenance costs, and difficulty adjusting to fluctuating energy demand [1]. In an era defined by scalable operation, remote monitoring and plant forecasting requirements, nuclear systems must evolve to become more adaptive and responsive.

Modern reactor concepts are currently being developed to overcome these limitations. One way is by incorporating digital instrumentation and control (I&C) systems, advanced sensors, and remote monitoring capabilities that could enable load-following operation, smart-grid integration, and operational flexibility [2, 3]. However, the complexity of reactor physics, limited availability of in-core measurements, communication latency, and stringent safety constraints hinder effective use of these technologies and access to the data they could provide in the absence of an appropriate integration framework [4]. Such a framework could be provided by Digital Twin (DT) technologies assimilating real-time operational data with computational models.

A DT is a virtual replica of a physical asset that evolves dynamically with its physical counterpart through continuous, bidirectional data exchange [5, 6]. Unlike static simulations or digital shadows with one-way data flow, DTs can both reconstruct hidden system states and recommend control actions to achieve desired operating conditions. These capabilities can be categorized into forward problems, where unmeasured states such as flux distributions or thermal fields are inferred from available data, and inverse problems, where control strategies such as optimal rod movements are derived to meet prescribed objectives [7, 8]. Therefore, DTs extend system observability, enable predictive maintenance, and support operator decision-making in real time.

Although DTs have already brought benefits to several domains such as aerospace [9–11], manufacturing [12], and healthcare [13–15], their application to nuclear systems remains relatively nascent [16, 17]. National laboratories have pioneered DT tools and frameworks, including INL's NRIC testbeds for advanced reactors [16, 18], ORNL's Virtual Environment for Reactor Applications (VERA), and ANL's reduced-order methods for forecasting and system diagnostics [19]. In industry, initiatives such as the twinning tool for the BWRX-300 SMR [20], Framatome's simulators for operator training in collaboration with EDF, and tools from X-Energy and Kairos Power for maintainability optimization highlight the growing industrial adoption [21]. Finally, academic efforts at North Carolina State University, University of Texas at Austin, Massachusetts Institute of Technology, and Idaho State University, demonstrate further the potential of DT tools in modern reactor designs [22–26]. Despite these advances, the majority of the proposed work remains conceptual, fragmented, or limited to non-nuclear test environments, with only a few validated in operating reactors [7, 27].

In this paper, we present the design, implementation, and experimental validation of a real-time DT for Purdue University Reactor One (PUR-1). PUR-1 is the first reactor in the U.S. licensed to operate with a fully digital I&C system, making it an ideal testbed for the development of a digital twin. The DT leverages the PUR-1 digital I&C capabilities to establish bilateral communication between the physical and digital assets [28]. By coupling multiphysics forward models, inverse optimization routines, and machine-learning surrogate models, the PUR-1 DT operates in synchronization with the PUR-1 operation cycle. The contributions of this work are (i) a set of DT requirements tailored for nuclear environments, (ii) a full-scale demonstration of a cyber-physical DT in a nuclear reactor with real-time synchronization, state estimation, anomaly detection, and forecasting through filtering and surrogate models trained on high-fidelity simulations, (iii) experimental benchmarking of the principal neutronics, thermal-hydraulics, and ML-based surrogate models and (iv) evaluation of the integrated DT over complete reactor operating cycles, including quantitative assessment of model accuracy and end-to-end latency. Through these contributions, this study provides one of the first validated DT frameworks for nuclear systems. The results establish the feasibility of real-time, cyber-physical DTs in research reactors and lay the groundwork for evaluating various DT processes such as anomaly detection, forecasting, cybersecurity, and autonomous control.

The remainder of the paper is organized as follows. Section 2 presents DT definition and requirements for nuclear applications. Section 3 introduces the architecture and development approach of Digital Twins. Section 4 and Section 5 present the development steps and the high-level configuration of PUR-1 DT, respectively. Section 6 describes the physical system, Section 7 presents its virtual system, and Section 8 describes the cyber-physical integration framework linking the two systems. Section 9 reports validation and accuracy results for the integrated simulation and AI/ML models. Section 10 demonstrates a representative use case of PUR-1 DT operation over a reactor operational cycle. Finally, Section 11 describes the lessons learned throughout PUR-1 DT development and Section 12 summarizes the main findings and outlines a path forward for further development of the presented work.

# 2 Background

## 2.1 Definitions

The broad adoption of DTs across engineering domains has resulted in variability in their definition with differences in coupling integration level, model fidelity, data integration, and adaptability. DT definitions are often tailored to each specific application and shaped by the intended operational role of each system [16, 29] (Table I). Recognizing this variation, several agencies propose a generic application-agnostic definition covering all definition discrepancies. For example, the U.S. Nuclear Regulatory Commission (U.S. NRC) defines a DT as a *"…virtual representation of an entity, process, or system, synchronized at a frequency and fidelity sufficient to maintain state concurrence"* [30]. Similarly, the DT Consortium describes it as *"…an integrated data-driven virtual representation of real-world entities and processes, with synchronized interaction at a specified frequency and fidelity"* [31]. Despite differences in phrasing, these definitions converge on a common concept: A Digital Twin is a high-fidelity digital representation of a physical asset with dynamic data exchange between the two systems capable of evolving over time.

A more in depth classification of DTs has been proposed by Kritzinger et al., distinguishing DTs by levels of integration (Figure 1) [5]. At the lowest level, digital models are static simulation models replicating a single state of the physical asset through manual update. At an intermediate level, a digital shadow incorporates one-way automatic data flow from the physical to the digital system, as is typical in cyber-isolated nuclear facilities. At the highest level, a digital twin is bidirectionally coupled, such that changes in the physical system are continuously mirrored in the digital representation and can be communicated back to the physical twin [32].

Table I: Digital Twin definitions proposed in literature

| Authors | Definition Focus | Objective | Industry |
|---|---|---|---|
| Tuegel et al. [11] | Geometric detail up to microstructure level along the lifecycle of the physical system | High Fidelity FEM and CFD simulation for structural degradation and predictive maintenance | Aviation |
| Glaessgen and Stargel [33] | Multiphysics simulation with probabilistic quantification of inputs uncertainty | High fidelity simulation enhancing compliance with safety regulation | Aerospace |
| Boschert and Rosen [34] | Multi dimensionality on simulations for different components and processes | Simulation model significancy across physical system life cycle | Manufacturing |
| Chen [35] | Computational capabilities simulating all subsystems and counterparts | Intelligence system integration into design, operation and maintenance process | Manufacturing |
| Schluse et al. [36] | Testbed capabilities for physical system response evaluation | System optimization by evaluating system behavior under future states | Manufacturing |
| Tao et al. [12] | Dynamic multiscale simulations capability | Simulation integration and synchronization with CPS | Manufacturing |
| Liu et al. [37] | Continuous updated "living model" of the parent system | Predictive maintenance of physical system based on near real-time data fusion | Aviation |
| Mohammadi & Taylor [38] | Parent system state identification and evolution prediction | Continuous analysis and monitoring for long-term prediction and real-time decision making | Smart Cities |
| Gahlot et al. [15] | Diagnostics capabilities enabled by continuous physical system monitoring. | Data analysis and forecasting through real-time data from IoT-based framework | Healthcare |

Beyond integrational nuance, DTs are defined by core properties that enable a comprehensive representation of their physical counterparts: (i) continuous, bidirectional data exchange between the virtual and physical systems, (ii) synchronization in real or near real time, (iii) sufficient fidelity in the digital model to capture governing physical processes, and (iv) adaptability, where the twin evolves along its physical counterpart [6, 9].

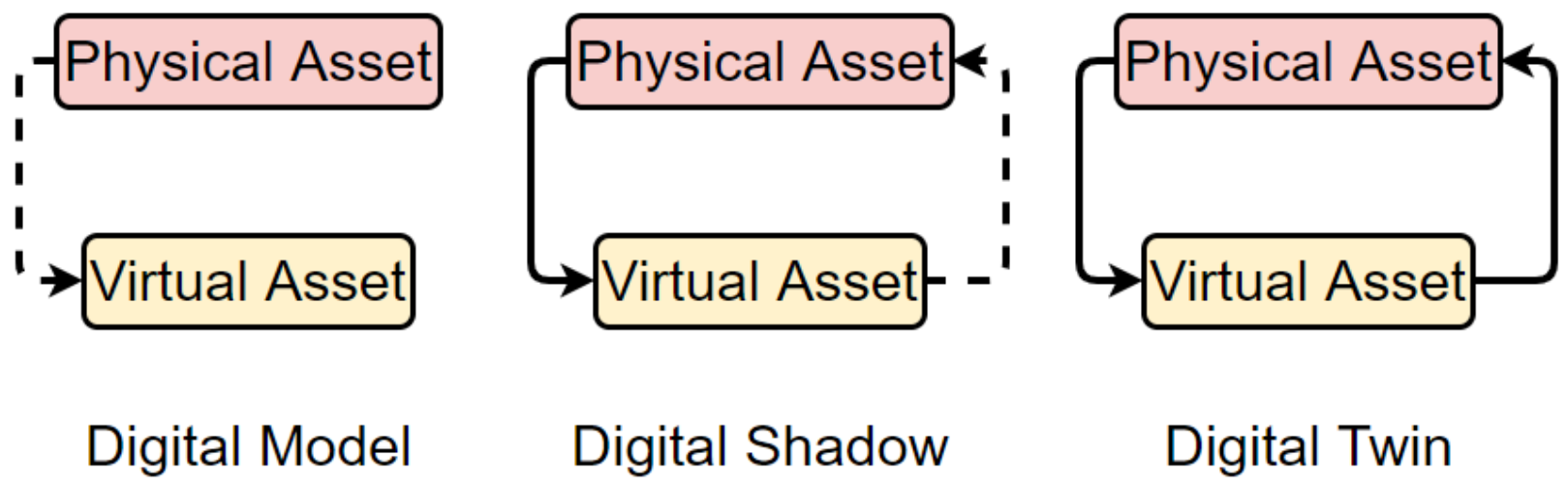

Figure 1: The three levels of DT integration

### 2.2 Digital Twin for Nuclear Environment Applications

Despite the widespread adoption of DTs across industrial sectors, each system must be tailored to its specific application [9]. Similarly to definition deviations, DT requirements vary according to the intended purpose of use. Beyond the core requirements common to every DT system, as discussed in the previous section, additional requirements emerge based on the specific objective and the characteristics of the physical system it represents.

This work focuses on DTs for nuclear reactor applications, with particular attention to the requirements imposed by such safety-critical, high-consequence environments. In addition to core DT requirements, the nuclear reactor environment introduces additional requirements:

(1) A DT should be implemented with real-time data acquisition and data processing capabilities.
(2) A DT should be designed to update dynamically with the evolution of the system state.
(3) A DT should be designed to support bidirectional communication flow with the physical system.
(4) A DT operation should be fully synchronized with the operational cycle of its physical counterpart.
(5) A DT should be designed with state estimation, forecasting, decision support, predictive control, and anomaly detection capabilities.
(6) A DT should be designed to support both short- and long-term hidden-state estimation and forecasting.
(7) A DT should be designed to provide high-fidelity three-dimensional reconstruction of neutron flux, temperature and flow rate field with adequate spatial and temporal resolution.
(8) A DT should be designed to achieve high accuracy (> 90%), a forecasting horizon adequate to enable recommendation evaluation before action, and anomaly detection across multiple concurrent signals.
(9) A DT should be designed to be easily calibrated, debugged, maintained, updated, and adapted to changing reactor conditions in real time.
(10) A DT should be designed to be explainable, supporting uncertainty quantification.
(11) A DT should be designed with an agentic AI-driven graphical user interface and AR/VR visualization to support system-user interaction.
(12) A DT should be designed to be cyber secure, ensuring the integrity, availability, and confidentiality of generated data.

These requirements are fundamental for the development of DT for nuclear reactor environments. The degree of satisfaction of all these requirements might differ based on the objective of application.

### *2.2.1 Related Work*

DT applications are now commonplace in aerospace, automotive, manufacturing, maritime, and industrial processes at various integration levels [11, 12, 15, 33, 39–41]. They provide predictive maintenance, design optimization, and real-time support. Nuclear applications are less evolved, though significant progress has been made. Idaho National Lab's NRIC digital framework is developed with commercial off-the-shelf tools and coupled to testbeds for advanced designs at different power capabilities [21]. The MAGNET digital model is proposed to mirror microreactor operation by combining real-time sensor data with high-fidelity simulations to evaluate remote monitoring and unsupervised operation [3, 42]. Oak Ridge's VERA is intended for simulating reactor behavior to the molecular scale and is paired with a virtual platform for rapid prototyping, quality evaluation, and lifecycle analysis, including predictive maintenance using high accuracy ML and AI models [19, 43]. Beyond national laboratories, industry is also exploring nuclear DT tools. Exelon applies methods by training a natural-language model on historical OT incident reports to identify failure modes [2], and Framatome, in collaboration with Électricité de France (EDF), employs DT tools for workforce education and operator training [44, 45].

In academia, several research groups have proposed DT tools and frameworks for conceptual and modern reactor designs. However, real-time operational data are often unavailable, most implementations rely on synthetic, simulated, or non-synchronized datasets making it challenging to satisfy all of the DT requirements (Table II). These efforts include demonstrations for molten-salt systems and various DT-oriented capabilities, such as autonomous control, predictive maintenance, ML-based operational prediction, and uncertainty quantification, as well as high-fidelity multiphysics simulators for lifecycle assessment and education purposes [16, 23, 46, 47]. More complete deployments in research reactors remain uncommon but have been demonstrated in a limited number of cases, including near-real-time twinning of the AGN-201 reactor to support state identification, anomaly detection, and advanced visualization [25, 26]. In contrast, other efforts have produced high-fidelity web-based digital shadows for research reactors that provide predictive capabilities (e.g., $k_{eff}$, flux distributions, and reactivity changes) but remain non-synchronized and do not allow bidirectional interaction with the physical system [48, 49].

# 3 Digital Twin Architecture

A key question arises: How should one approach the development of a digital twin? The development of a digital twin requires the translation of system purpose functional, computational, and integration requirements that define the architecture. For nuclear reactor applications, these requirements must account for model fidelity, data availability, reactor dynamics, latency, uncertainty, and cybersecurity. Accordingly, the development process can be organized into three stages: requirement-driven design, framework development, and operational implementation.

## 3.1 Design Approach

The requirements identified in Section 2.2 define the capabilities expected from a DT developed for nuclear reactor applications. Translating these high-level requirements into an implementable framework requires identifying both the underlying design basis for each requirement and the associated resulting implication. The design basis represents the physical, operational, computational, safety, or human-system characteristic of the nuclear environment that motivates a requirement, whereas the framework implication identifies the architectural capability or process required to satisfy it. This requirement-driven approach establishes traceability between the characteristics of the physical system and the design of the DT architecture, as summarized in Table III.

Table II: Instance of nuclear-related DT efforts in academia as of 2026.

| Digital Twin | Capabilities | Requirements Fulfilled |
|---|---|---|
| Nearly Autonomous Management and Control System (NCSU) [22] | Plant-referenced simulators, status monitoring, analysis of historical plant data, predictive maintenance | 5,6,8,9,10 |
| "SAFARI"- Secure Automation For Advanced Reactor Innovation (University of Michigan) [16, 24] | Status monitoring, predictive maintenance, design, prototype development, and rapid demonstration | 5,9 |
| MIT [50] | High fidelity neutronics and thermal-hydraulics simulations, predictive maintenance | 5,6,9 |
| Idaho State University [25] | Advanced monitoring, abnormal states development, forecasting, high fidelity neutronics simulation | 5,8,9,10,11 |
| The University of Texas at Austin [48] | Advanced monitoring, high fidelity simulation, predictions | 5,6,7,9, |
| Shanghai Jiao Tong University [51] | Advanced monitoring, high fidelity simulations, state identification | 4,5,8,10 |

## 3.2 Development Process

Once the design basis and applicable requirements have been established, DT development proceeds through progressively increasing levels of fidelity definition and integration. Specifically, the DT development can be categorized into three distinct phases: conceptual, preliminary and final development phase.

The *conceptual development phase* establishes the physical-system boundaries, intended DT functions, information flows, and physical–virtual interactions. It also identifies observable and unobservable state variables, required data sources, sampling frequency, spatial and temporal fidelity, prediction horizons, and performance acceptance criteria, and selects the modeling approach. Physics-based methods directly represent governing phenomena and offer strong interpretability but can be computationally expensive [9], data-driven methods instead learn from operational data using machine learning to infer patterns, predict states, and detect anomalies, often without relying on prior physical knowledge [9, 52], though at the cost of generalization error and dataset bias [7]. Hybrid approaches mitigate both limitations by combining the interpretability of physics-based models with the predictive strength of data-driven techniques. The choice among these should be driven by each DT process's requirements rather than by the modeling technique itself.

Table III: Design basis and framework implications of nuclear digital twin requirements

| Req. | Design basis | Framework implication |
|---|---|---|
| 1 | Continuously evolving reactor conditions characterized from fast transients | Real-time data acquisition, preprocessing, and transmission to the virtual asset |
| 2 | Physical system state evolves during operation alternating system characteristics and configuration | Continuous state assimilation and dynamic updating of the digital representation |
| 3 | DT outputs may support decision-making, automated control, or autonomous operation requiring a closed communication loop | Closed-loop bidirectional communication between physical and virtual assets |
| 4 | DT operation must maintain the synchronization with reactor operation. | Time-constrained execution and synchronization of the complete DT workflow |
| 5 | Limited system observability and the need for anticipatory operational support | State estimation, forecasting, anomaly detection, decision support, and control capabilities |
| 6 | Reactor phenomena evolve over different characteristic time scales | Multi-timescale model execution, state updating, and forecasting |
| 7 | Compact geometry, harsh environment and limited direct accessibility to critical system-state information is often present in modern reactor designs | Implementation of high-fidelity simulation models replicating the detailed reactor geometry. |
| 8 | DT outputs may influence operational decisions to provide information for a sufficient operational regime | Quantitative acceptance criteria for accuracy, forecasting, latency, anomaly detection, and robustness |
| 9 | Model performance can degrade as operating conditions or available data evolve | Modular calibration, validation, retraining, updating, and model replacement |
| 10 | Operator trust requires access to output confidence and decision basis | Uncertainty quantification, confidence assessment, interpretability, and traceability |
| 11 | Complex DT information must be communicated efficiently to users | Integrated GUI and advanced visualization capabilities, including AR/VR |
| 12 | DT data and recommendations may be vulnerable to malicious interference | Cybersecure communication, authentication, integrity protection, and security monitoring |

During the *preliminary development phase*, the conceptual architecture is decomposed into individual models, processes, and interfaces. Data-acquisition and preprocessing requirements are translated into data structures and communication interfaces, while physics-based and data-driven models are developed, calibrated, and independently evaluated. DT operation at this stage is often asynchronous with that of the physical counterpart. High-fidelity models can establish reference solutions and generate training data, whereas reduced-order or surrogate models can be introduced where simulation execution time compromises synchronized operation requirement. State-estimation, forecasting, anomaly-detection, optimization, control, uncertainty-quantification, and visualization functions are subsequently developed according to their allocated requirements. At this stage, component-level verification and validation are essential to establish model accuracy, numerical stability, and uncertainty before individual processes are incorporated into the complete DT workflow.

The *final development phase* integrates the validated components into the complete cyber-physical framework. At this stage, attention shifts from the performance of individual models to DT behavior as an interconnected system. Data dependencies, communication pathways, model-execution order, update frequencies, and interactions between processes must be coordinated such that the outputs of one component remain temporally and physically consistent with the inputs required by subsequent components. End-to-end testing should assess synchronization, accumulated computational and communication latency, robustness to incomplete or corrupted data, uncertainty propagation, cybersecurity overhead, and human–system interaction. The integrated DT should then be evaluated against the acceptance criteria established during conceptual design, providing a traceable connection between the original requirements and the final operational architecture.

## 3.3 Operational Framework

During operation, the DT functions as a continuous physical–virtual loop in which reactor measurements are acquired, processed, and assimilated into the digital representation. The updated state is then used for state estimation, forecasting, anomaly detection, and decision support, with resulting information communicated to the operator or, where permitted, returned to the physical system through a bidirectional control loop. Because individual reactor phenomena evolve over different time scales, DT processes may operate at different update frequencies. However, the critical workflow must remain synchronized with the physical system [53].

Maintaining this awareness requires a reliable, continuous data flow between assets, enabling timely assimilation and state updates. Although high-fidelity twinning extends beyond data streaming alone; it also relies on the coordinated integration of multiple processes and models to infer and update the physical system state. Therefore, DT workflows are commonly organized into discrete architectural layers (Figure 2) that acquire and post-process physical system observations (data layer), utilize these inputs for modeling and simulation of state reconstruction and prediction (modeling and simulation layer), and communicate virtual system outputs to the operator/user (visualization layer).

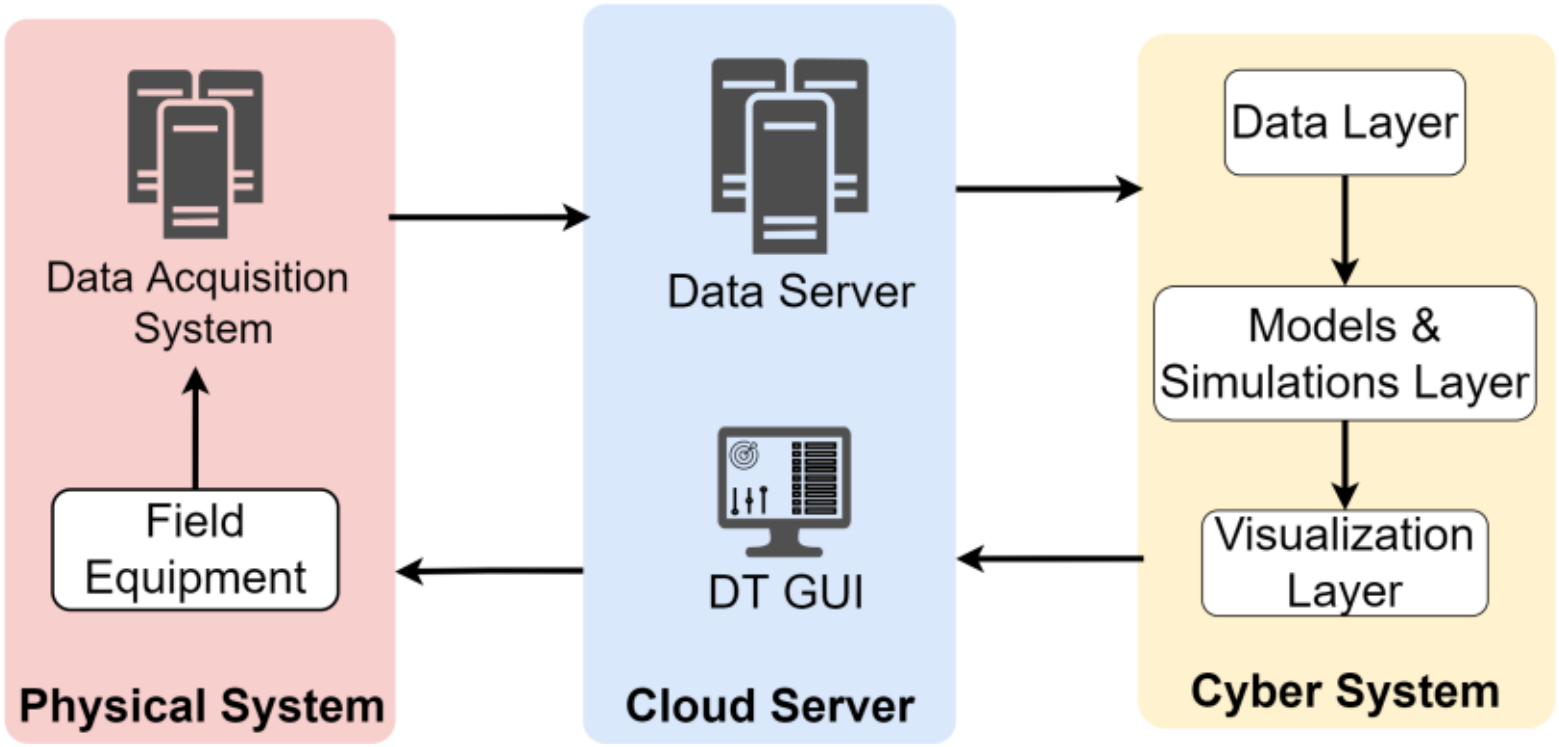

Figure 2: Overview of a high-level Digital Twin architecture

## 4 PUR-1 DT Development Approach

The PUR-1 DT directly addresses the nuclear DT requirements identified in the previous section. The PUR-1 characteristics enable the establishment of real-time data acquisition and bidirectional physical–virtual interaction (Reqs. 1–3), allowing reactor measurements to continuously update the virtual asset while DT-generated actions are evaluated through an independent actuation pathway. To satisfy the synchronization requirement (Req. 4), the operational cycle of PUR-1 was first identified, allowing the selection of a modeling strategy that would yield the highest accuracy given the available computational resources. Given system control guidelines, dedicated ML-based estimation, prediction, PID, and MPC–MHE modules were implemented for state estimation, forecasting, anomaly detection, and predictive control (Reqs. 5 & 6), operating within the bounds of the reactor safety system.

High-fidelity neutronic and thermal-hydraulic models were developed to reproduce the detailed reactor geometry and physical phenomena that cannot be directly observed through instrumentation, addressing the physical-fidelity requirement (Req. 7). Accuracy, explainability, and confidence in DT outputs (Reqs. 8 & 10) were addressed by benchmarking each model individually, quantifying their associated uncertainties, and including alternative models with different levels of accuracy and computational cost. The selected approach aimed for an overall accuracy above 90% with a forecasting horizon of 10 s and monitoring of at least 5 concurrent signals. In order to support efficient interaction between user and system, a unique user interface was implemented for easy adaptability and maintainability of the modules (Reqs. 9 & 11). Finally, cybersecurity (Req. 12) is incorporated through a communication architecture capable of supporting authentication, encryption, and QKD-assisted cryptographic schemes without compromising synchronization.

## 5 PUR-1 DT Configuration

The PUR-1 DT integrates two independently controlled physical subsystems with a multilayer virtual asset. The first physical subsystem is the PUR-1 reactor and its fully digital instrumentation and control system, which provides real-time operational measurements through a unidirectional data pathway. The second is the CERVEROS cyber-physical testbed, which provides additional sensing and a bidirectional actuation interface. Data from both subsystems update the virtual asset, whereas control commands generated or evaluated by the DT are returned only to CERVEROS, preserving the isolation of the PUR-1 control system.

This configuration enables the PUR-1 DT to function as an integrated real-time decision-support environment rather than a standalone simulation tool. By combining live reactor measurements, cyber-physical testbed data, high-fidelity physics models, surrogate models, AI/ML-based diagnostics, and visualization interfaces, the DT provides an expanded representation of reactor behavior beyond directly measured quantities. Its capabilities include online state awareness, estimation of unmeasured neutronic and thermal fields, short-term prediction of reactor response, anomaly and cybersecurity-event detection, and visualization of reactor conditions through both graphical and augmented-reality interfaces. These capabilities are realized through the architecture shown in Figure 3, in which the physical subsystems, data-processing functions, modeling and simulation modules, surrogate models, visualization interfaces, and action-evaluation pathway are combined to support synchronized operation of the PUR-1 DT.

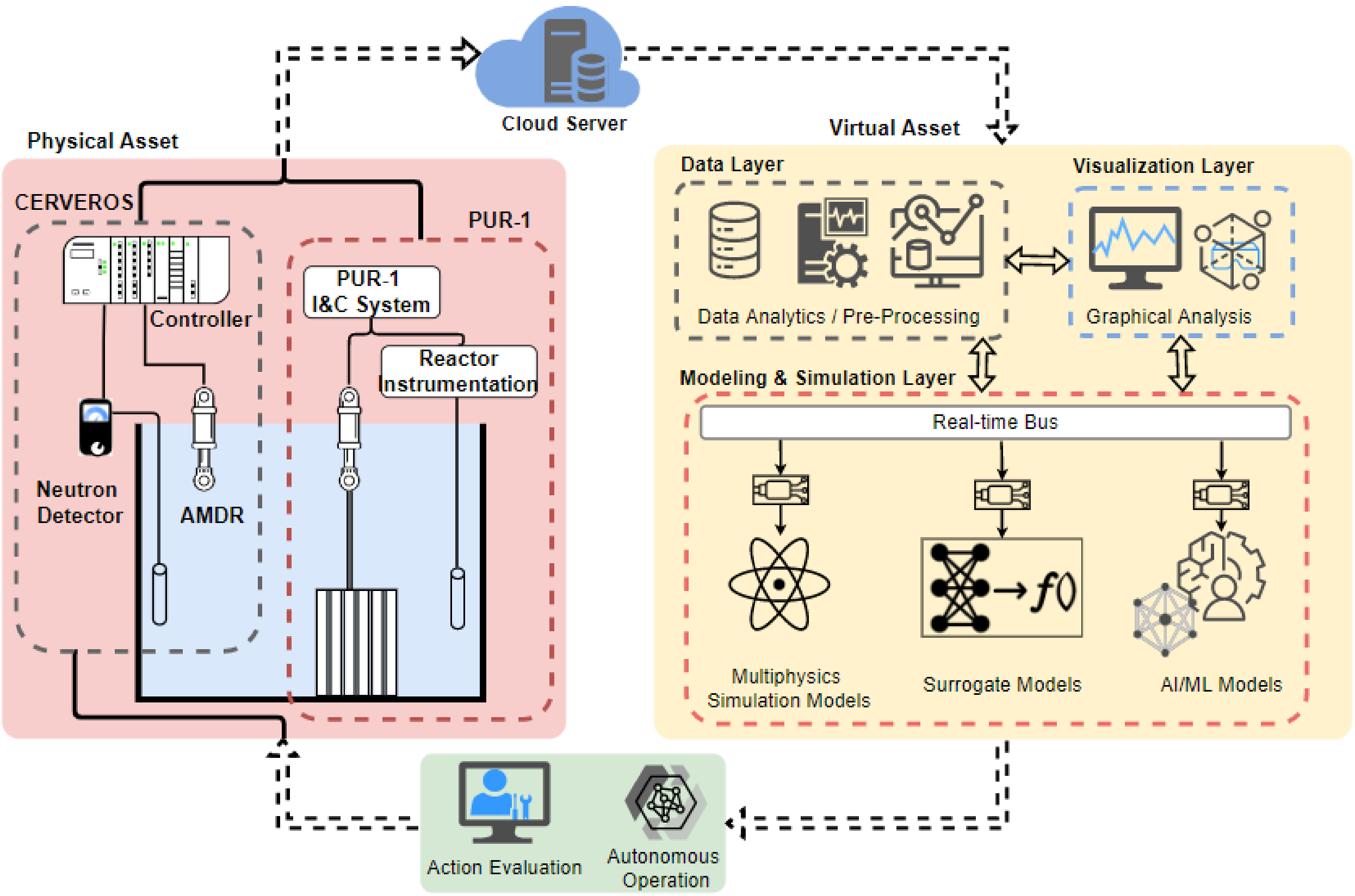

Figure 3: PUR-1 DT configuration

# 6 PUR-1 DT Physical System

As introduced in Section 5, the physical asset of the PUR-1 DT comprises the PUR-1 reactor and the CERVEROS cyber-physical testbed. The following subsections describe each system in detail, along with the data acquisition infrastructure that links them to the virtual asset.

## 6.1 Purdue University Reactor One (PUR-1)

PUR-1 is the first nuclear reactor in the U.S. licensed by the U.S. NRC to operate with a fully digital I&C system. The reactor is mainly used for education purposes and as a gamma and neutron source for various research applications across different scientific fields, such as nuclear engineering, health sciences, agriculture and nanotechnology (Figure 4).

PUR-1 is a heterogeneous non-power reactor cooled by natural circulation, moderated by light water, and surrounded by graphite reflectors. The reactor core consists of sixteen (16) fuel elements contained in aluminum containers and controlled by three (3) control rods (CRs). PUR-1 core is fueled with MTR-type plate fuel of High-Assay Low Enriched Uranium (HALEU) positioned in the fuel assembly (FA) containers. The reactor core is surrounded by Graphite Assemblies (GAs) that constitute the reflector, while control rods are installed within fuel plates in modified FAs [54]. Six of the GAs contain an additional cylindrical tube facilitating radiation exposure experiments. This configuration allows experimental capsules to be positioned at the core boundary and to be activated (Figure 5).

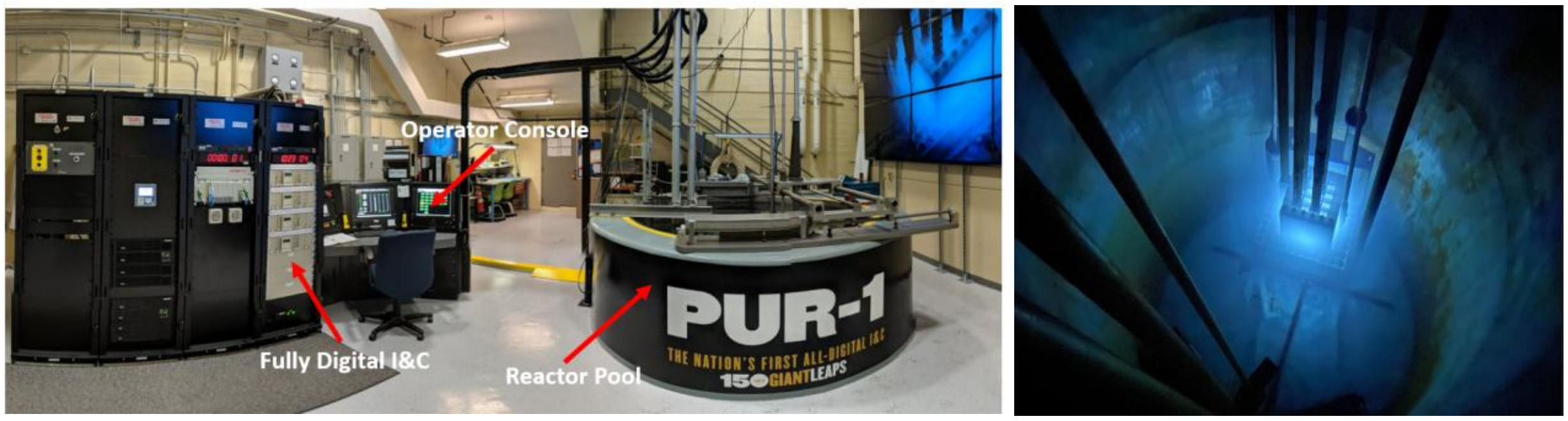

Figure 4: Purdue University Reactor One (PUR-1) facility

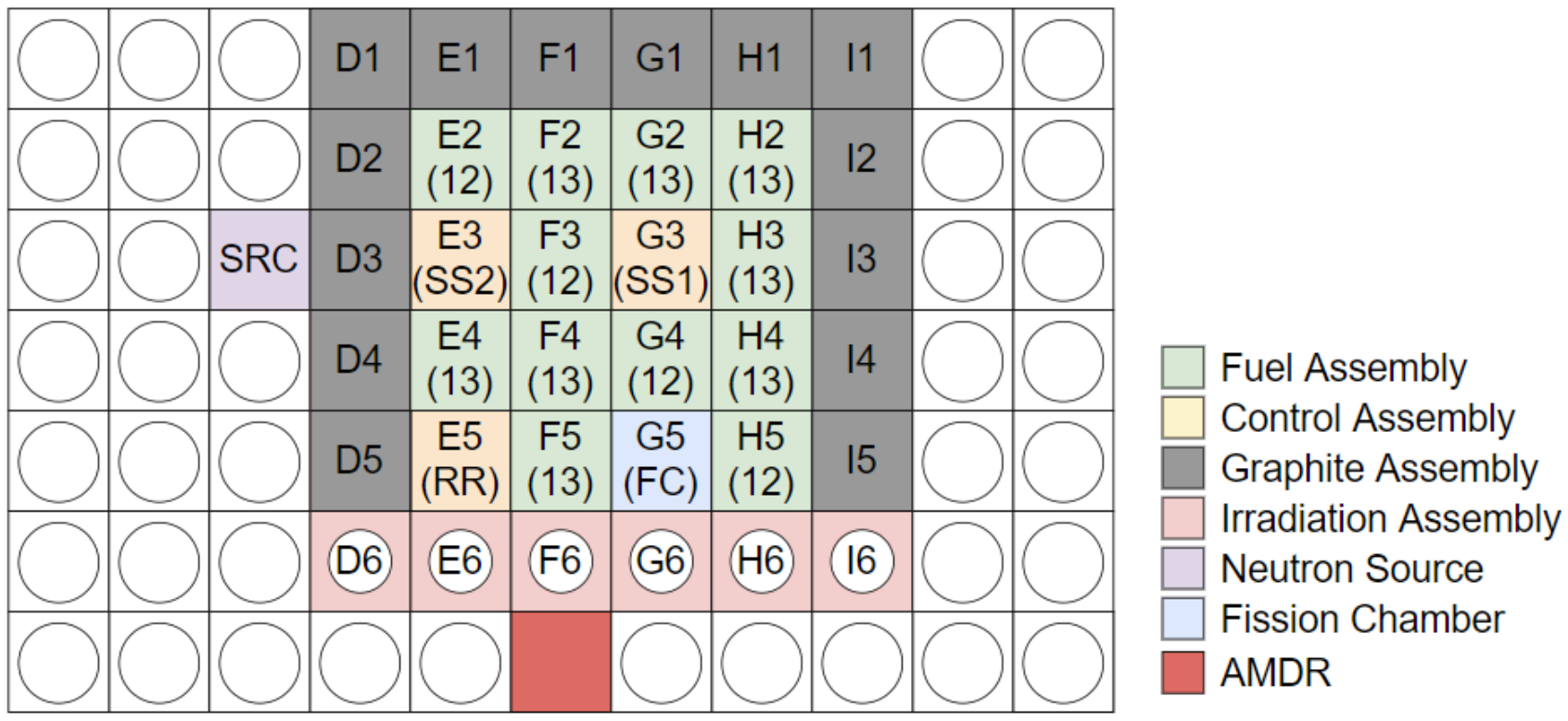

Figure 5: PUR-1 core layout

## 6.2 Cyber-Physical Testbed (CERVEROS)

The Purdue reactor facility includes a cyber-physical testbed installed in the reactor pool called the Cybersecurity Experimental-setup for Remote superVision and Evaluation of Reactor Operation Systems (CERVEROS).

CERVEROS constitutes an additional mechanism of inserting external perturbations into the system by emulating CR movements. The testbed is equipped with an Auxiliary Moderator Displacement Rod (AMDR) moved by a linear actuator (Figure 6), along with several temperature sensors, part of the testbed's instrumentation system. Despite the fact that CERVEROS is installed in the reactor pool, it operates independently without interfering with the reactor I&C system [28]. This enables the bilateral communication with the business network and allows the application and evaluation of actions originating from the DT.

Similarly to conventional control rods, AMDR motion perturbs neutron population by introducing small reactivity changes through moderator displacement. It is constructed out of indium foils enclosed within a sealed plastic container, mounted on an aluminum extension which is connected to the actuator (Figure 6). AMDR's influence on the neutron population is observable through the reactor instrumentation system, allowing the induced perturbations to be detected by the existing neutron detectors [55].

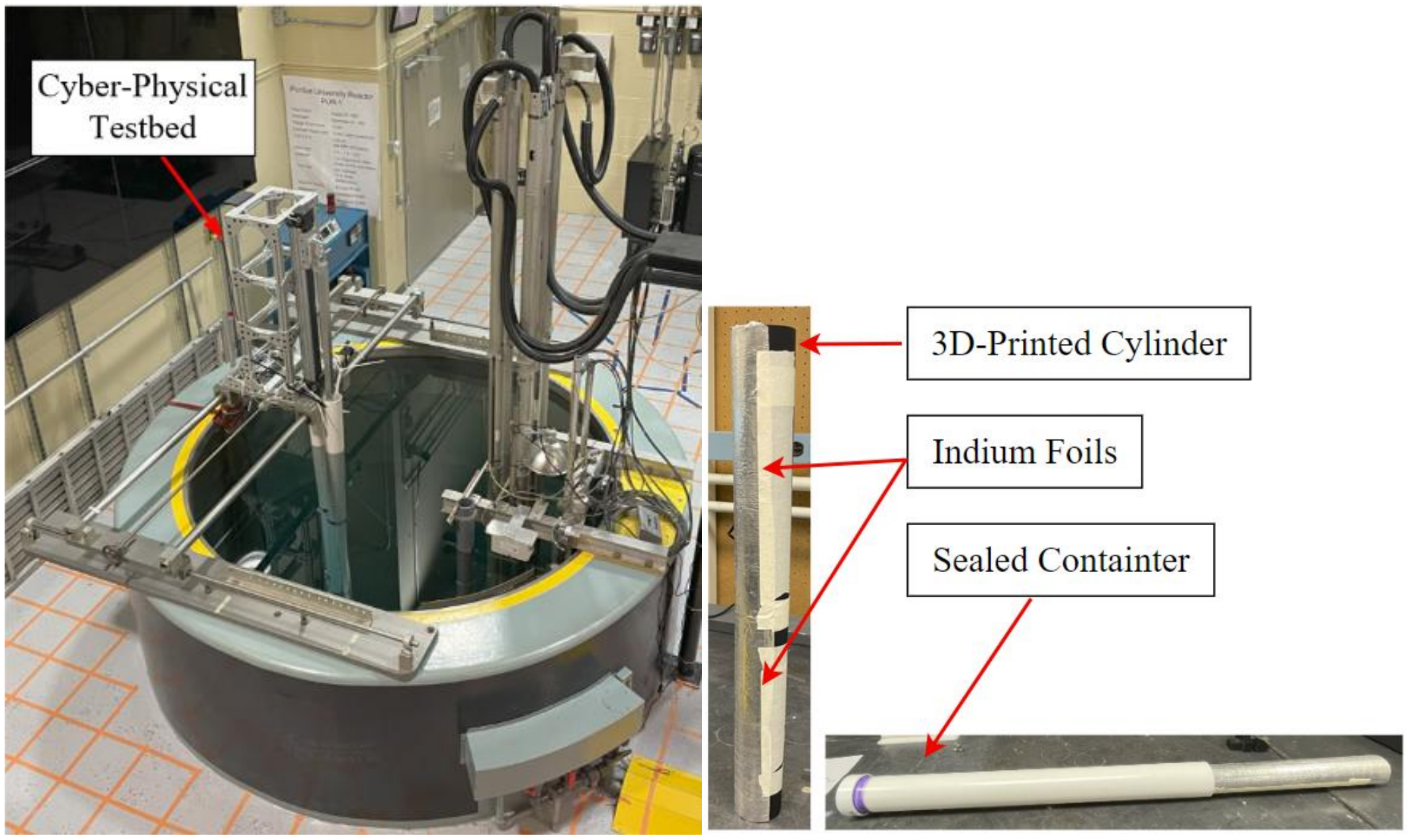

Figure 6: CERVEROS testbed installed in PUR-1 (right) and the AMDR assembly with an indium-filled configuration (left)

The effect of the AMDR is described through the reactivity worth curves [56], characterizing its influence on the inserted reactivity as a function of its position [57]. The AMDR reactivity worth curve is calculated through the In-hour equation as a function of the reactor period calculated from the measured neutron flux [58].

$$\rho = \frac{l_p}{T} + \sum_{i=1}^{6} \frac{\beta_i}{1 + \lambda_i \cdot T} \tag{1}$$

Where $l_p$ is the effective prompt neutron lifetime, $\beta_i$ the fraction of delayed neutrons of $i^{th}$ precursor, $\lambda_i$ the decay constant of $i^{th}$ precursor. The reactor period $T$ is given as:

$$T = \frac{1}{\ln(1 + \Delta CR)} \tag{2}$$

Where $\Delta CR$ the neutron flux change rate collected from PUR-1 instrumentation system. The reactivity worth curve was determined by moving AMDR in small increments and calculating the neutron change rate prior and post movement. During this procedure, the reactor was maintained at critical operation, while the operator applied compensatory actions to offset the reactivity perturbation introduced by the AMDR displacement.

The resulting dataset was fitted using nonlinear least squares with a two-term sinusoidal basis. The dominant sinusoidal component captures the overall variation in reactivity insertion across AMDR travel range, while the second component accounts for smaller deviations from the experimental data. As shown in Figure 7, the fitted curve lies within the experimental uncertainty bounds. The inferred reactivity depends on the neutron population change rate associated with each AMDR displacement (Eqs. 1& 2). When the resulting change in count rate is small relative to the approximately constant uncertainty of the neutron detector, the measurement uncertainty constitutes a larger fraction of the observed response and consequently propagates into a larger uncertainty in the calculated reactivity worth. This behavior is particularly evident at AMDR positions above 30 cm, where the differential worth decreases and successive AMDR displacements induce progressively smaller changes in neutron population.

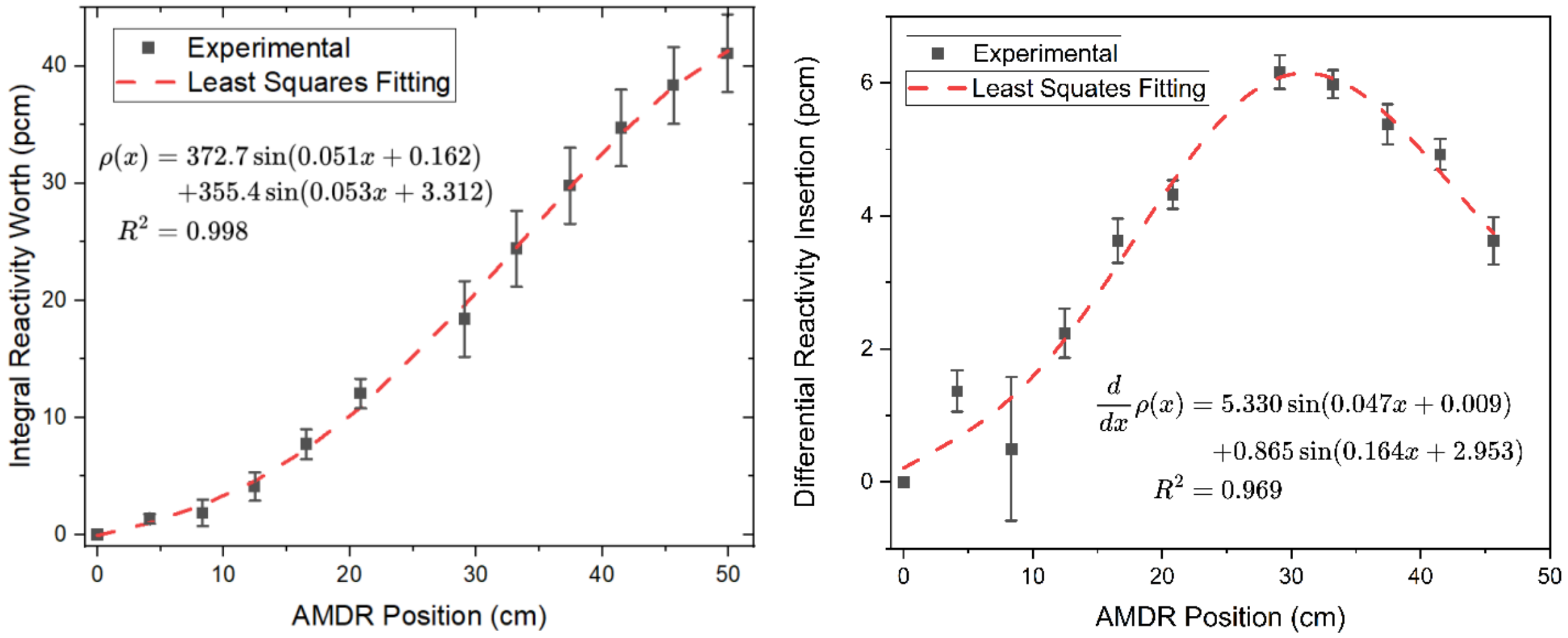

Figure 7: Integral (left) and differential (right) AMDR reactivity worth curve for experimental and fitted values with $3\sigma$ uncertainty

## 6.3 Data Acquisition System

PUR-1 data acquisition system is based on a nuclear-grade digital controller monitoring and controlling more than 2,000 parameters. These include both digital and analog operational signals as well as controller self-diagnostics. Sensors are directly connected to the controller's I/O cards minimizing latencies introduced through intermediate data transfer. Under this configuration, the principal factor governing the operational cycle is the controller frequency. All data are stored in real-time at a historian server. To ensure concurrence with the physical counterpart, the virtual system must operate at the same frequency, enabling real-time state awareness and decision-support for reactor control. Reactor I&C system is subject to regulatory restrictions on information flow implementing a data diode in network architecture [59].

In contrast, CERVEROS cyber-physical testbed operates under an independent digital control architecture that supports bidirectional data exchange with the virtual system. This capability is enabled via an OPC UA server integrated in testbed's controller. Beyond operational considerations, the protocol and its security features provide a practical basis for cybersecurity research, since authentication and encryption schemes can be systematically evaluated within a controlled environment [60]. The testbed I&C architecture adopts a hybrid Purdue model approach to maintain a realistic and defensible security posture consistent with industrial practice [61].

# 7 PUR-1 DT Virtual System

The virtual system of the PUR-1 DT comprises of three main layers: the data layer, modeling and simulation layer, as well as the visualization layer.

## 7.1 Data layer

Prior studies using PUR-1 operational data indicated that real-world data often contain artifacts and irregularities from sensor response times, network latency, calibration drift, or transient variability [62]. Such issues introduce noise, outliers, missing values, and anomalies, which can vary in frequency and amplitude (Figure 8). Outliers can distort normalization and lead to misleading conclusions. Missing or null values pose significant problems for AI/ML models, which are typically not resilient to irregular datasets.

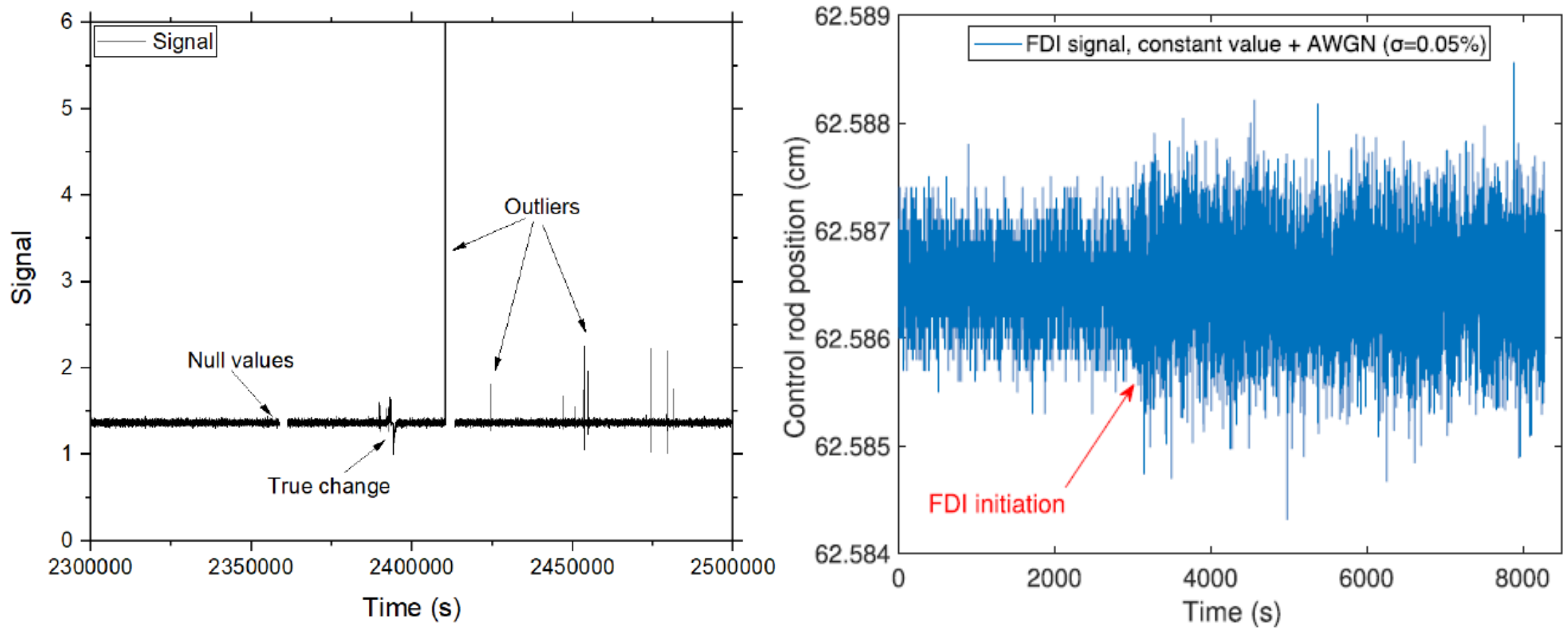

Figure 8: Operational data artifacts demonstrating null values and outliers in comparison with true changes in the physical system [63] (left) and noise analysis for cybersecurity evaluation on operational data [64] (right)

To address these challenges, the data layer includes a suite of preprocessing techniques such as filtering, interpolation, and classification to refine the collected datasets. Electronic noise, for example, may require advanced digital filtering to recover meaningful signals and support cybersecurity evaluations by detecting abnormal noise patterns potentially resulting from interference or sensor malfunctions [64]. Apart from improving measurement quality, the data layer extracts preliminary system insights, including anomaly and fault detection, state identification, and estimation of key operational parameters, thereby ensuring that higher DT layers operate on accurate, consistent, and trustworthy data streams.

## 7.2 Modelling and Simulation layer

The modelling and simulation layer includes a suite of computational models to solve forward and inverse problems.

### 7.2.1 Forward Problem Approach

The finite number and spatial coverage of sensors in PUR-1 limit direct observation of several reactor quantities, including spatial neutron-flux, temperature, and coolant-flow distributions. The forward-model framework therefore combines measured reactor data with physics-based models to reconstruct quantities that are not directly available through the reactor instrumentation system. As shown in Figure 9, the framework couples Multiphysics, Monte Carlo neutron-transport, and point-kinetics models to represent the thermal-hydraulic, neutronic, and time-dependent behavior of the reactor. CR positions, reactor power, and bulk pool temperature are acquired from the installed sensors and used to define the current model state, initial and boundary conditions. Processed inputs are initially fed into the CFD module to capture thermal hydraulics effects. Since PUR-1 does not employ a forced cooling system [65], the governing field equations (Eqs. 3, 4 and 5) [1] are solved under the Boussinesq approximation (Eq. 4), where temperature-induced density variations generate buoyancy forces that drive natural circulation within the reactor pool:

$$\nabla \cdot \mathrm{u} = 0 \tag{3}$$

$$\rho_0 \left( \frac{\partial \mathrm{u}}{\partial t} + \mathrm{u} \cdot \nabla \mathrm{u} \right) = -\nabla p + \mu \nabla^2 u - \rho_0 \beta (T - T_0) g \tag{4}$$

$$\rho_0 \left( \frac{\partial T}{\partial t} + u \cdot \nabla \mathrm{T} \right) = \mathrm{k} \nabla^2 T + \dot{q}''' \tag{5}$$

where

$$\beta = -\frac{1}{\rho_0} \left( \frac{\partial \rho}{\partial T} \right)_p \tag{6}$$

is the isobaric thermal expansion coefficient.

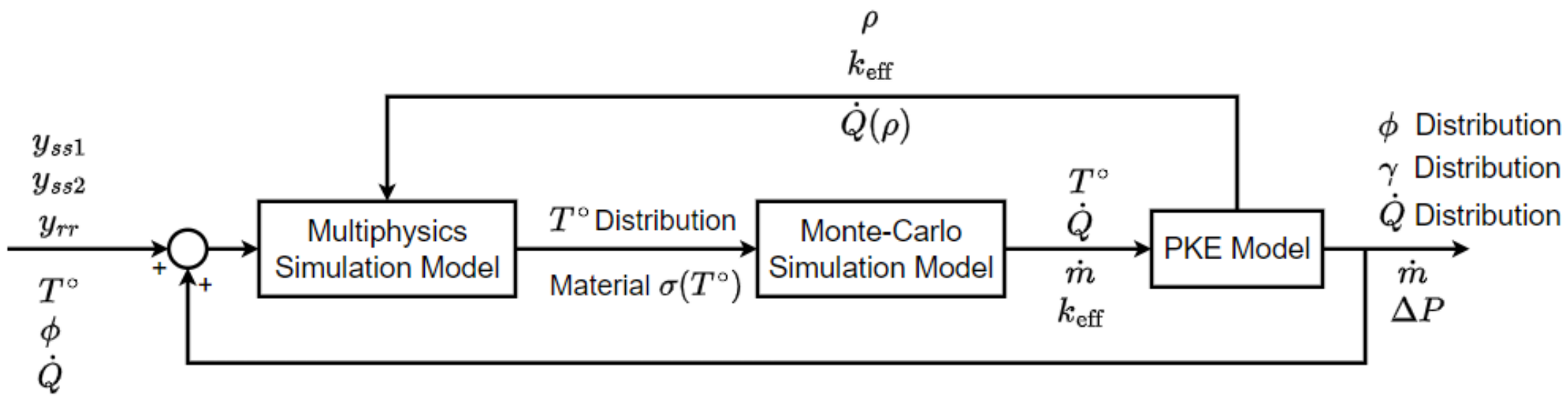

Figure 9: Forward problem approach of PUR-1 DT

The solutions obtained from a Multiphysics CFD model compensate for limited in-core temperature sensing by collocating quantities of interest such as component-wide temperature, heat, and coolant flow distributions, along with localized thermal hotspots. The resulting thermal field mapped on a geometric representation of PUR-1 [2] at the current timestep informs a Monte Carlo neutron and photon transport module. Since all control rod (SS1, SS2, and RR) positions and material temperatures influence neutron and gamma interaction (e.g. radiative capture, elastic/ inelastic scattering, etc.) probabilities, temperature-dependent cross sections are employed to stochastically solve the Eigenvalue formulation of the stationary Boltzmann transport equation (Eq. 7) [58, 66].

---

[1] Mass (Continuity), Momentum (Navier Stokes), and Energy Balance Equations respectively

[2] As necessitated by the heterogeneity of PUR-1's core, the Boltzmann equation governing neutron transport and distribution is solved with unambiguous accounting of all geometric domains (volumes and surfaces) through constructive solid geometry and repeated random sampling (Monte Carlo).

$$\mathbb{L}\varphi = \frac{1}{k_{eff}} \mathbb{F}\varphi \tag{7}$$

$$\mathbb{L}\varphi(r, E, \Omega) = \Omega \cdot \nabla\varphi(r, E, \Omega) + \Sigma_t(r, E)\varphi(r, E, \Omega) \tag{8}$$

$$\mathbb{F}\varphi(r, E, \Omega) = \chi(\mathrm{E}) \int_0^{\infty} \oint_{4\pi} \nu\Sigma_f(\mathrm{r}, \mathrm{E}')\phi(\mathrm{r}, \mathrm{E}', \Omega')\mathrm{d}\Omega'\mathrm{d}\mathrm{E}' \tag{9}$$

Where Eqs. 8 and 9 are the transport (loss) and fission (source) operators, respectively. The open-source OpenMC [67] Monte Carlo code is leveraged to solve this yielding effective multiplication factor ($k_{eff}$), neutron, gamma, and heating distributions. The corresponding power profile at a given node or voxel is then computed from [68, 69]:

$$P_i = \int_{V_i} q'''(r)dV \tag{10}$$

with

$$q'''(r) = \int_0^{\infty} \phi(r, E)\Sigma_H(r, E)dE \tag{11}$$

and

$$\Sigma_H(r, E) = \sum_i N_i(r) \sum_{\alpha} \kappa_{i,\alpha}(E) \tag{12}$$

The power profile is then returned to the Multiphysics model to update the thermal field. The updated temperatures are then fed back to Monte Carlo model, and the coupling is iterated until convergence. While these models capture steady-state spatial behavior, temporal solutions during reactor transients are resolved through the Point Kinetics Equations (PKE) model. Triggered manually per use-case or automatically when PUR-1's measured power ramp exceeds 0.5%/s, this module employs a deterministic system of coupled first-order linear differential equations (Eqs. 13 & 14) [58, 70].

$$\frac{dn}{dt} = \left[\frac{\rho(t) - \beta}{\Lambda}\right] n(t) + \sum_{i=1}^{6} \lambda_i C_i(t) \tag{13}$$

$$\frac{dC_i}{dt} = \frac{\beta_i}{\Lambda} n(t) - \lambda_i C_i(t) \tag{14}$$

This module advances the in-core neutron population in time by resolving prompt and delayed contributions. Once inferred upon, provides a dynamic solution to $k_{eff}$, nominal power, and reactivity based on CR motion, reactivity worth curves, and delayed neutron precursor parameters. The PKE module then feeds this information back to the OpenMC-COMSOL convergence loop to refine spatial resolution of updated physical parameters (Figure 10). In other words, the PKE module regulates the frequency of a quasi-steady-state coupling scheme during modeled and physical reactor transients.

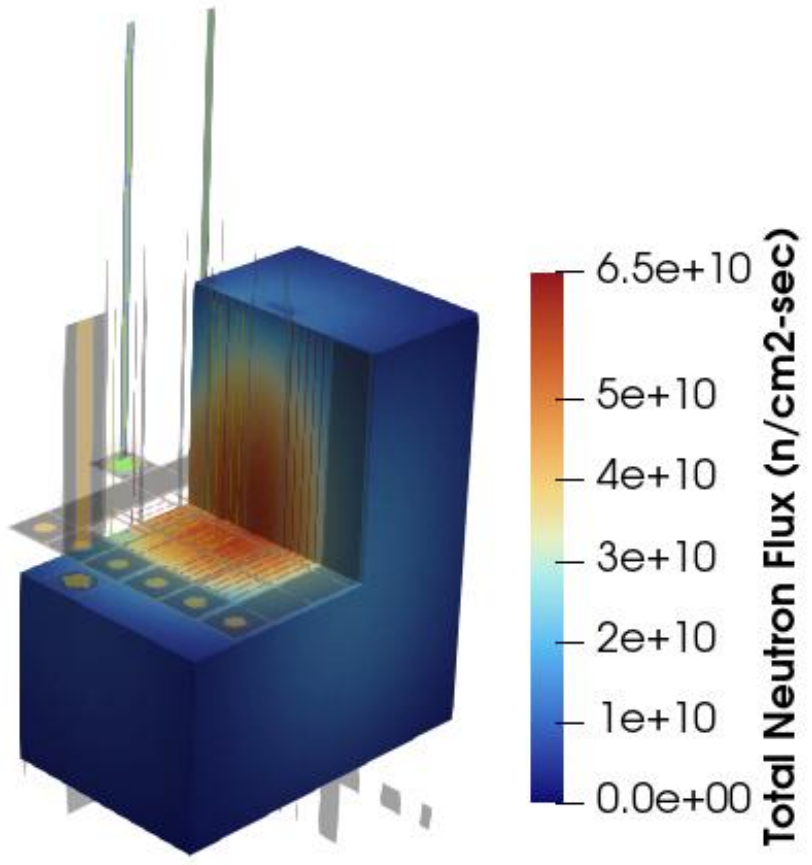

a)

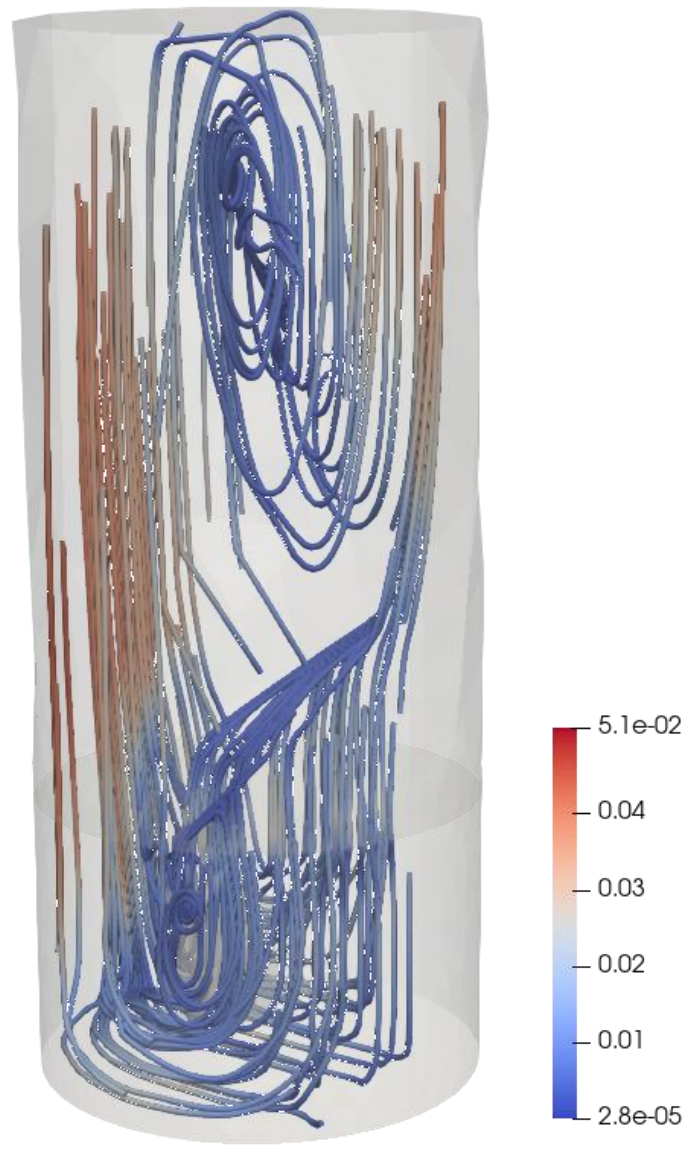

b)

Figure 10: Forward problem approach outcome with a) neutron flux distribution prediction and b) fluid flow development in the reactor pool at a 10% power

### *7.2.2 Inverse Problem Approach*

Reactor operations progress a physical state of PUR-1 towards its desired state. PUR-1 DT plays a supporting role by evaluating action consequences via data-assimilated predictive and diagnostic analytics modules with forward problem solutions. These operations estimate and solve inverse problems, thus prescribing optimal actions in support. First, the forward problem estimates steady-state or transient geometric boundary conditions of the core from an operator-defined target state and an initial digital state (e.g., nominal power level or neutron and gamma flux distributions). Alternatively, the initial digital state may be defined by reactor instrumentation, representing the observed physical state of PUR-1 at the current timestep and satisfying all assumptions and boundary conditions required to solve the inverse problem

As shown in Figure 11, the inverse model provides a pathway toward achieving this target state. The manner of achieving this state optimally with short- and long-term consequence awareness is the desired output of the inverse model. A simple example of this can be the knowledge of how multiple time-dependent PUR-1 CR trajectories may achieve a final physical state that is conditionally similar within its initially prescribed target state. Here, the inverse-model is deployed in parallel with the predictive forecasting and prognostics modules as described in the following subsections. Hence, PUR-1 DT resolves the target state as a constraint on the solution space and employs objective functions to discriminate among feasible trajectories, either derived from downstream reactivity effects (e.g., minimizing reactor poisoning, reducing fuel depletion rate, and maximizing refueling cycle length) or reactor safeguards (e.g., cybersecurity, defense-in-depth, etc.).

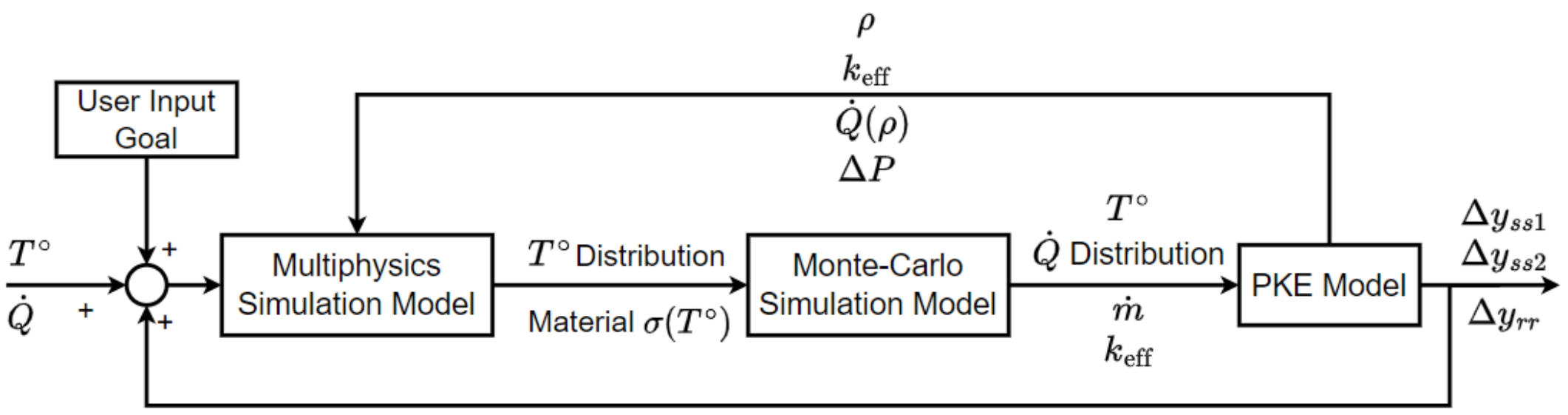

Figure 11: Inverse Model of PUR-1 DT

### 7.2.3 *Anomaly Detection, Explainability, and Control Modules*

The PUR-1 DT employs a cybersecurity module which includes artificial intelligence and machine learning (AI/ML) tools for real-time anomaly detection. For example, using the cybersecurity module we showed that Random Forest (RF) models can distinguish between normal, abnormal, and cybersecurity-related events monitoring operational (OT) and information technology (IT) data [63]. In addition, an eXplainable AI (XAI) module based on the SHAP framework and adapted to multivariate time-series data is also included to provide temporally informed explanations of reconstruction-based or predictive model outputs, supporting the localization and interpretation of abnormal system behavior (e.g., Figure 12) [71, 72].

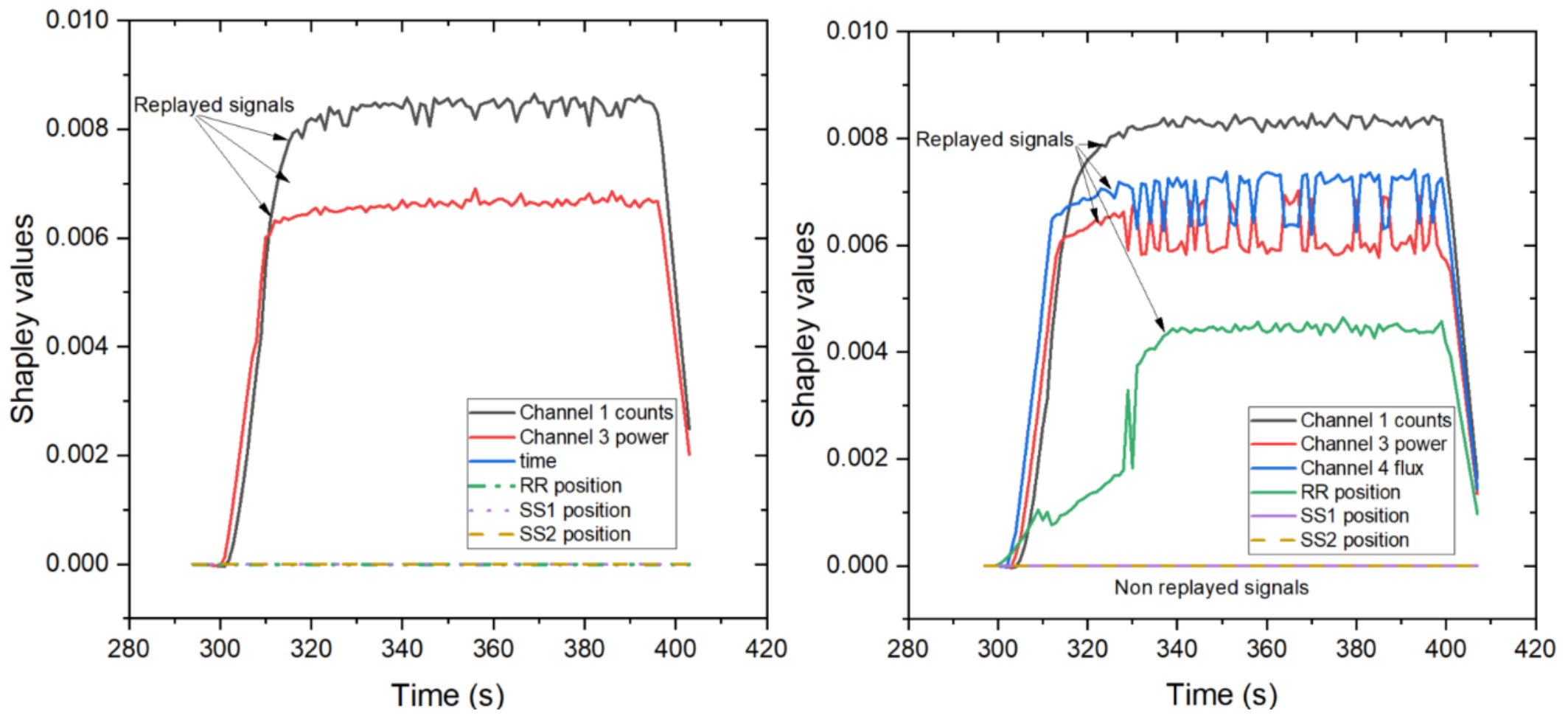

Figure 12: Shapley values profiles per feature during an FDI attack interfering with 2 signals (left), and 4 signals (right) as reported in [72].

The DT supports automated and semi-autonomous operation of the AMDR through the integration of various control strategies, e.g., Proportional Integral Derivative (PID), Reinforcement Learning, and model predictive control with moving horizon estimation (MPC–MHE) [73]. Estimated and predicted reactor states derived from the digital system and operational data, are provided to the controllers to determine the optimal actuation required to track a predefined power demand (Figure 13). The PID controller generates AMDR commands from the deviation between measured and target power, w hereas MPC–MHE predicts the reactor response over a finite horizon (Figure 14). The resulting AMDR trajectory is optimized subject to limits on reactor power, reactivity insertion, and allowable power-change rate. Through this capability, the PUR-1 DT provides a framework for evaluating and comparing supervised automated control and progressively autonomous reactor operation strategies.

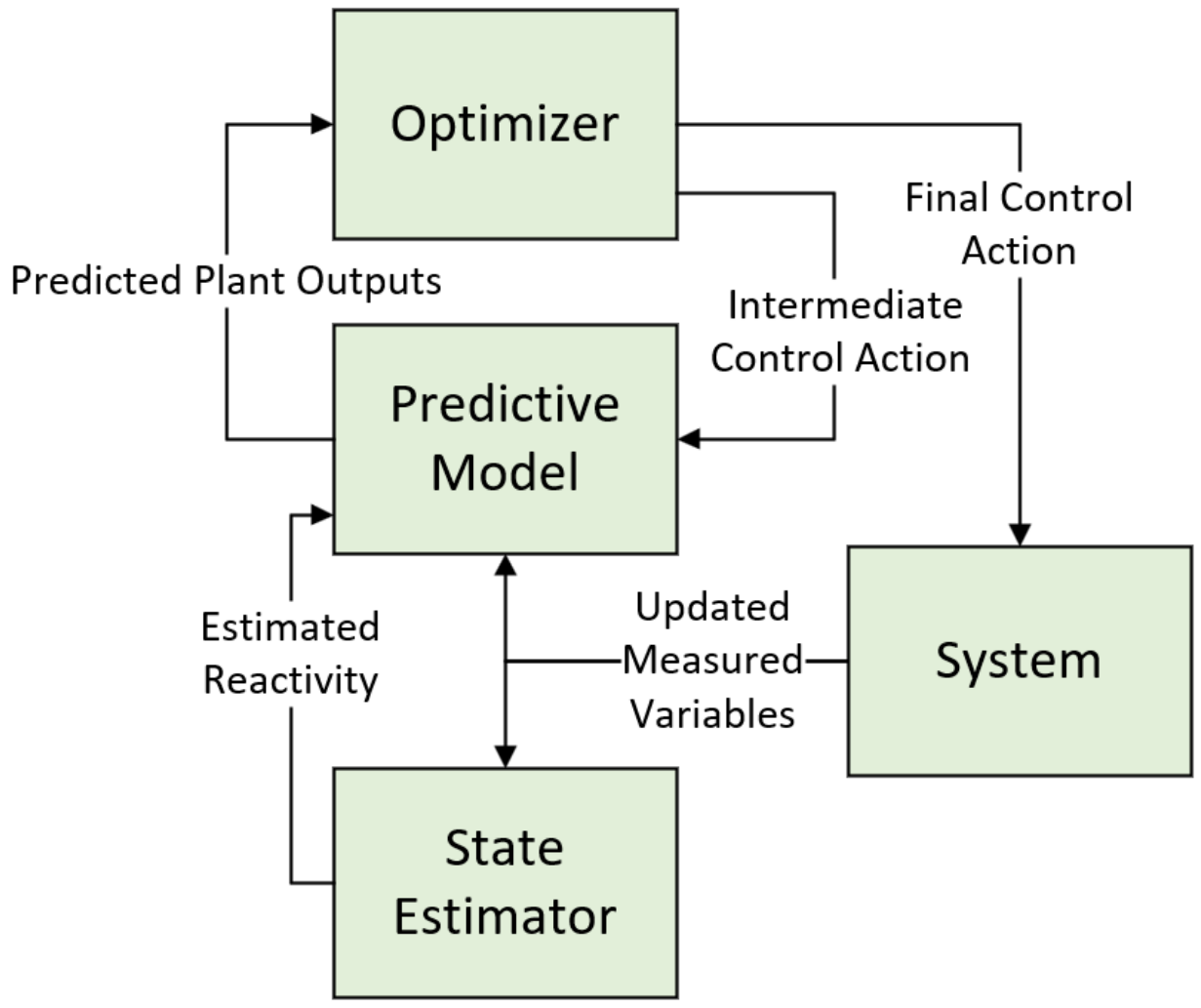

Figure 13: Graphical description of control framework [73]

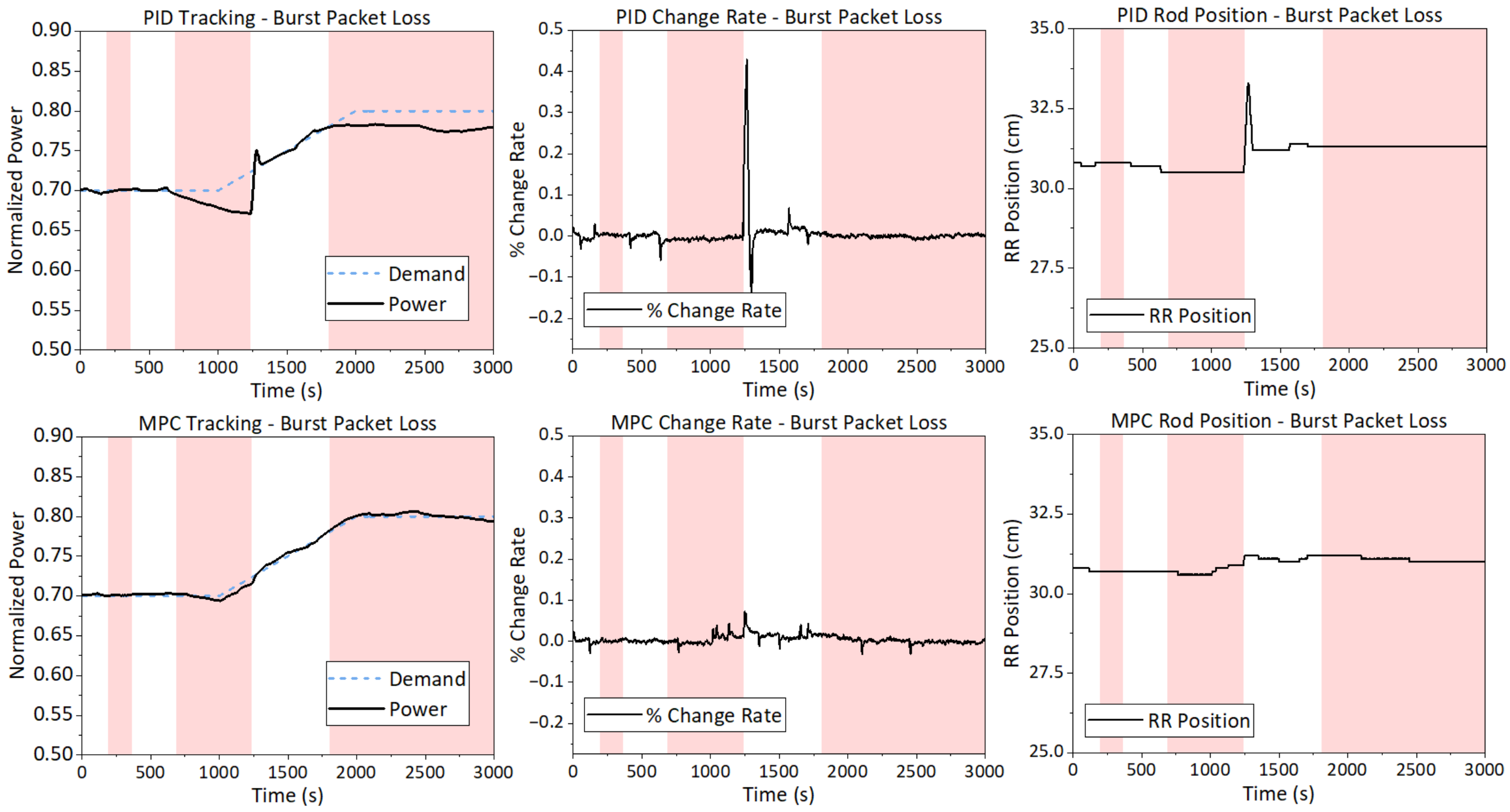

Figure 14: PID (top) and MPC (bottom) during burst packet loss with a change in setpoint (areas shaded in red indicate packet loss bursts)

*State estimation*

To explore reactor-state awareness strategies during temporary communication losses, the PUR-1 DT employs a Kalman-filter-based state estimator. Under normal operation, incoming measurements continuously correct the estimated neutron density and delayed-neutron precursor concentrations [74].

During packet loss or communication outages, the estimator module propagates these states using the reactor dynamic model until measurements are restored. The KF is formulated as:

$$x_{k+1} = A(\rho_k)x_k + w_k, y_k = Cx_k + v_k \quad (15)$$

with process noise $w_k \sim N(0, Q)$ and measurement noise $v_k \sim N(0, R)$. The KF performs prediction, innovation computation, and update at each scan:

$$\hat{x}_{k+1|k} = A_{d,k}\hat{x}_{k|k}, P_{k+1|k} = A_{d,k}P_{k|k}A_{d,k}^T + Q_{d,k} \quad (16)$$

$$K_{k+1} = P_{k+1|k}C^T\left(CP_{k+1|k}C^T + R\right)^{-1} \quad (17)$$

$$\hat{x}_{k+1|k+1} = \hat{x}_{k+1|k} + K_{k+1}\left(\tilde{\tilde{y}}_{k+1} - C\hat{x}_{k+1|k}\right), \qquad P_{k+1|k+1} = (I - K_{k+1}C)P_{k+1|k} \quad (18)$$

When measurements are unavailable due to packet loss, the KF propagates the state prediction as $\hat{x}_{k+1|k+1} = \hat{x}_{k+1|k}$. The KF module can also be integrated into the automated PID controller operation serving as a measurement proxy during communication dropouts. Anomalies are detected using a chi-square test, while the innovation sequence is normalized by its innovation covariance following a chi-square distribution [75–77]:

$$g_k = v_k S_k^{-1} v_k^T \sim \chi^2(n) \quad (19)$$

where $v_k$ is the innovation, $S_k$ is its covariance, and $n$ are degrees of freedom. Thresholds are set based on desired false alarm probability (e.g., 95% confidence). The anomaly detector produces binary alarms that can be aggregated over time to compute an anomaly rate, distinguishing transient setpoint changes (nominal) from persistent deviations due to network faults or adversarial actions. An alarm is raised if:

$$g_k > \chi_{1-\alpha}^2 + \delta_a, \qquad for \;\; \delta_a \in \mathbb{R}^+ \quad (20)$$

For example, for α=0.05 (5% false alarm probability) and one degree of freedom, then $\chi_{1-\alpha}^2 = 3.84$. State estimation module operation and anomaly detection feature were evaluated under real operational scenarios indicating high alignment with real data (Figure 15) and higher accuracy modules included in the virtual system (e.g., PKE model, SMs).

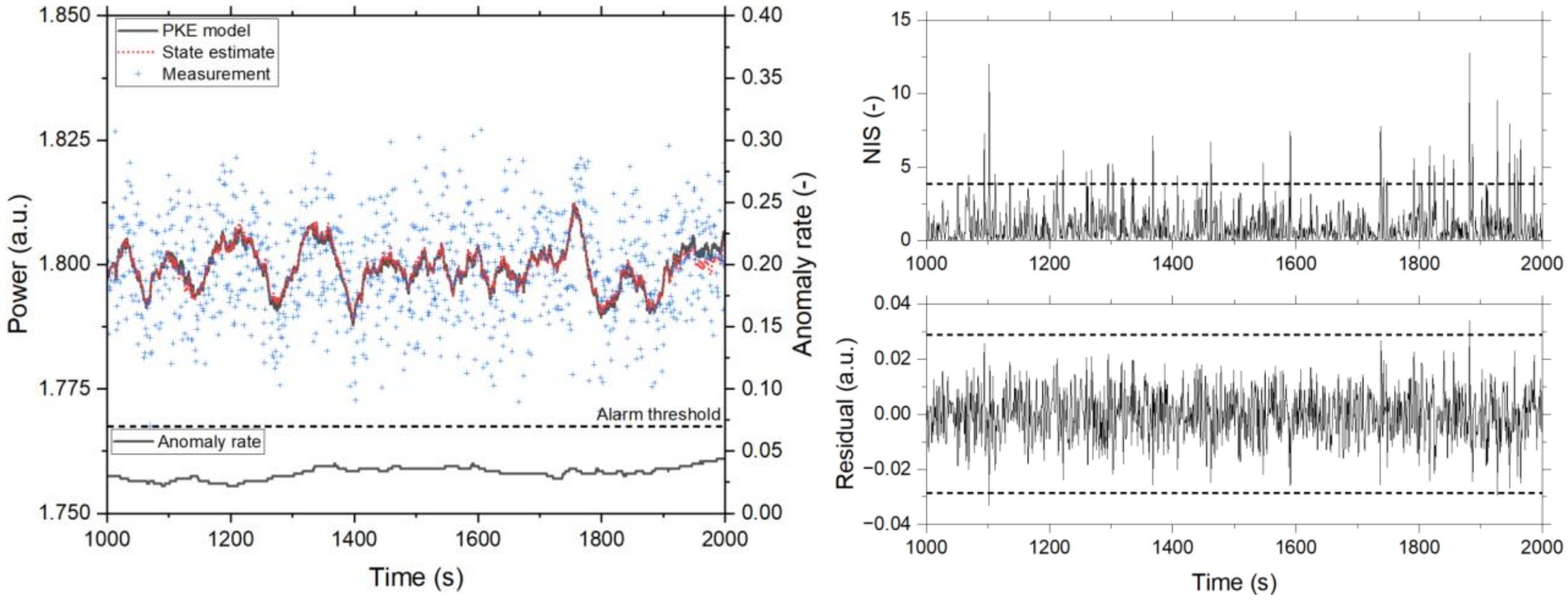

Figure 15: System response and anomaly detection under nominal conditions.

*Network communications modelling*

Supplementary to the state estimation module, an additional model replicating physical system internal communications is also included in the PUR-1 DT. This module emulates all communication between sensors, the controller, actuators and communication protocols (Figure 16) as a discrete stochastic process providing anomaly detection for internal communication discrepancies such as packet loss, latency, and quantization (Figure 17). The communication is modeled as:

$$\tilde{y}_k = y_{k-d_k} + h_k^{(net)} \tag{21}$$

$$\tilde{u}_k = u_{k-d_k} + h_k^{(net)} \tag{22}$$

where $h_k^{(net)} \sim \mathcal{N}(0, \sigma_{net}^2)$ models additive network noise, and $d_k$ varies per packet due to serialization, propagation, and queuing effects. Jitter is represented as stochastic variability in $d_k$. Each device (actuator, controller, KF, etc.) receives a signal with its own $(p_l, d_k, \sigma_{net}^2)$. Table IV shows typical properties of various communication protocols commonly found in nuclear instrumentation and control systems.

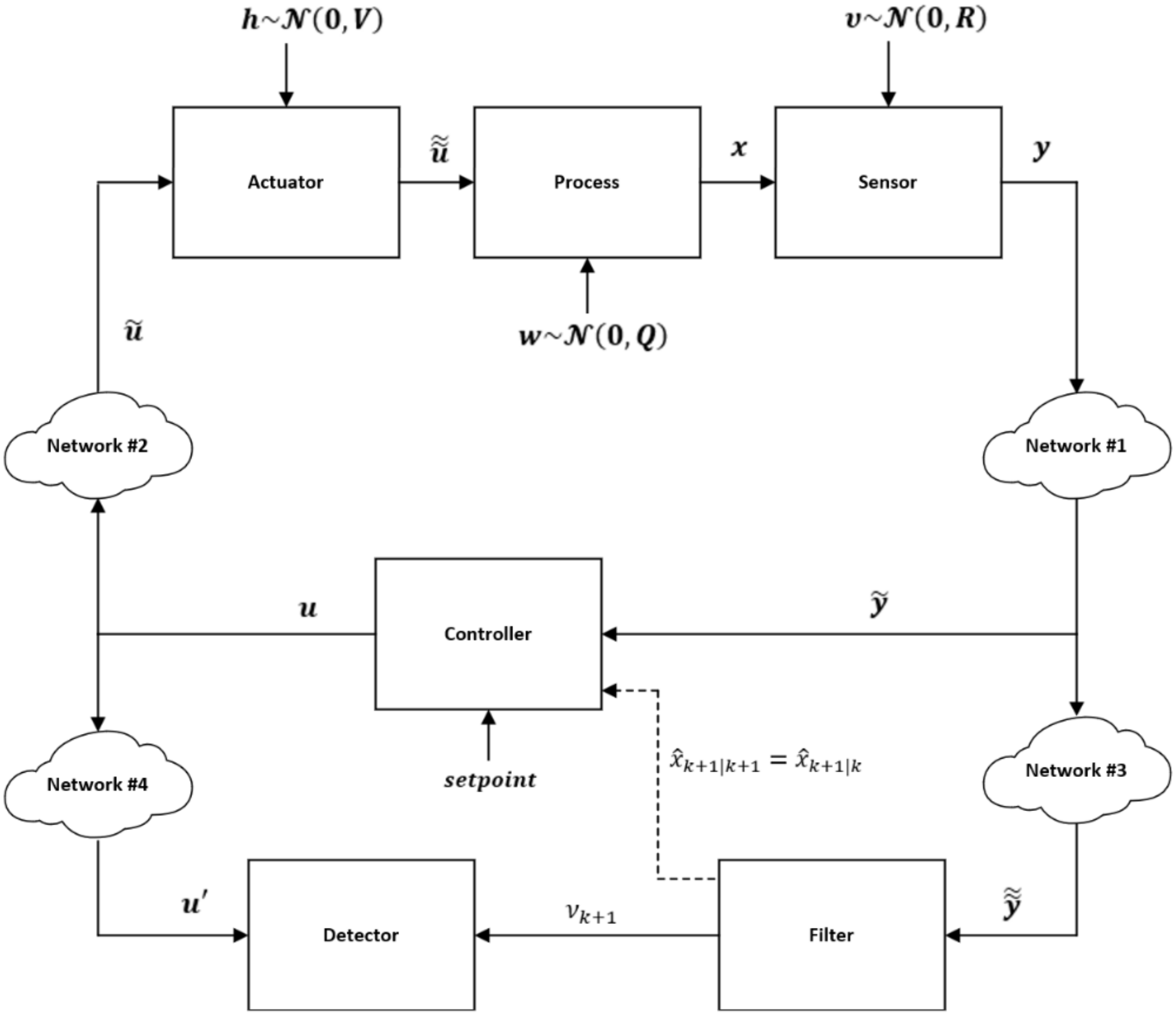

Figure 16: Schematic description of the networked reactor control system model.

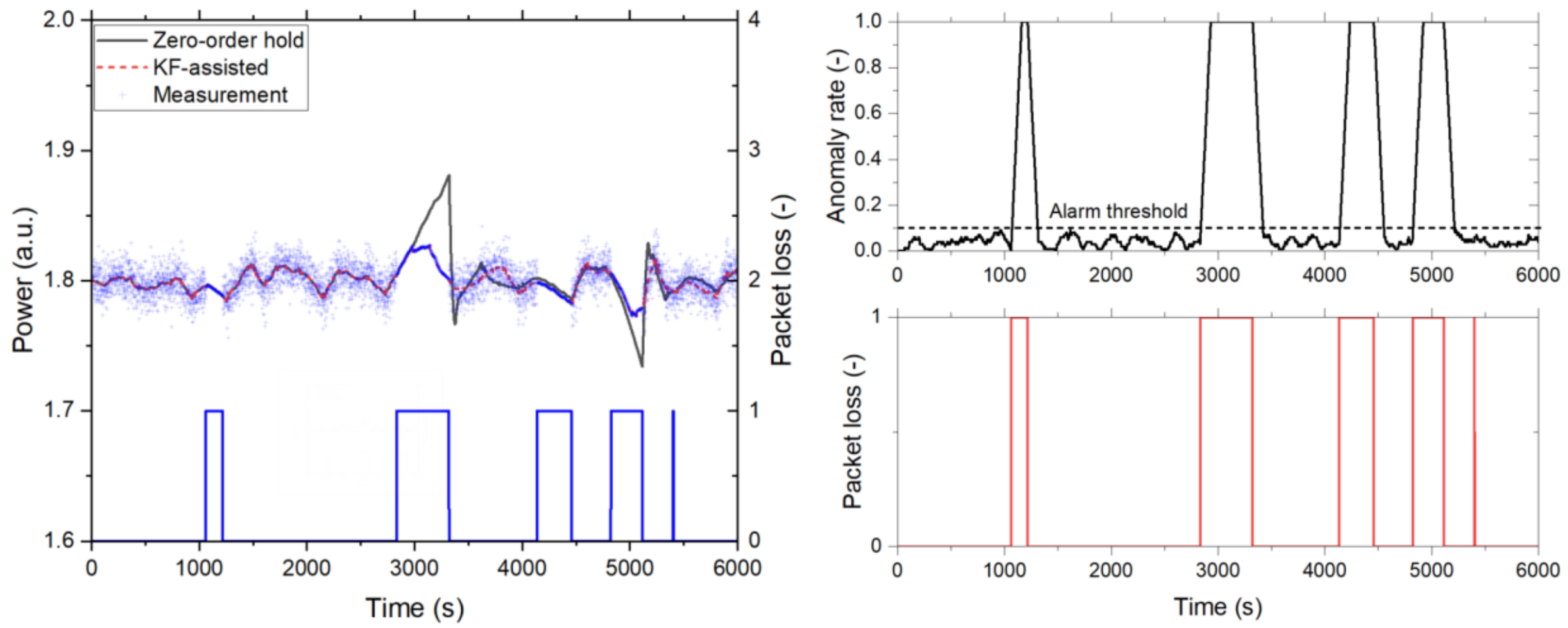

Figure 17: System response and anomaly detection under steady state conditions for medium burst packet loss. Blue vertical lines indicate instances of packet loss.

Table IV: Commonly used industrial communications protocols and properties.

| Protocol | Physical layer | Max devices | Max speed | Typical latency | Max distance | Bit error rate | Quantization Resolution |
|---|---|---|---|---|---|---|---|
| Modbus RTU | RS-485 | 32-128 | 115.2 kbps | 6-20 ms | 1200 m | $10^{-9}$ | 12-16 bits |
| Modbus TCP | Ethernet | 255+ | 100/1000 Mbps | 1-5 ms | 100 m | $10^{-12}$ | 14-16 bits |
| EtherCAT | Ethernet | 1000+ | 100 Mbps | 0.01-1 ms | 100 m | $10^{-12}$ | 16-24 bits |
| Profibus DP | RS-485 | 126 | 12 Mbps | 1-5 ms | 100 m | $10^{-9}$ | 12-16 bits |

#### *7.2.4 Forecasting*

Detailed high fidelity physics-based simulations hinder DT operations due to the long elapsed computational time. As an example, Monte Carlo and Multiphysics simulation models typically require 2 hours and 8.6 hours respectively when executed in high performance computing clusters, indicating that simulation output cannot be implemented along virtual asset operation in real time. To avoid the inherent delay introduced by simulations, Reduced Order Models (ROMs) and Surrogate Models (SMs) are employed in the virtual system to estimate various quantities of interest such as the effective multiplication factor ($k_{eff}$) and the neutron flux distribution in the reactor core for the current and future system states.

To create a dataset capable of representing all operational scenarios, a series of simulation runs were conducted under varying operational conditions. Multiple runs were executed modifying CR heights in increments of 0.1 cm to sample the configuration space with sufficient resolution and capture the corresponding changes in core neutronics. Similarly, for adequate representation of thermal state in the PUR-1 pool, a separate sweep was performed over a range of reactor power levels and initial conditions, providing information for components temperature.

## 7.3 Visualization layer

*The visualization layer* includes the user interface, the link for users to interact with the virtual system. User interface in a DT is the primary pathway for an operator to access, interpret, and interact with complex data and simulations incorporated into the virtual system [78, 79]. Through this layer, all available information is organized in a way that is fully comprehensive and intuitive to the operator employing graphical analysis, dynamically updated 3D models, or even augmented and mixed reality technologies. Overall, visualization layer acts as the backend of the user interface which bridges the gap between data and actionable insights and accelerates decision processes.

All DT information is provided to the operator through an interactive graphical user interface (GUI) that enables monitoring and supervisory control of the PUR-1 DT. The GUI is organized into architecture-specific displays selectable from a summary menu and provides a unified operational view by integrating real-time OT data with DT module outputs, thereby supporting consistent situational awareness of both reactor and virtual asset state (Figure 18). A dedicated panel presents consequence analysis and recommended actions derived from DT modules for operator decision support, with the option to apply selected actions to the CERVEROS testbed for controlled evaluation. The interface also supports model selection within the DT workflow, allowing the operator to switch between alternative simulation models (e.g., replacing the default OpenMC neutronics model with MCNP-based) and SMs according to the scenario of interest and associated accuracy of the selected model (Figure 19).

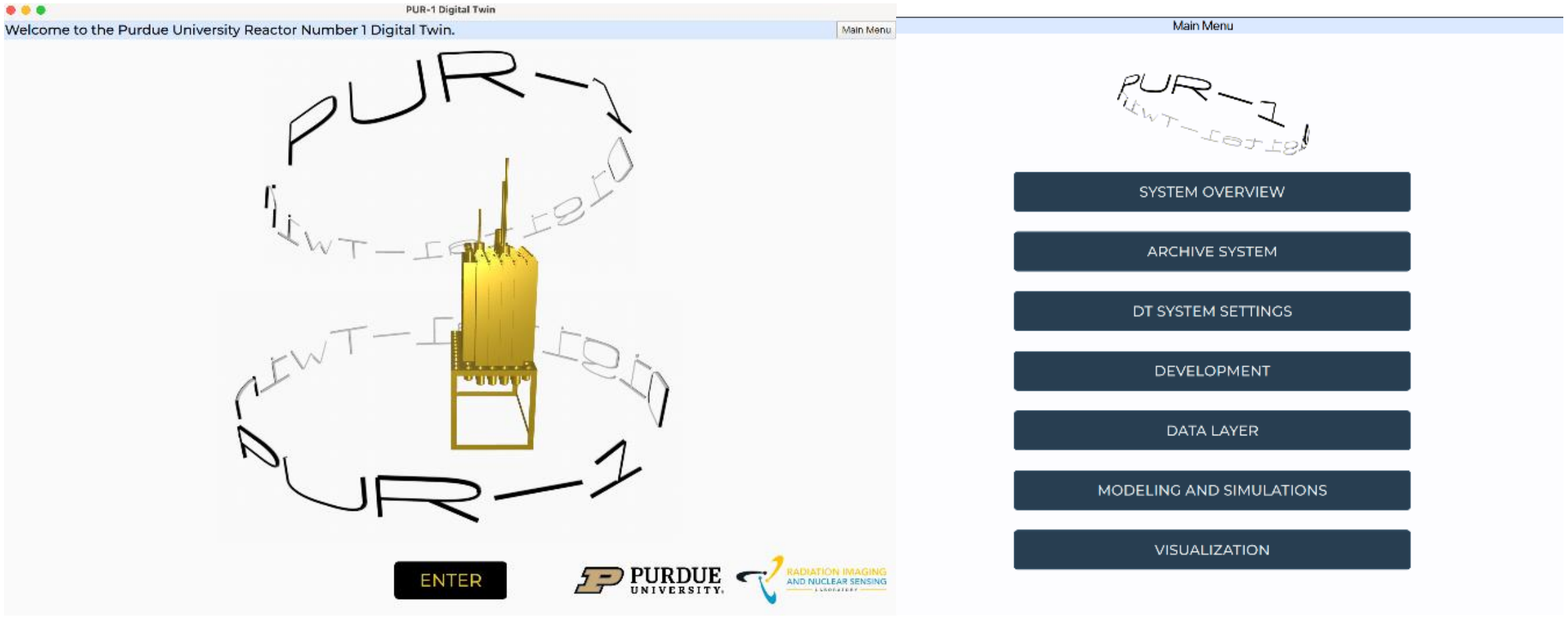

Figure 18: PUR-1 DT main display of user interface.

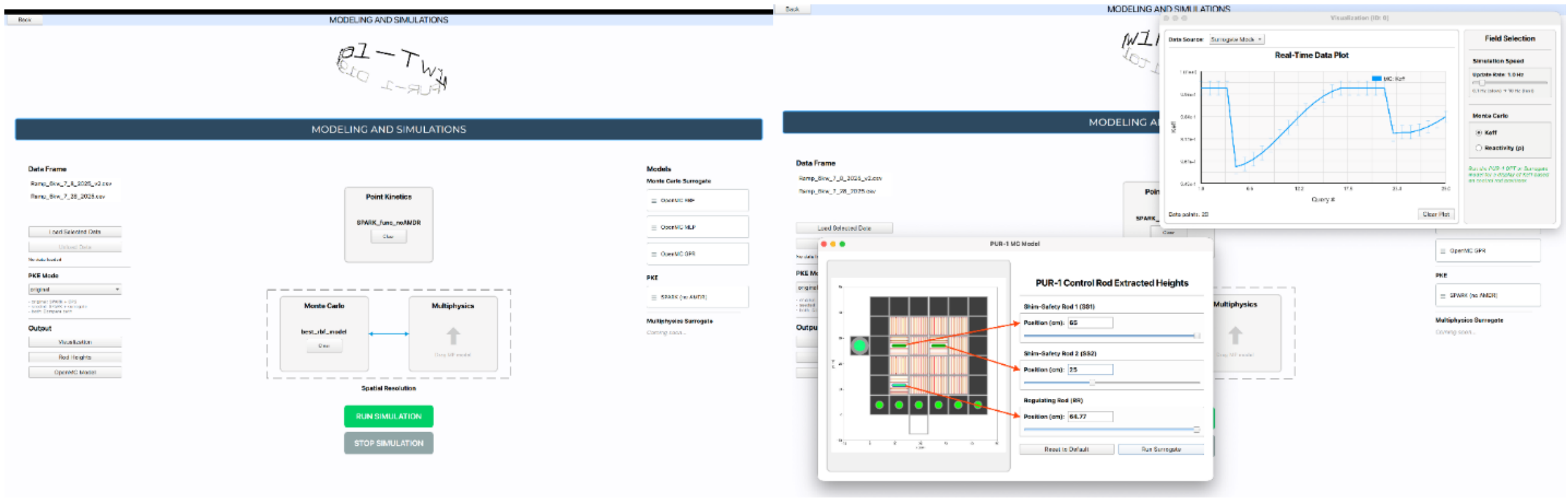

Figure 19: Modeling and Simulation layer display of user interface.

Besides user interface capabilities, the developed DT can be accessed through glass-top display screens located in the DT demonstration room, enabling a more intuitive interaction with the system (Figure 20). The room is configured to resemble a modern nuclear reactor control room and allows users to interact with the DT both online, during reactor operation, and offline, for educational and training purposes. Furthermore, the PUR-1 DT has been integrated into an augmented-reality environment using dedicated Meta Quest Pro headsets. Using the high-fidelity virtual representation of the reactor and CAD models of PUR-1 components, the AR environment replicates the reactor facility configuration and provides an immersive visualization of the system (Figure 21). By integrating PUR-1 DT in the AR environment, control-rod positions, neutron flux distributions, and temperature fields can be dynamically updated and visualized as they evolve during reactor operation.

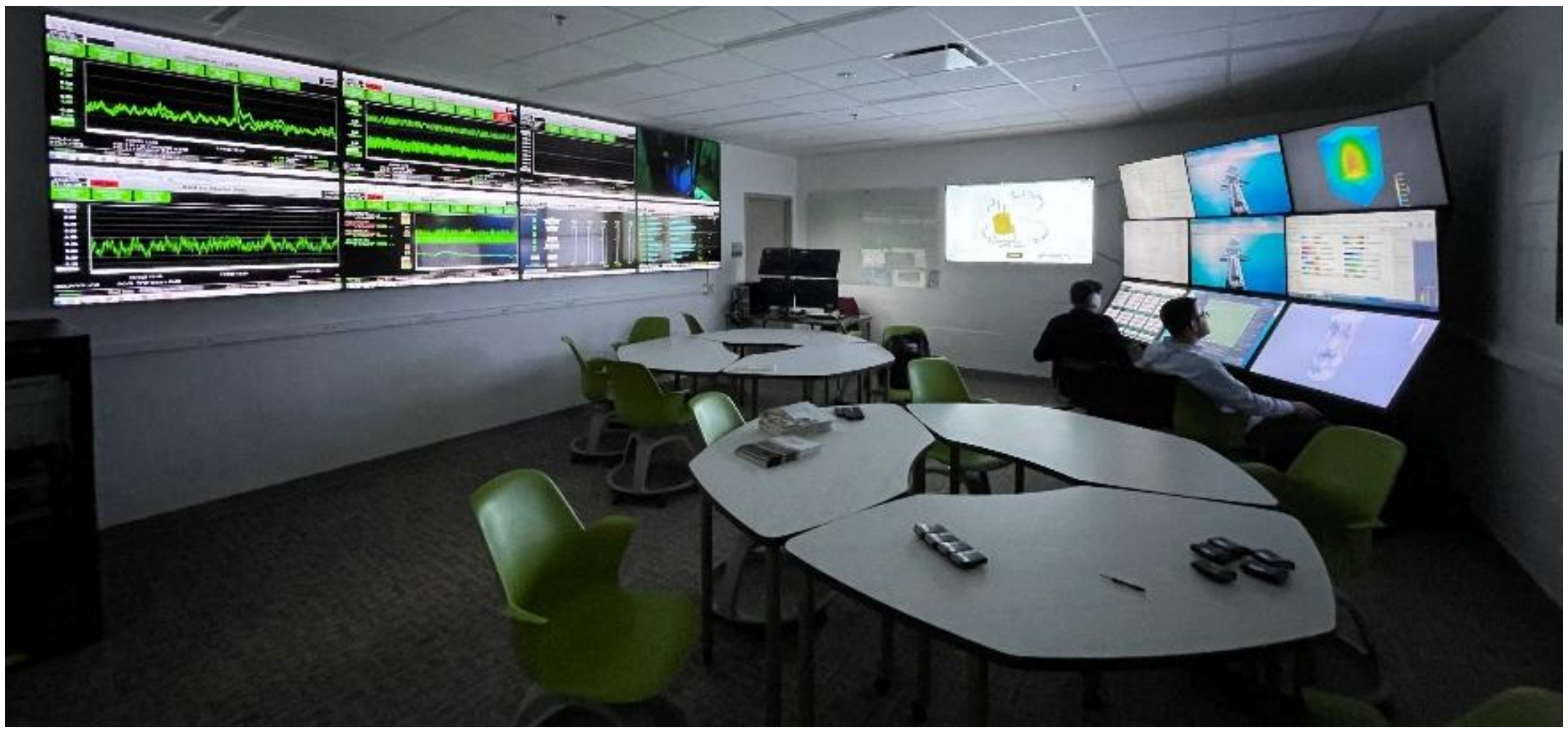

Figure 20: PUR-1 DT demonstration room.

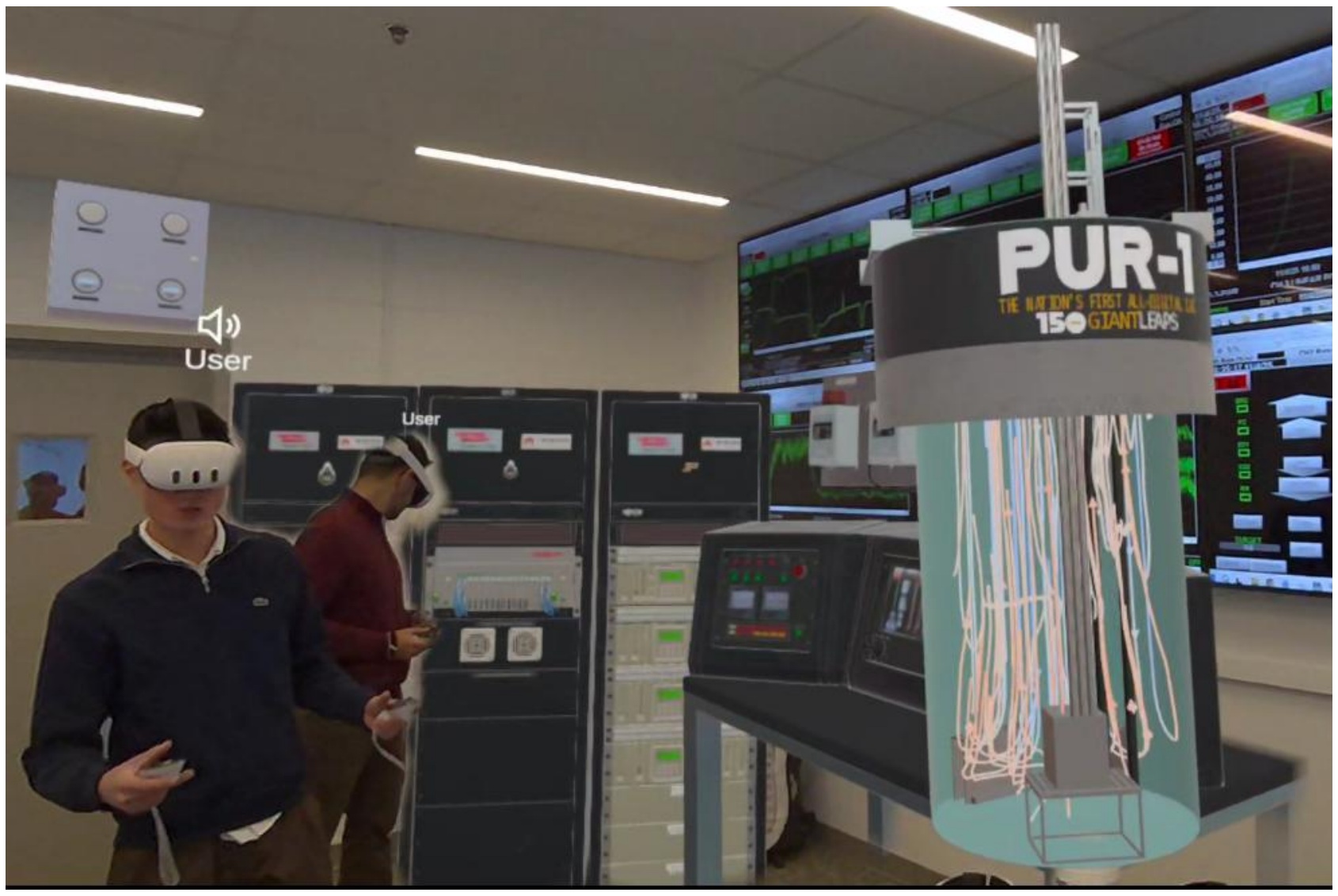

Figure 21: PUR-1 DT integration in AR environment.

A well-designed digital twin should transform raw measurements into operator-actionable knowledge while preserving data integrity and uncertainty bounds across relevant operating histories [10, 80]. Given the limited effective information throughput of human operators (~10 bits/s), the PUR-1 DT incorporates an agentic retrieval-augmented generation (RAG) capability within the Visualization layer to streamline knowledge extraction for decision support [81]. The LangChain-based interface (Figure 22) enables natural-language querying over curated PUR-1 documentation (e.g., operator logs, technical specifications, and SOPs) together with historical context and live data streams, with configurable parameters including Top-K retrieval, maximum response length, and temperature [82]. The interface provides three synchronized outputs: the generated response, the retrieved source excerpts, and the ReAct-based reasoning trace produced by the agentic workflow [83].

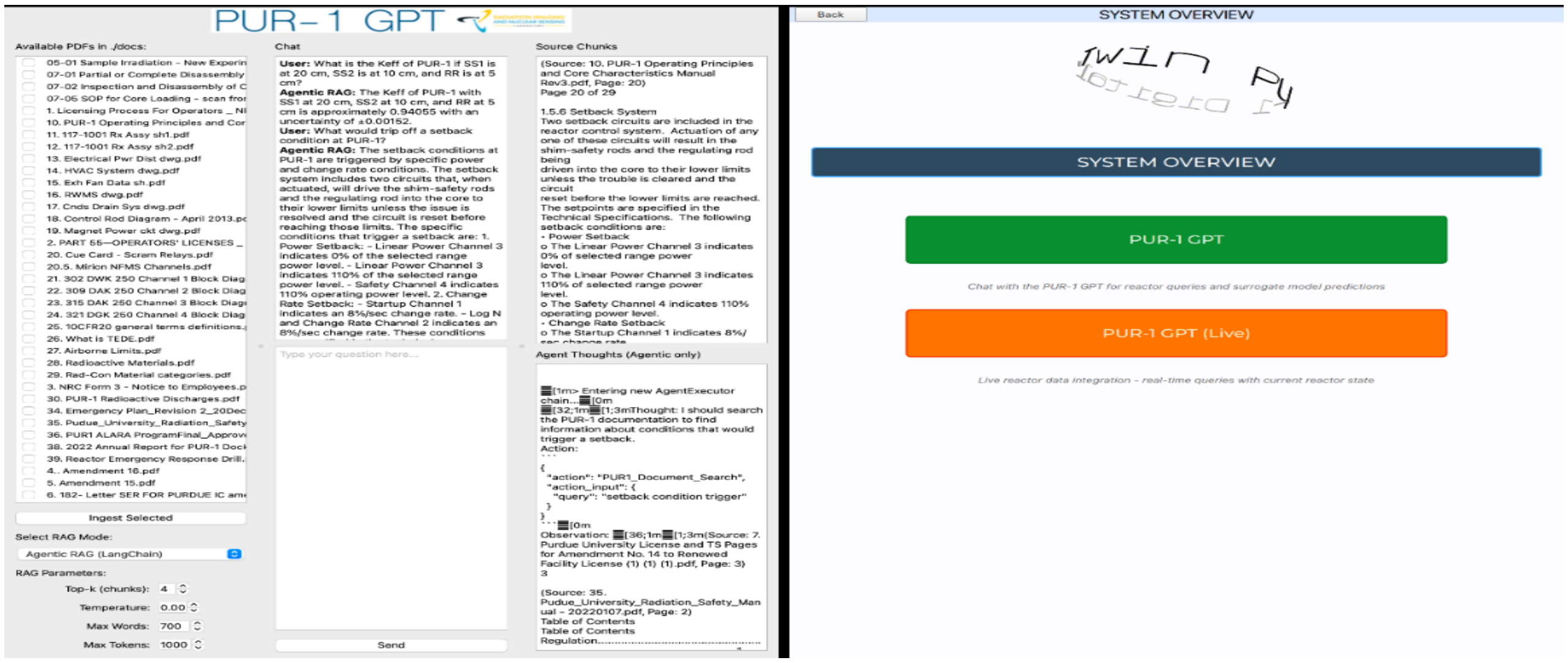

Figure 22: PUR-1 GPT Agentic RAG GUI with query-response pertaining PUR-1 interlock conditions.

# 8 Cyber-physical Integration

To coordinate data exchange across the DT layers while satisfying the synchronization requirement, a dedicated real-time communication bus was implemented within the virtual system (Figure 23). This custom network configuration was integrated along with the DeepLynx platform to minimize transmission latency and organize I/O flows across processes [84]. When the output of one process is required as input for another, the subsequent process retrieves the necessary data directly from the bus, rather than through the cloud server.

All signals are collected and consolidated within a local data warehouse implemented using the DeepLynx platform. This repository provides a unified, structured representation of data originating from both the physical system and the virtual asset. Organizing heterogeneous measurements, commands, and model outputs within a single database is essential since DT workflows depend not only on the availability of individual time series, but also on the ability to interpret their provenance and context. DeepLynx is used to store signal metadata (e.g., tag identity, units, sampling rate, subsystem association, and time-of-acquisition) and to preserve information that links each variable to its generating component, communication pathway, and downstream consumers.

A key capability of this architecture is that the unified database enables systematic investigation of interconnections, origins, and dependencies across DT components through graph-based exploration. By representing signals and assets as nodes and their relationships as edges, the platform supports traceability of how measurements propagate through preprocessing, state-estimation, and modeling modules and how resulting outputs inform the Visualization Layer.

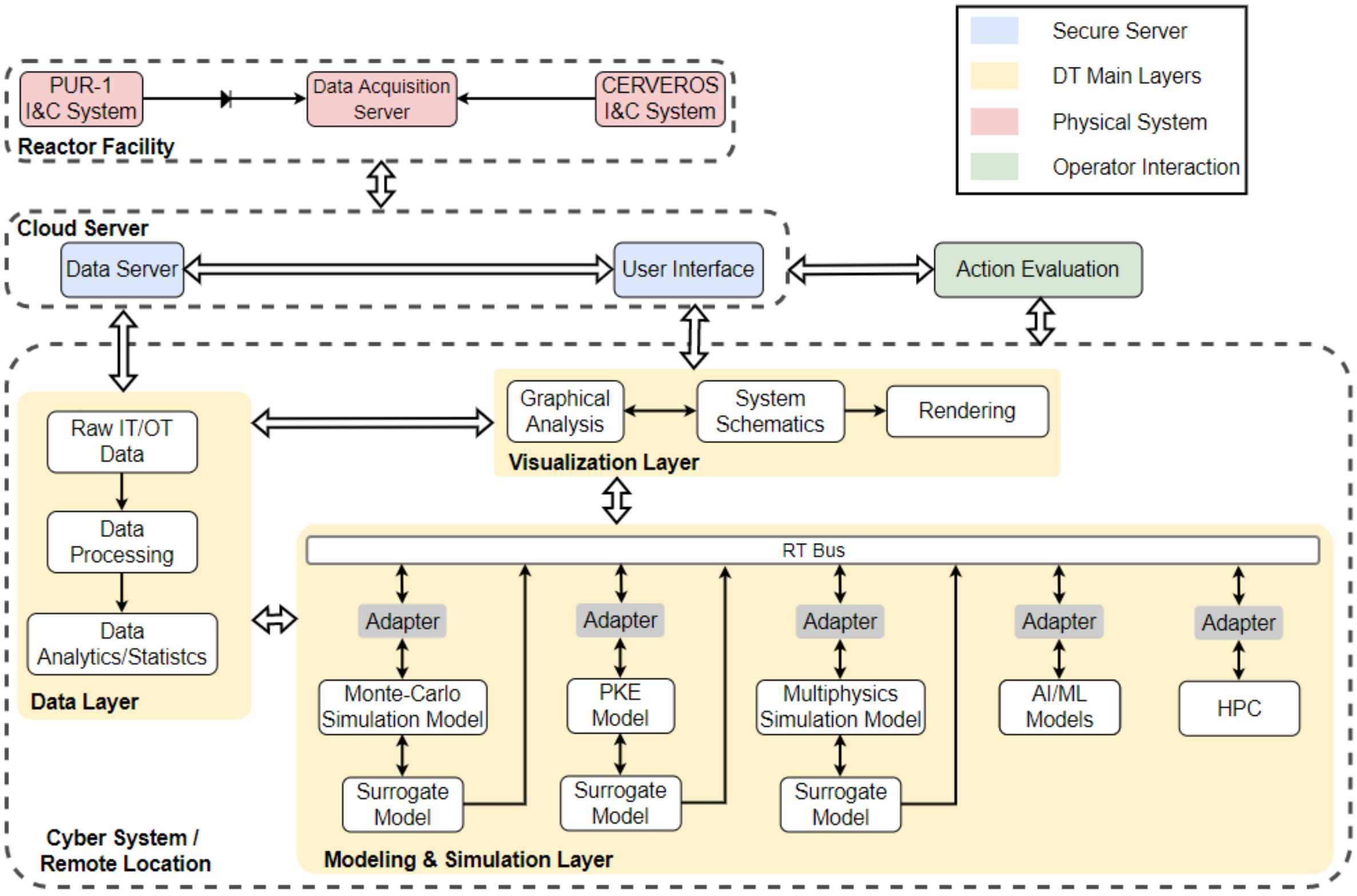

Figure 23: PUR-1 DT architecture.

Although initial data processing takes places in the Data Layer, each module often requires data input in specific formats. To address this, a data adapter was implemented before each process to refine, validate, and standardize incoming data. These Python-based adapters mitigate errors that could arise due to inconsistent formatting or abnormal data characteristics. Lastly, a surrogate model is introduced in simulation models to overcome the inherently long runtimes of high-fidelity simulations, as discussed in the previous section.

## 8.1 Network Communications

Reliable operation of the PUR-1 DT depends on a low-latency communication architecture that supports continuous data exchange between the physical and digital systems while enabling authentication and encryption without disrupting synchronization. To support this requirement, multiple communication links have been established between the physical and virtual systems. Within the PUR-1 network, operational data collected by the RTP 3000 TAS controller are transmitted to the historian using the User Datagram Protocol (UDP). The historian subsequently forwards these data to the virtual system through a TCP/IP socket connection (Figure 24).

The CERVEROS testbed supports bidirectional exchange with the virtual system through an OPC UA interface. The Schneider Modicon M241 PLC hosts an OPC UA server that exposes selected measurements to the virtual system and receives authorized AMDR actuation commands. As part of the experimental evaluation, such communication links can be secured by applying symmetric encryption algorithms enhanced with QKD-generated keys. The quantum communication layer supplies truly random and continuously refreshed cryptographic keys to the secure application, while the encrypted reactor data are subsequently transmitted over the authenticated classical channels (Figure 24). Once received by the digital system and decrypted, the data are distributed among the Data, Modeling and Simulation, and Visualization Layers through the DeepLynx integration framework.

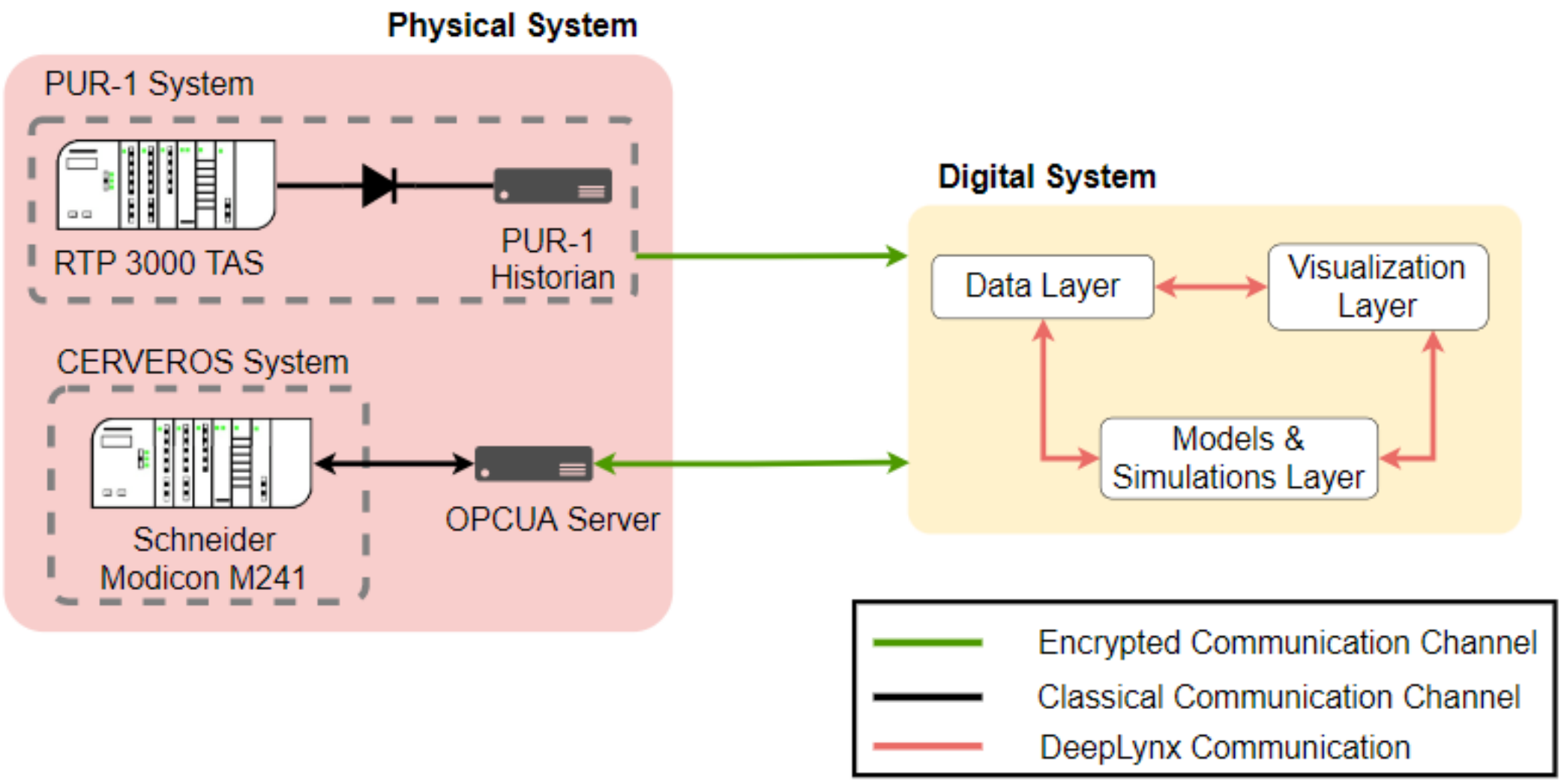

Figure 24: PUR-1 DT network architecture

End-to-end latency is calculated as the combined time required for key retrieval, encryption, data transmission, decryption, and delivery to the receiving application. Independent tests in PUR-1 showed that for the transmission of up to 2,000 reactor signals, the experimentally measured mean latency was approximately 390 ms across the evaluated OTP, AES-256, and ASCON configurations (Figure 25). The low encryption-decryption and data transmission latency ($t_{trans} = t_{encrypt} + t_{broadcast} + t_{decrypt}$) demonstrate that secure communication can be incorporated without disrupting real-time data availability [85]. The reported latency accounts only for communication and cryptographic operations, while model-execution time is evaluated separately within the DT computational architecture.

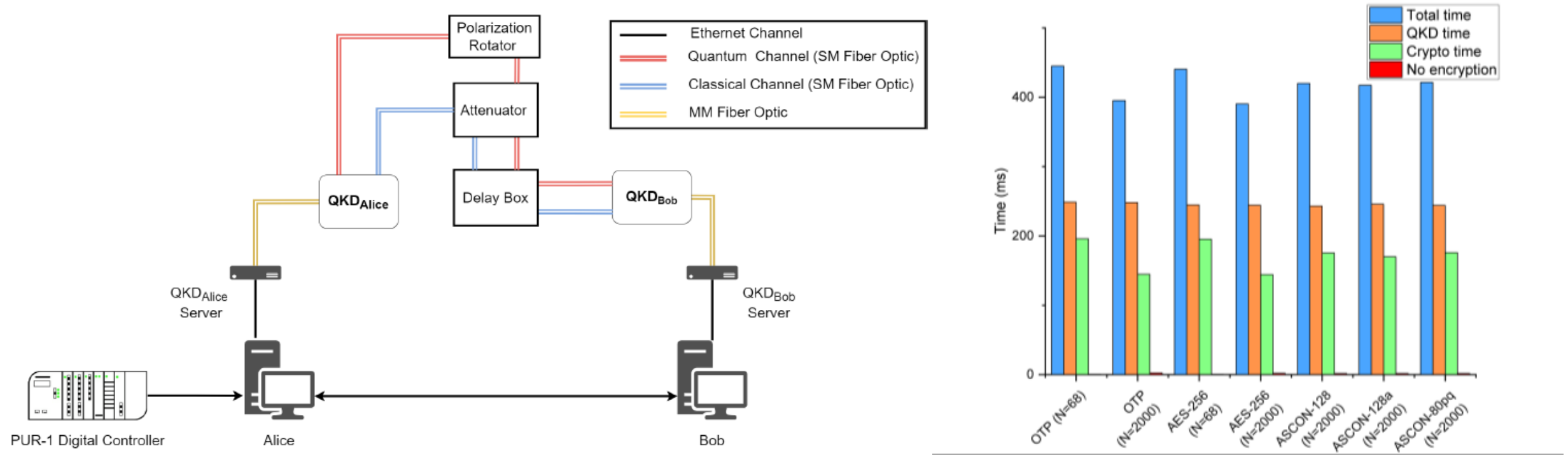

Figure 25: Schematic of PUR-1 QKD experiment setup (left) and average latency (right) for evaluated encryption algorithms during PUR-1 DT operation as described in [85].

## 8.2 Internal Communication Latency Analysis

Apart from monitoring the communication latency between the physical and virtual systems, internal virtual system latencies are quantified to verify that the complete workflow remains within the physical-system operational cycle (<1 s ). Two internal latency metrics were considered: the module execution time, $t_{exec}$, and the presentation time, $t_{print}$. For each DT module, $t_{exec}$ represents the elapsed time between receipt of the required input OT data and generation of a new output. Given that several DT processes are sequentially coupled, with the output of one module serving as input to another, execution delays can propagate through the DT workflow. The cumulative latency along each dependency path must remain compatible with the required synchronization interval. For DT modules that are executed in parallel, the total internal execution latency is determined by the longest computational dependency path required to generate a particular output as:

$$t_{\mathrm{excec,crit}} = \max_{N \in \mathbb{P}} \left( \sum_{i \in N} t_{\mathrm{exec},i} \right), \tag{23}$$

where $\mathbb{P}$ denotes the set of computational dependency paths associated with the considered DT output and $N$ represents an individual path. For a fully sequential workflow, $t_{\mathrm{excec,crit}}$ reduces to the sum of the execution times of the participating modules, whereas for parallel processes, it is determined by the longest dependent execution path. DT outputs are therefore first propagated to dependent computational processes before being presented to the user.

The second metric, $t_{print}$, quantifies the time between generation of a module output and its presentation to the user through the graphical user interface. Although the visualization of scalar quantities and alerts introduces negligible delay, the reconstruction and rendering of three-dimensional fields, such as neutron-flux and temperature distributions, can introduce significant latency. This latency depends primarily on the spatial resolution of the reconstructed field, mesh density, and computational resources available to the workstation hosting the GUI. Monitoring $t_{print}$ provides a measure of how soon users are notified for newly generated DT information.

For workflows in which DT information is communicated to the operator, the end-to-end latency can be expressed as

$$t_{\text{total, open-loop}} = t_{\text{trans}}^{P\to V} + \underset{i\in N}{t_{\text{excec,crit}}} + t_{\text{print}}, \tag{24}$$

where $t_{trans}^{P\to V}$ is the physical-to-virtual communication latency and denotes the sequential dependency path required to generate the corresponding output. For DT actions transmitted back to the physical system,

$$t_{\text{total, closed-loop}} = t_{\text{trans}}^{P\to V} + \underset{i\in N}{t_{\text{excec,crit}}} + t_{\text{trans}}^{V\to P} \tag{25}$$

Synchronization is maintained when the relevant end-to-end latency is below 1 second ($t_{total} < 1\ s$) for any operational mode.

# 9 Benchmarking

The PUR-1 DT was benchmarked against a series of experimental measurements.

## 9.1 Monte Carlo Benchmarking

A campaign of irradiation experiments was performed to map the neutron flux distribution in PUR-1 and serve as OpenMC model benchmarking and validation. The experimental campaign focused on irradiating more than 60 gold foils at several axial and radial locations at the PUR-1 core boundary using high purity gold foils (Figure 26). Gold foils are widely used in nuclear industry for calibration leveraging $^{198}$Au stable decay properties [86, 87]. The experimental procedure included bringing the reactor at critical power and irradiating the gold foils for a specific period in the core. After removal from the reactor core, the gold foils were placed in a high-purity germanium (HPGe) detector to measure the gamma emission rate. Following Neutron Activation Analysis (NAA) calculations, the neutron flux each foil was exposed to, can be accurately calculated.

Neutron flux calculation involves monitoring multiple parameters such as gold foils characteristics, irradiation time, cooling and measuring time, absorption cross-section for different neutron energies, self-shielding factor and detector efficiency. In addition, a correction factor was introduced describing the reactor power level during the experiment. Following IAEA and ASTM standards, total neutron flux is calculated as the summation of physical thermal neutron flux, physical epithermal neutron flux and sub-cadmium cutoff neutron flux, as it has been presented in [88–94]. Therefore, the final total flux can be calculated as:

$$\phi_{tot} = \phi_{thermal} + \phi_{overlap} + \phi_{epithermal} \tag{26}$$

Where $\phi_{thermal}$ is the physical thermal flux, $\phi_{epithermal}$ the physical epithermal neutron flux with energy range below 0.1 MeV and above Cadmium cutoff (0.55 eV) and $\phi_{overlap}$ the true epithermal in epithermal-thermal overlap component. These values are described through the following equations:

$$\phi_{thermal} = \frac{2}{\sqrt{\pi}} \cdot \sqrt{\frac{T}{T_0}} \cdot \phi_0 \tag{27}$$

$$\phi_{epithermal} = \frac{r \cdot \phi_0}{(1 - b_4 r)} \cdot \ln\left(\frac{0.1\ MeV}{0.55\ eV}\right) \tag{28}$$

$$\phi_{overlap} = \frac{r \cdot \phi_0}{(1 - b_4 r)} \cdot \ln\left(\frac{0.55\ eV}{\mu kT}\right)\left(f_1 + \frac{w'}{g}\right) \tag{29}$$

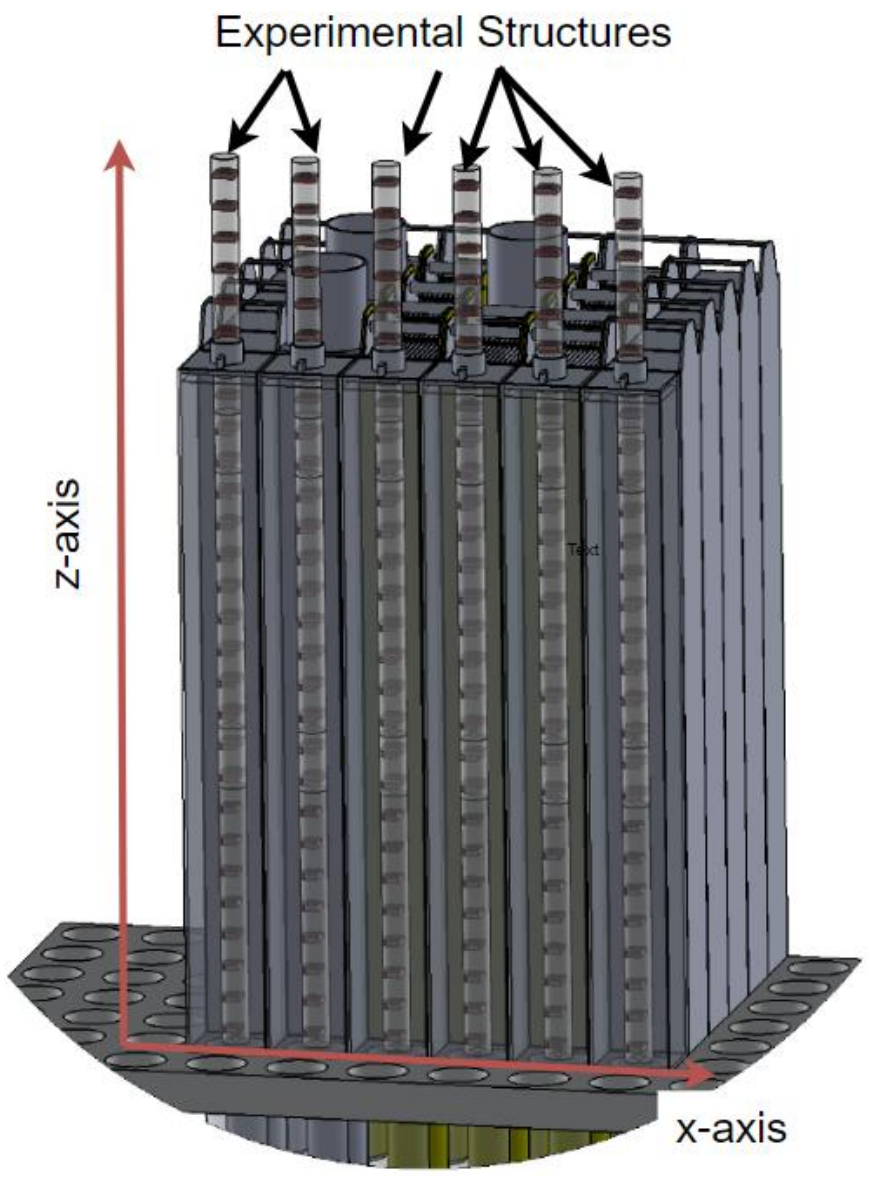

Figure 26: Experimental structures for neutron distribution experiment in PUR-1 reactor core.

Assuming a conventional 2200 m/s thermal flux, the proportionate flux accounting for thermal range only using the 1/v law is defined as a function of explicitly thermal effective cross section described by Westcott flux, epithermal to thermal ratio (r) and factors accounting departure of cross-section from 1/v law in thermal (g) and epithermal (s) regions. Therefore, the neutron flux for 1/v approximation is calculated as:

$$\phi_0 = \phi_{westcott} \frac{(g + rs)}{g} \tag{30}$$

$$\phi_{westcott} = \frac{G \cdot \lambda_{198}}{\epsilon \cdot N_{197} \cdot \sigma_{effective,197} \cdot (1 - e^{-\lambda_{198} t_0}) \cdot e^{-\lambda_{198} t_w} \cdot (1 - e^{-\lambda_{198} t_c})} \cdot \frac{P_{max,\text{avg}}}{P_{t=0 \to t=t_0}} \tag{31}$$

Benchmarking of a Monte Carlo simulation code requires precise modeling of system geometry, materials composition and CR configuration. For this reason, a detailed CAD model of the experimental structure was implemented in OpenMC along with the material properties of all experimental components. To improve accuracy of calculations, distinct tallies were defined around each foil segregating model outputs to distinctive volumes of interest to facilitate the output normalization procedure [95].

Table V: Parameters definition of Equation 26-31

| Parameter | Description |
|---|---|
| $G$ | Gold foil counts corrected for background counts (counts) |
| $\lambda_{198}$ | Au-198 decay constant ($\mathrm{sec}^{-1}$) |
| $\rho_{197}$ | Au-197 density ($g/cm^3$) |
| $\epsilon_{det}$ | Detector photo-peak efficiency |
| $\epsilon_{ep}$ | γ-branching ratio |
| $\epsilon_{pm}$ | Photon escape (non-absorption) probability |
| $S_{th}$ | Thermal-neutron self-shielding factor |
| $\sigma_{\mathrm{eff},197}$ | Effective cross section of Au-197 |
| $N_{197}$ | Atomic density ($atoms/cm^3$) |
| $m_{foil}$ | Foil mass ($g$) |
| $t_w$ | Time interval between foil removal and the onset of counting ($sec$) |
| $t_0$ | Duration of foil irradiation ($sec$) |
| $t_c$ | Duration of counts measuring ($sec$) |
| $T$ | Assumed neutron temperature best fitting PUR-1 Maxwellian neutron spectrum |
| $T_0$ | 2200 m/s thermal neutron temperature |

Flux normalization yielded accurate total neutron flux estimation on the gold foils based on the selected tallies. As shown in Figure 27, flux values peaked in assemblies at the core center and attenuated towards the edges. The comparison of simulated and experimentally calculated total neutron flux indicated good alignment between the two methods. The percentage difference between experimental and simulated flux for each foil location, resulted in an average deviation of 6.75%.

It is worth noting that this evaluation demonstrates good accuracy in medium and high flux location but presents lower correlation at lower flux levels. The average percentage difference considering only medium (position E, H) and high (position F, G) flux assembly positions is 2.89%. Although a lower correlation is observed at low flux levels, the overall low percentage error indicates that the OpenMC model is adequately benchmarked and supports the validity of the simulation results.

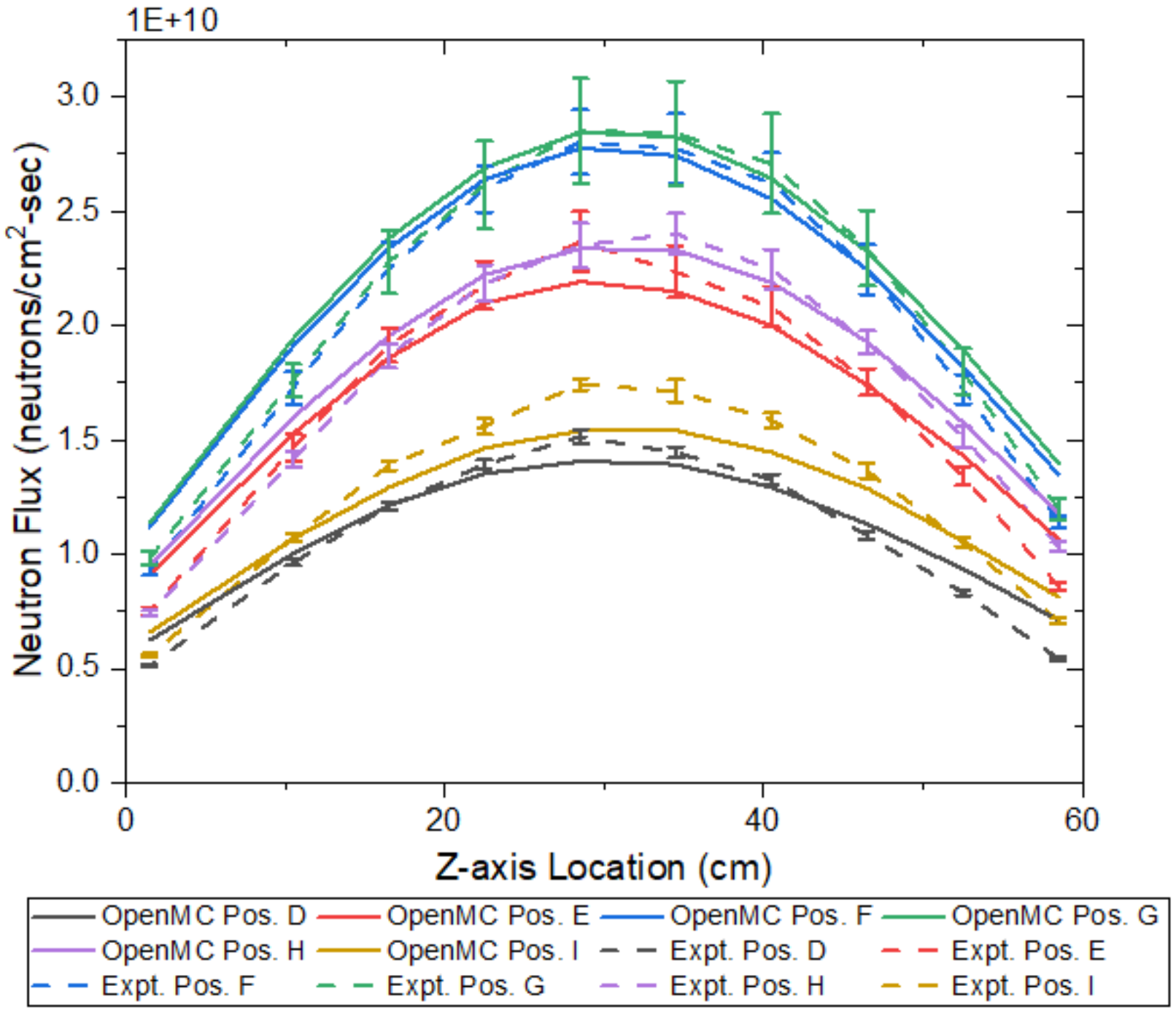

Figure 27: Comparison of experimental and simulated neutron flux of gold foils

## 9.2 Multiphysics Benchmarking

The heat distribution model was benchmarked against previous experimental studies performed in PUR-1 pool. Miller et al. conducted experiments investigating the correlation between average pool temperature, reactor power and thermocouple location under varying thermal power level over a 5-hour period [96]. This study reported that the average temperature close to pool surface is location- independent and can be described as a linear relationship as a function time by the equation:

$$T\ (°C) = 0.2017 \cdot t + 26.158\ for\ t \in [0,5]\ hrs \tag{32}$$

Simulation data from the COMSOL model were extracted at 15-minute intervals using the same time span and power inputs as the experimental study. From the computational mesh of 18901 points, a subset of 153 points located within 20 in of the pool surface was selected for comparison resulting in an average percentage difference of 1.37% (Figure 28).

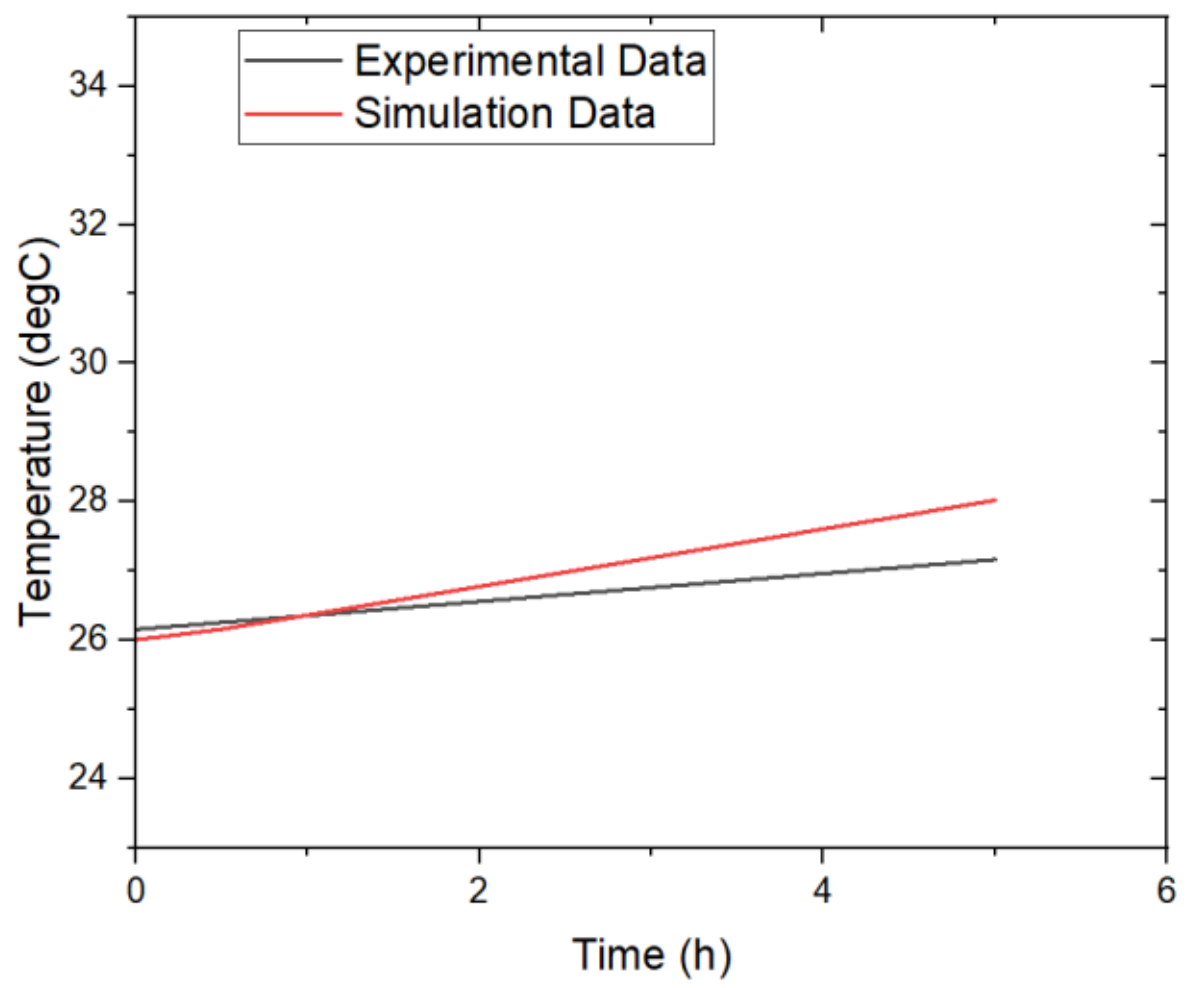

Figure 28: Pool temperature estimations from simulation model compared with experimental data

## 9.3 Point Kinetics Equations

The time-dependent neutron population in PUR-1 was modeled using point reactor kinetics, in which spatial effects are collapsed into the fundamental mode of the neutron flux distribution. Reactivity feedback is incorporated through the principal contributors, namely CR positions and temperature-dependent effects, alongside the delayed neutron fractions and their associated precursor concentrations. Beyond modelling transients, the role of delayed neutrons is fundamental to reactor controllability. From a DT forecasting standpoint, the finite delayed fraction imposes a strict bound on the permissible neutron multiplication factor:

$$k - 1 < \beta \tag{33}$$

This constraint prevents operation in a prompt-critical regime, where criticality would rely exclusively on prompt neutrons and the neutron lifetime would be on the order of ~$10^{-6}$-$10^{-4}$ sec. For PUR-1, six delayed neutron groups are considered, with group-dependent fractions and decay constants obtained from the OpenMC model. These parameters are summarized in Table VI.

Table VI: OpenMC-derived delayed neutron parameters for PUR-1

| Group | Decay constant $\lambda_i$(sec$^{-1}$) | Delayed neutron yield per fission | Fraction $\beta_i$ |
|---|---|---|---|
| 1 | 1.3337E-02 | 0.00052 | 2.6141E-04 |
| 2 | 3.2732E-02 | 0.00346 | 1.3439E-03 |
| 3 | 1.2080E-01 | 0.00310 | 1.3137E-03 |
| 4 | 3.0295E-01 | 0.00624 | 2.9103E-03 |
| 5 | 8.5024E-01 | 0.00182 | 1.1985E-03 |
| 6 | 2.8555E+00 | 0.00066 | 5.0087E-04 |
| Total Yield: 0.0158 | | Total delayed fraction ($\beta$): 7.5286E-03 | |

To benchmark the model, PKE predictions were compared against recorded operational data to assess accuracy. The forecasting capability was not activated, since the analysis was performed offline using historical reactor data (Figure 29). The average percentage difference between model predictions and measured neutron population over a range of operational scenarios (from startup to shutdown) was found to be 6.89%, indicating good overall agreement between the predicted and observed neutron behavior in the PUR-1 core.

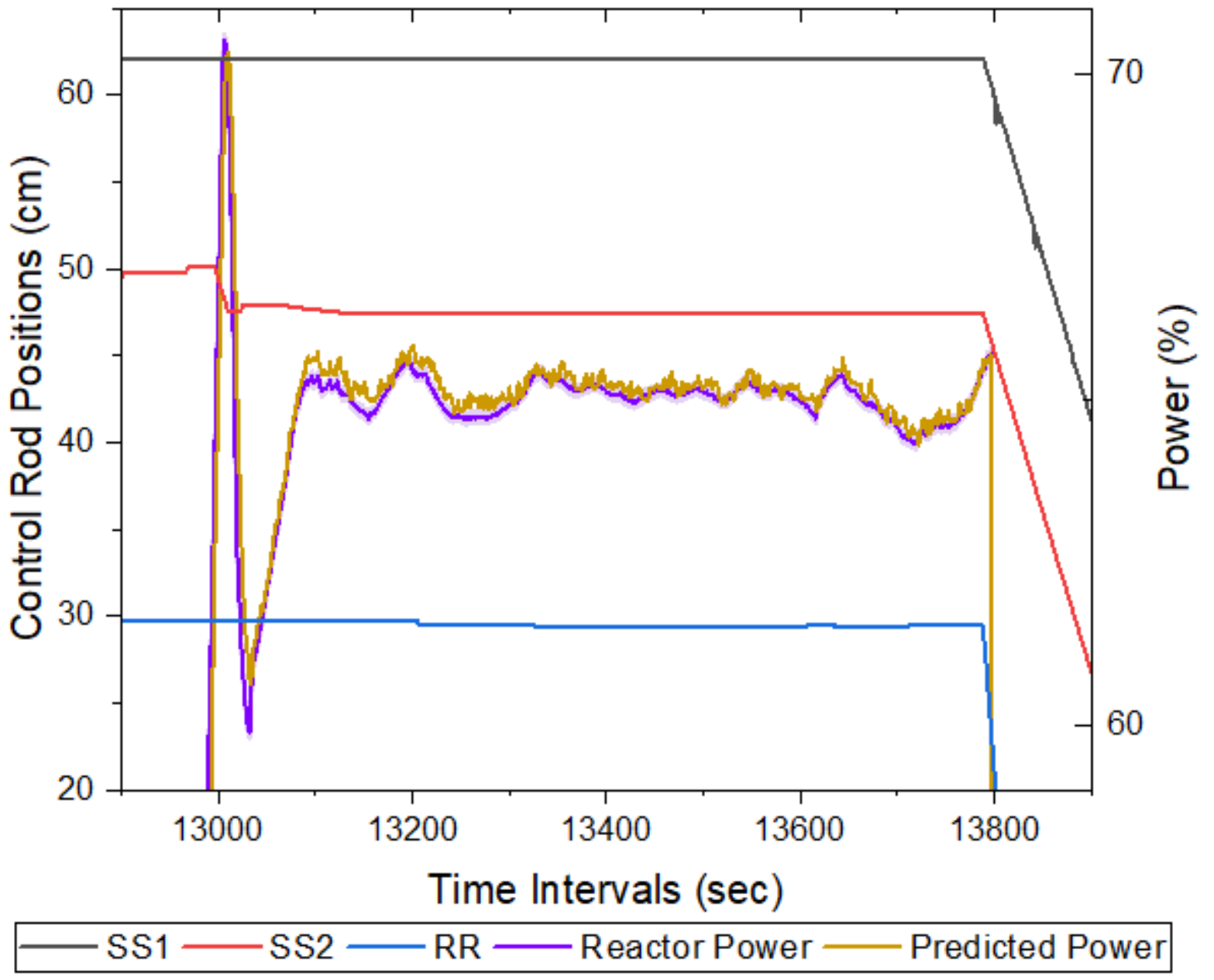

Figure 29: PKE power estimation in comparison with PUR-1 power at steady-state operation

## 9.4 Effective Multiplication Factor Prediction

A SM is implemented predicting the effective multiplication factor ($k_{eff}$) as a function of control-rod heights. Through the SM, the operator is supplied with real-time state information supporting enhanced situational awareness and decision-making.

Gaussian Process (GP) regression was selected for describing input-output relation. GP is a nonparametric, Bayesian statistical learning method designed for modeling complex functions. It assumes that function values of any input vector follow a Gaussian distribution, fully specified by a mean function and a covariance kernel [97]. Unlike linear or polynomial regression, which impose rigid functional forms, GP adapts to the data structure via the choice of kernel, enabling both local smoothness and global trend capture [97]. For this reason, multiple kernels with different accuracies were developed enabling the user to select along various levels of confidence.

The GP can be described as function of the mean function, $m(x_i)$, and the covariance kernel function $k(x_i, x_j)$, modeling the correlation between any finite set of input vectors (Eq. 34). Therefore, assuming a known input vector $x_i$, the known target vectors $y_i$ and $C_i$ the covariance matrix of the training data, GP mean value is described from the Eq. 35.

$$f(x_i) \sim GP\left(m(x_i), k(x_i, x_j)\right) \tag{34}$$

$$m(x_i) = \mathrm{k}^T \cdot \mathrm{C}_i^{-1} \cdot \mathrm{y}_i \tag{35}$$

In the scope of this work three different kernels were evaluated to accurately describe the correlation between CRs configuration and $k_{eff}$. In particular, the selected kernels are:

- *Radial Basis Function (RBF) kernel*

  This kernel corresponds to a stationary, isotropic covariance function that depends on the Euclidean distance between input vectors [97].

$$k(x_i, x_j) = \exp\left(-\frac{\|x_i - x_j\|^2}{2\sigma^2}\right) \tag{36}$$

  Where $\sigma$ is the length scale parameters.

- *Matérn kernel*

  This kernel is based on a generalization of square exponential kernel, capable of adjusting the roughness of the modeled function. Increasing $\nu$ value avoids focusing on potential local maxima generating smoother gradients [98, 99]. For this reason, in this study it was selected $\nu = \frac{3}{2}$.

$$k(x_i, x_j) = \left(\frac{2^{1-\nu}}{\Gamma(\nu)}\right)\left[\frac{\sqrt{2\nu}|x_i - x_j|}{\ell}\right]^{\nu} K_{\nu}\left(\frac{\sqrt{2\nu}|x_i - x_j|}{\ell}\right) \tag{37}$$

  Where $\nu$ and $\ell$ are the positive parameter and $K_\nu$ a modified Bessel function.

- *Neural Network (NN) kernel*

  This kernel depends on the inner product between inputs providing a nonstationary covariance structure [97].

$$k(x_i, x_j) = \frac{2}{\pi}\sin^{-1}\left(\frac{2\tilde{x}_i^T \Sigma \tilde{x}_j}{\sqrt{(1 + 2\tilde{x}_i^T \Sigma \tilde{x}_i)(1 + 2\tilde{x}_j^T \Sigma \tilde{x}_j)}}\right) \tag{38}$$

  Where $\tilde{x}$ is the augmented input vector.

Kernel performance was evaluated using the Root Mean Square Error (RMSE) between predicted and known outputs. RMSE (Eq. 39) is a widely adopted metric that quantifies the typical magnitude of prediction error and is thus indicative of model accuracy. For surrogate-model selection, overall accuracy, goodness-of-fit, and the required training and tuning time were jointly considered (Table VII). Although the evaluated kernels exhibit similarly high $R^2$values (Figure 30), the RBF kernel achieved the lowest RMSE and is therefore selected as the default model in PUR-1 DT. Kernel uncertainty estimates are used to quantify prediction confidence and support model selection for each scenario.

$$RMSE = \sqrt{\frac{1}{M}\sum_{i=1}^{M} (\mathrm{y_i} - \hat{\mathrm{y}}_\mathrm{i})^2} \tag{39}$$

where M is the validation dataset size, $yi$ the known target from OpenMC testing dataset and $\hat{y}_1$ the predicted target.

Table VII: Regression algorithms metrics

| Model | $R^2$ | RMSE |
|---|---|---|
| GPR (RBF) | 0.9988 | $9.723 \cdot 10^{-3}$ |
| GPR (Matern) | 0.9982 | $1.212 \cdot 10^{-2}$ |
| GPR (NN) | 0.9908 | $1.962 \cdot 10^{-2}$ |

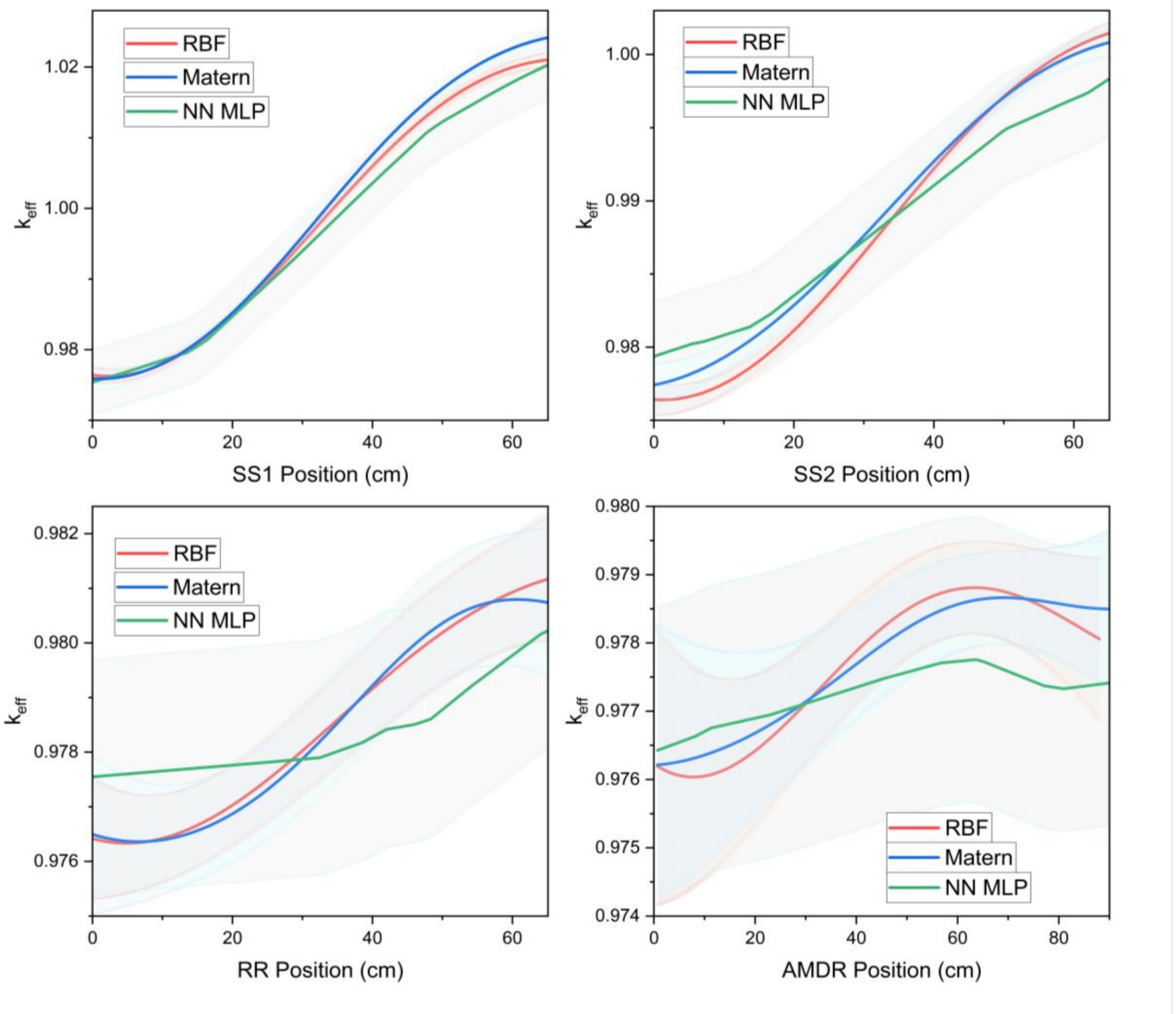

Figure 30: Estimation of $k_{eff}$ for different CR heights based on surrogate model

## 9.5 Neutron Flux Field Prediction

Through PUR-1 DT operation, a three-dimensional neutron flux field prediction is available to the user based on the current CR configuration. Utilizing the simulation dataset, the model is trained on instances of full-core flux field with the associated CR positions. To improve numerical stability and mitigate the strong dynamic range induced by localized flux peaks, the target fields are represented in a logarithmic domain.

$$x_i \in [X_{SS1}, X_{SS2}, X_{RR}, X_{AMDR}] \tag{40}$$

$$y_i = \log(1+\phi_i)\,, where\ \phi_i, y_i \in \mathbb{R}^{100\times 100\times 100} \tag{41}$$

The SM is formulated as a conditional decoder that maps the low-dimensional CR configuration to a full three-dimensional flux field. An initial multilayer perceptron (MLP) embeds the input vector into a high-dimensional latent representation. Each layer is defined as:

$$h_{\ell+1} = \sigma_i(W_\ell h_\ell + b_\ell) \tag{42}$$

, where $h_0 = x_i$, $W_\ell$ network weights and $b_\ell$ network biases and $\sigma_i(z) = z \cdot \frac{1}{1+e^{-z}}$ corresponds to SiLU activation function.

A linear layer maps the latent vector onto a low-resolution three-dimensional feature tensor of size 8×25×25×25, refined by a residual block and then progressively upsampled to 32×50×50×50 and 16×100×100×100, via transposed convolutions paired with residual blocks. Each residual block applies two GroupNorm–SiLU–Conv3d layers with a skip connection. A final two-layer convolutional head (3×3×3, then 1×1×1) maps the resulting features to the single-channel, log1p-transformed neutron flux field $\hat{y}$. Predictions are mapped back to physical units as $\hat{\phi} = \exp(\hat{y}) - 1$, clamped to non-negative values, before physical-domain accuracy is assessed.

Network parameters are optimized using the AdamW optimizer, which adapts the learning rate for each parameter based on moving averages of the gradients and squared gradients. Training seeks to minimize the mean squared error (MSE) between the predicted and target log1p flux fields through the defined loss function:

$$\text{Loss} = \frac{1}{N}\sum_{i=1}^{N}[\hat{y}(x_i) - y(x_i)]^2 \tag{43}$$

, where $\hat{y}(x_i)$ the predicted neutron flux field, $y(x_i)$ target field, $x_i$ the CR configuration input and $N = 100 \times 100 \times 100$ the total number of voxels. Given the size of the training dataset (~2,500 CR-configuration/flux field runs), generalization was assessed using 5-fold cross-validation. The overall model performance is evaluated using voxel MSE at log1p domain and relative L2 error (Eq. 44) which describes full filed reconstruction accuracy (Figure 31).

$$\text{RelL2}(\hat{y}, y) = \frac{\|\hat{y} - y\|_2}{\|y\|_2} \tag{44}$$

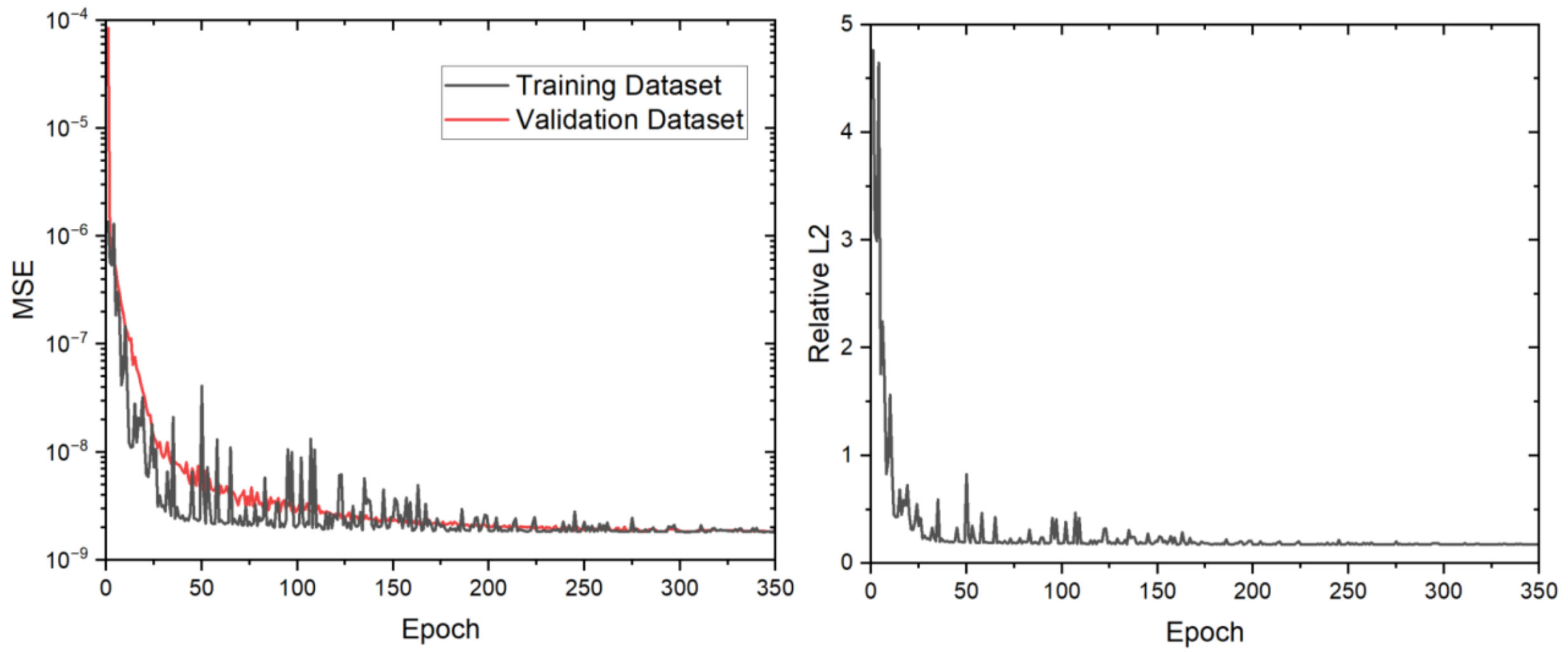

Figure 31: Neutron flux field prediction MSE (left) and relative L2 error (right) evolution with epochs

Model performance was assessed by comparing the predicted neutron flux field with the corresponding reference field from the simulation dataset. The average percentage difference was 3.62% over the entire neutron flux field. The surrogate can adequately reproduce the main spatial characteristics of the flux distribution, while the largest deviations, highlighted in the residual map (Figure 32), are concentrated near the CRs, source assembly location and field boundaries with maximum local discrepancies up to 70%. In these locations, small reference flux values amplify the relative percentage error, whereas the surrogate more accurately reproduces the dominant spatial flux distribution across the core.

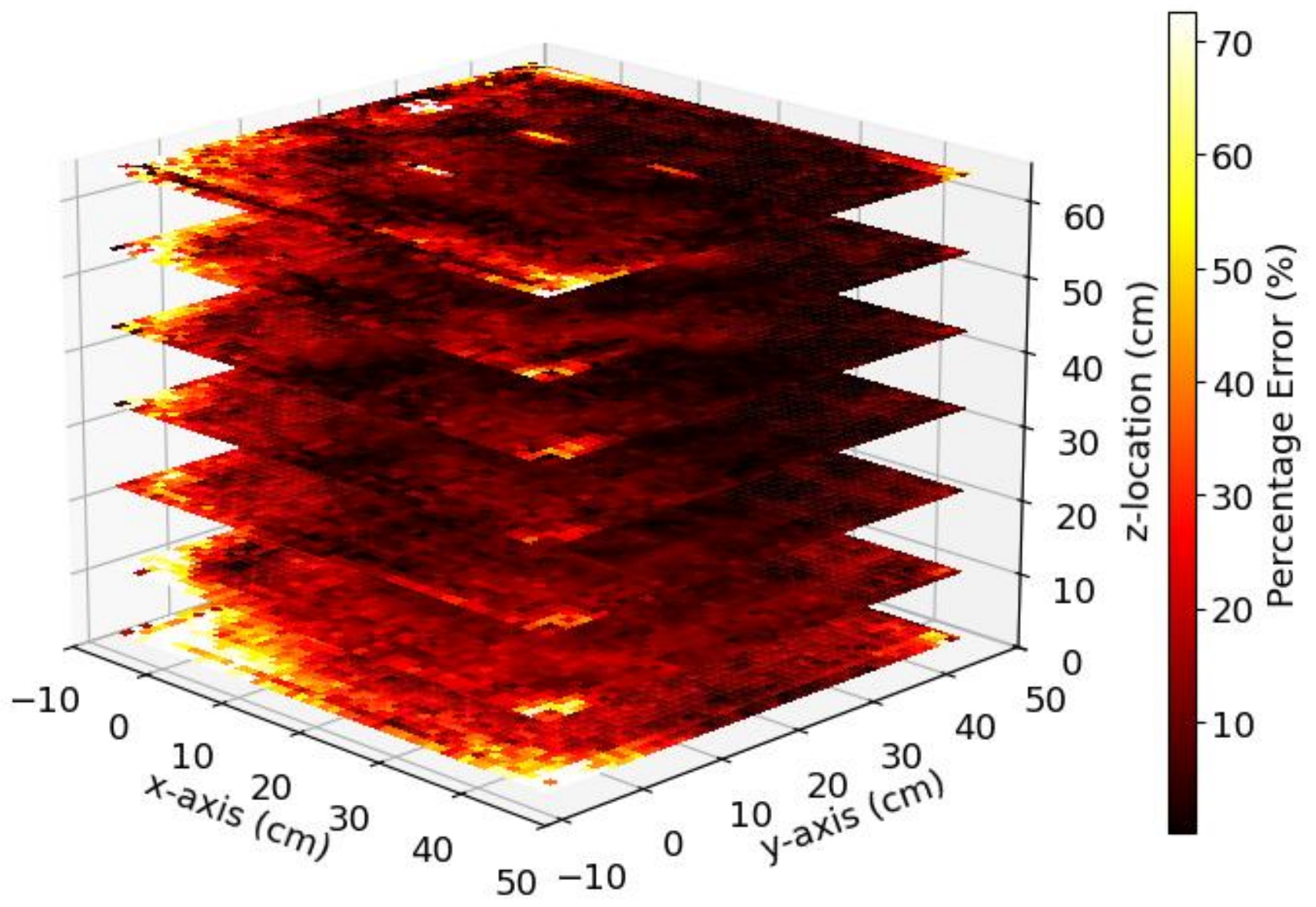

Figure 32: Percentage error between simulated and predicted neutron flux

To further evaluate the predictive capability of the surrogate model, predictions were compared with experimental measurements obtained during the Monte Carlo model benchmarking campaign. Specifically, the predicted flux field was post-processed to extract flux values at the gold foil locations in reactors irradiation assemblies, and these estimates were compared with the experimentally derived total neutron flux, yielding an average percentage error of 17.81% (Figure 33).

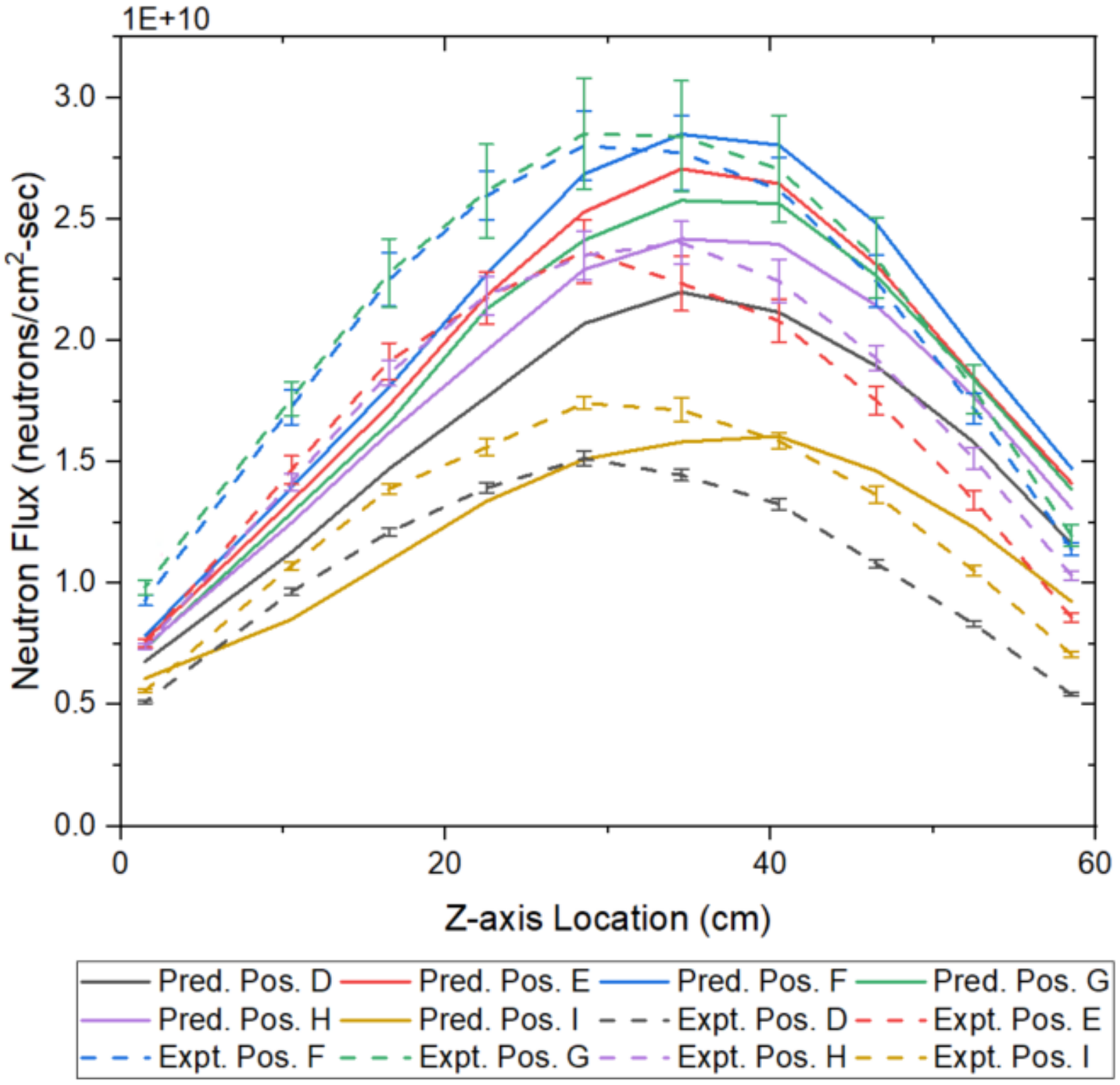

Figure 33: Predicted and experimental total neutron flux in gold foils irradiation positions

## 10 Use case

In the scope of this paper the DT framework is employed as a decision-support tool for operators, enabling real-time identification and supplemental information for reactor state and short-term forecasting of system behavior. The virtual system continuously enhances observability of the physical asset, while the prediction loop is initiated by operator input prior to CR movement.

To demonstrate this methodology, DT performance is evaluated over a complete reactor operational cycle involving power transitions from 0% to 10%, then to 65%, and back to 0%, repeated twice. Multiple modeling processes (PKE, Monte Carlo, and Multiphysics simulation models) along with the corresponding SMs are combined to replicate reactor current and future state. Depending on whether reactor operation is steady or transient, the virtual system executes processes in different order (Figure 34).

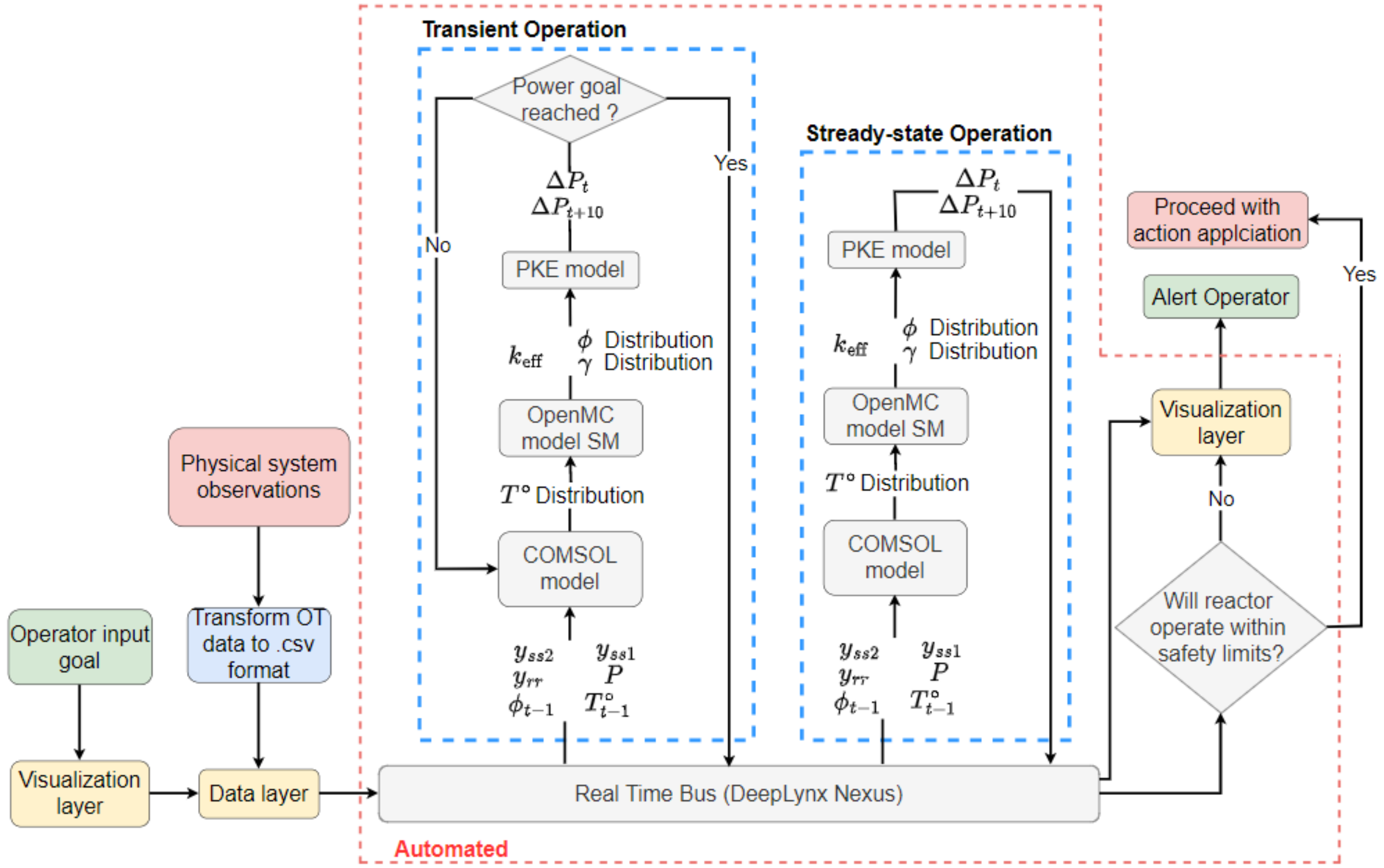

Figure 34: PUR-1 DT operation flow chart

Observations from the physical system are collected and transmitted to the virtual system, where preprocessed data from the Data layer are supplied to the main DT functions within the Modeling and Simulation (M&S) layer. The integration of these processes enables an enhanced representation of reactor state by providing estimates of quantities that are not directly available from plant instrumentation, including neutron population behavior and temperature distribution. Incoming information is initially utilized from the multiphysics model to provide a more detailed representation of temperature distribution in the reactor pool. Estimates for component temperatures are then supplied, along with CR positions, to the OpenMC-based SM to estimate neutron flux distribution and $k_{eff}$ in the core. Finally, physical-system observations combined with temperature predictions are used by the PKE model to provide real-time estimates of inserted reactivity and reactor power (Figure 35).

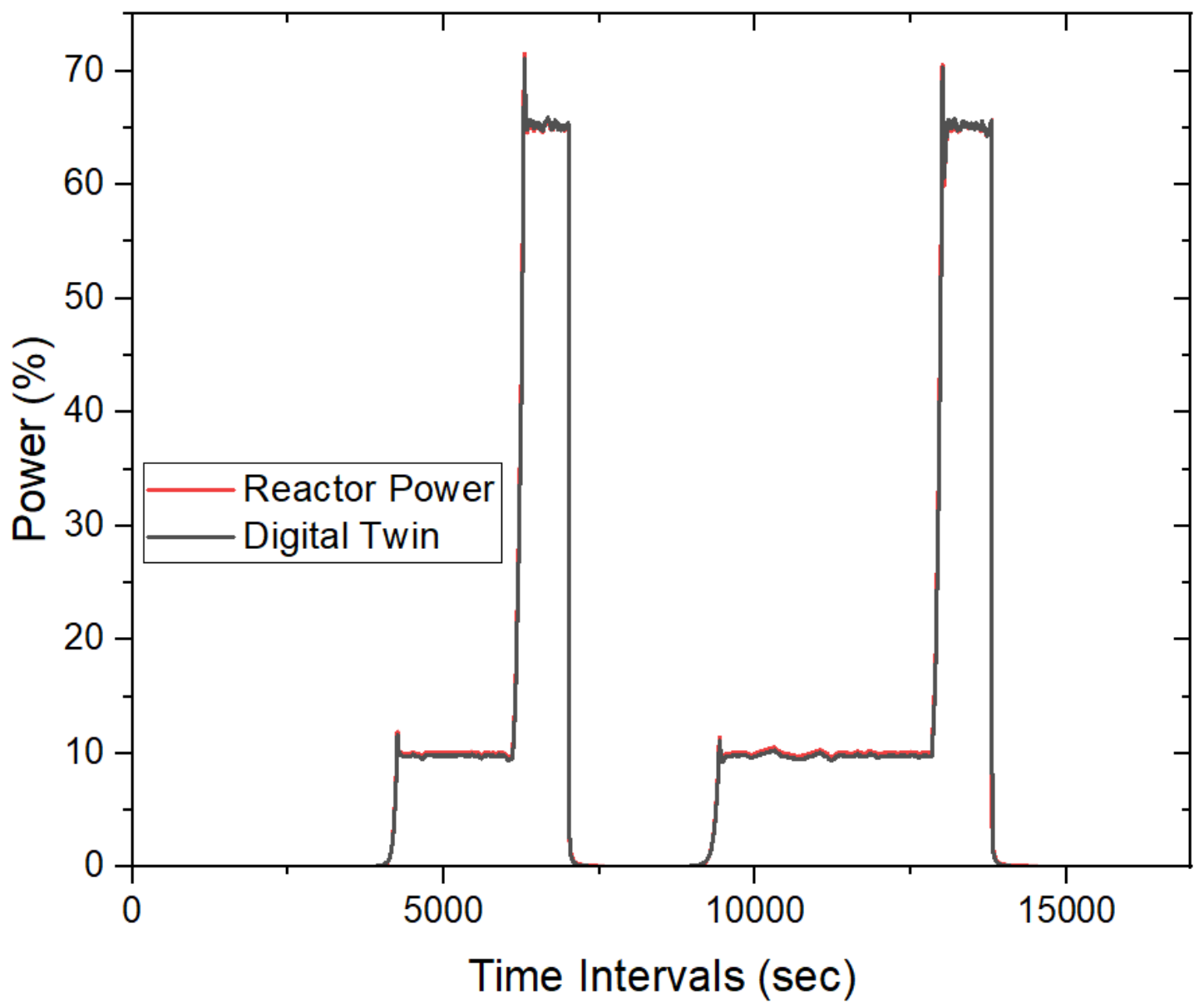

Figure 35: PUR-1 DT power predictions in comparison to power values collected from reactor instrumentation system

System state predictions are fully synchronized with physical asset operation through the SMs. In particular, the SM estimating $k_{eff}$ exhibits different sensitivities according to the reactivity worth of each rod, which is reflected in the varying slopes in Figure 36. During steady-state operation only RR was adjusted to compensate for power drifts illustrating negligible changes in $k_{eff}$ value, SS2 was employed for power level changes, and SS1, which has the highest reactivity worth, was only moved during startup and shutdown phases. Similarly, based on the CR configuration the neutron flux field in the core is continuously reconstructed and presented to the operator in concurrence with the system state.

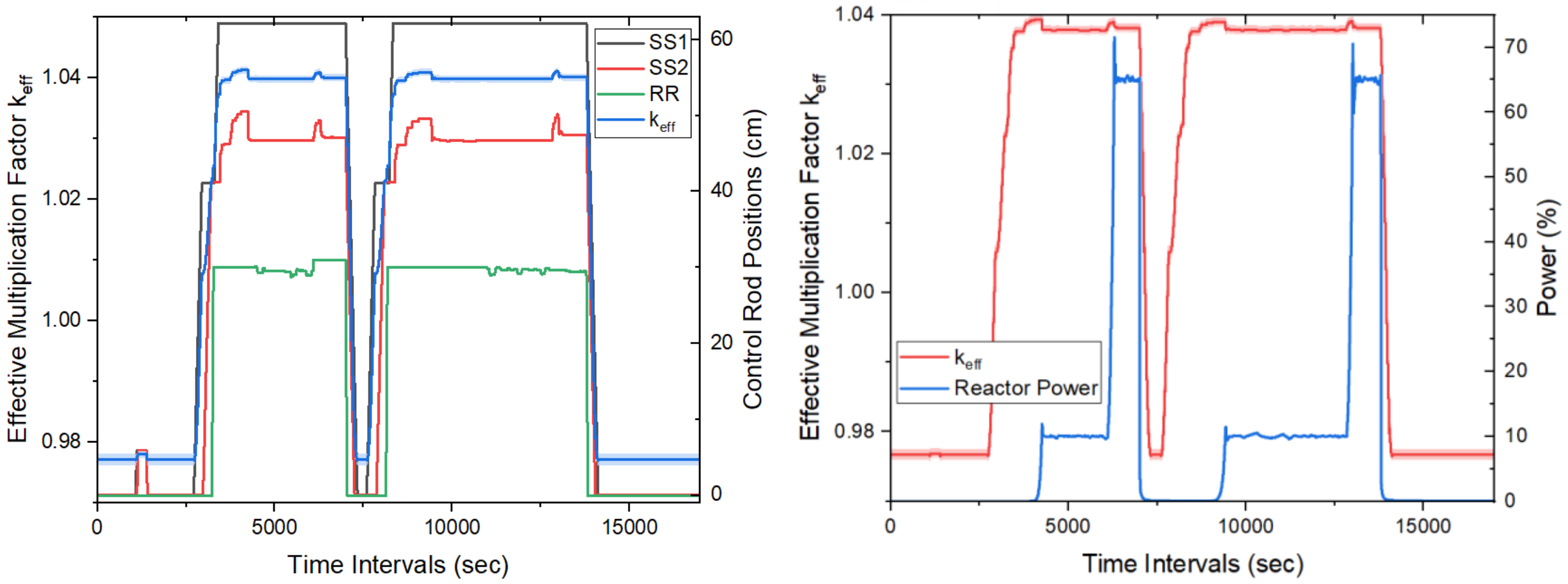

Figure 36: Effective multiplication factor in correlation with CR positions (left) and reactor power (right)

In contrast to the neutronic modules, the absence of a dedicated SM for the multiphysics simulation requires a different approach. Because thermal transients in the system evolve at a comparatively slower rate, the thermohydraulic predictions can be provided at a lower update frequency. Using the current reactor state as the initial condition, the COMSOL model predicts the evolution of heat flux and temperature distribution throughout the reactor pool over 300 time steps from the start of the simulation (Figure 37). Assuming simulation is initiated at time $t_N$ and requires approximately 100 s to complete, then at time $t_{N+100}$ the operator is informed both of how the applied action has affected the system over the preceding 100 s and of the projected system state at $t_{N+200}$. If additional operator actions occur during the simulation runtime, a new simulation is launched using updated initial conditions to preserve predictive consistency.

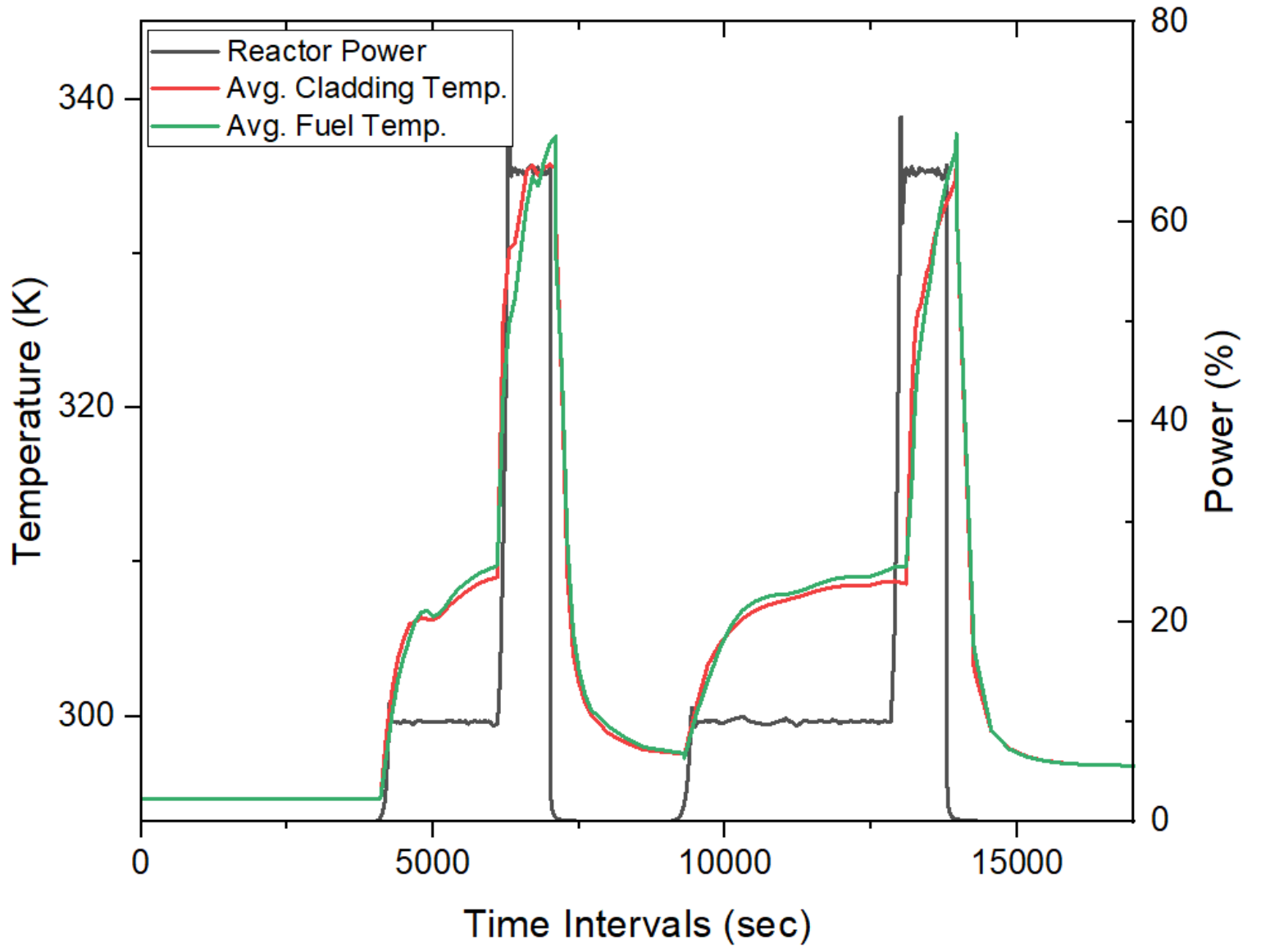

Figure 37: Average cladding and fuel temperature in PUR-1 core

The overall performance of the PUR-1 DT was assessed by comparing predicted reactor power against measurements collected from the physical system throughout the operational cycle. This comparison yielded an average percentage difference of 9.2%. The remaining discrepancy is attributed primarily to possible miscalibrations in the fitted CR worth curves and propagation of errors from the individual modules comprising the virtual asset. Nevertheless, the results demonstrate that the integrated DT framework can remain synchronized with reactor operation, provide expanded state awareness beyond directly measured quantities, and support short-term forecasting across startup, steady-state operation, and shutdown.

During the operational use case, the internal execution latencies of the active DT modules were monitored at each time step to evaluate whether synchronization with physical system was maintained. For each process, the total internal latency included both model execution and presentation of the generated output to the operator, as defined in Section 8.2. As shown in Figure 39, all three workflows remained below the no-delay threshold (1 s) while communication overhead was considered negligible. These results demonstrate that, under the baseline communication configuration, the evaluated virtual system operations can generate and present updated information through the GUI within the PUR-1 operational cycle.

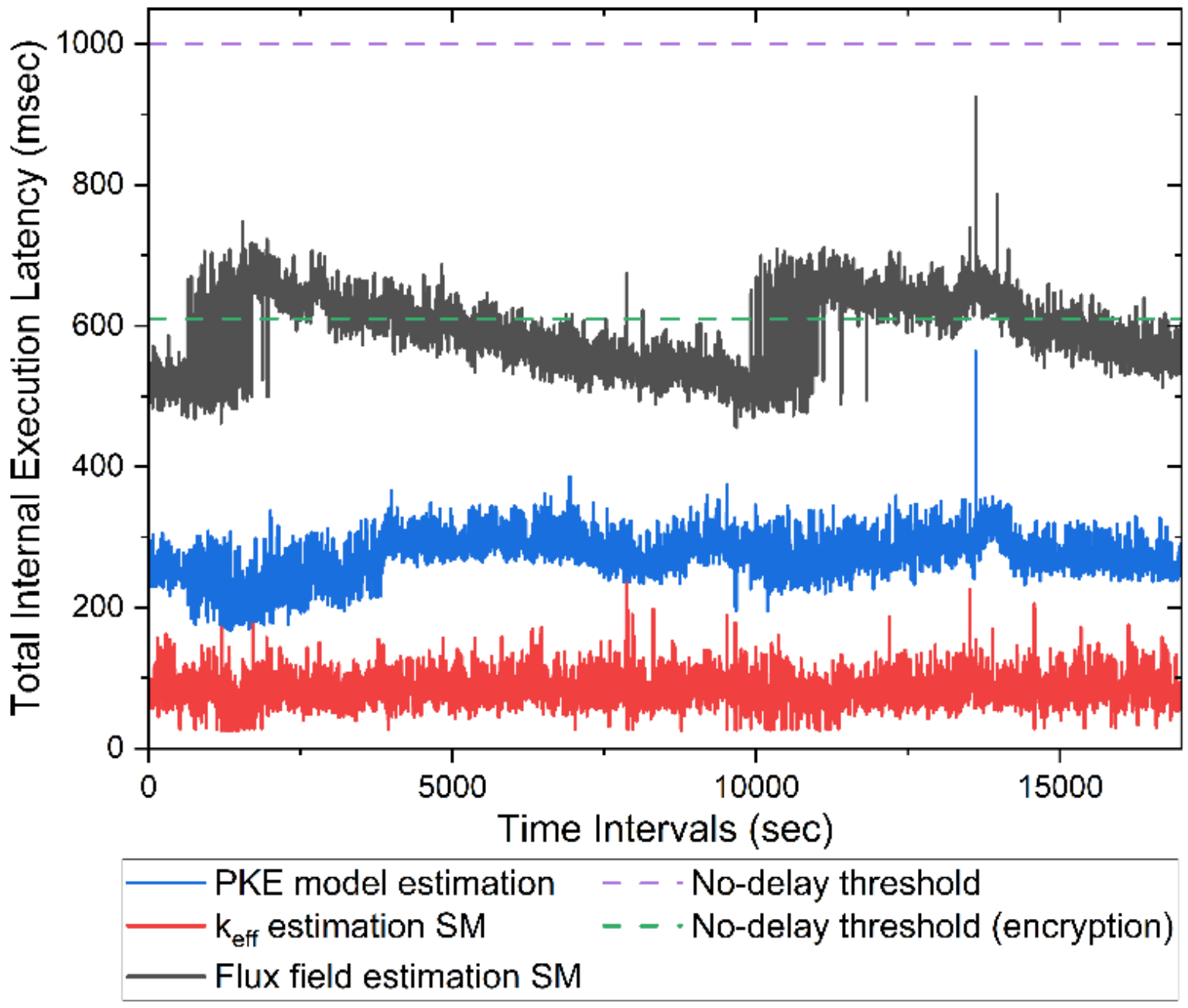

Figure 38: Total internal latency of the active PUR-1 DT modules throughout the evaluated reactor operational cycle. Dashed lines indicate the synchronization threshold and the internal-latency threshold after accounting for encrypted communication latency.

The effect of secure communication was subsequently evaluated by incorporating the experimentally measured (390 ms) physical-to-virtual transmission latency. Under this condition, the available latency budget for internal processing and presentation is reduced to

$$t_{total} = t_{cycle,PUR-1} - t_{\text{trans}}^{P\rightarrow V} = 1000 - 390 = 610\, ms \tag{45}$$

In this case, the PKE and $k_{eff}$ estimation workflows remain below this reduced threshold throughout the evaluated operational cycle. In contrast, the total latency associated with neutron-flux-field estimation periodically exceeds the (610 ms) budget, indicating that encrypted communication combined with high-resolution three-dimensional visualization can delay presentation of the reconstructed field to the operator (Figure 39).

To identify the source of this delay, the neutron-flux surrogate was evaluated separately without the GUI rendering and presentation latency. As shown in Figure 39, the SM execution time remains well below both the sub 1 s baseline threshold and the 610 ms threshold associated with encrypted communication. Eliminating visualization feature, the reconstructed flux field remains computationally available to other DT modules within the required synchronization interval. This analysis indicates that the additional latency originates primarily from rendering and displaying the high-resolution 3D field. Neutron flux field visualization can be updated at a lower frequency without compromising the timely availability of the underlying model output to other DT processes. These findings confirm the feasibility of a real-time DT operation for assisted state identification and forecasting in PUR-1.

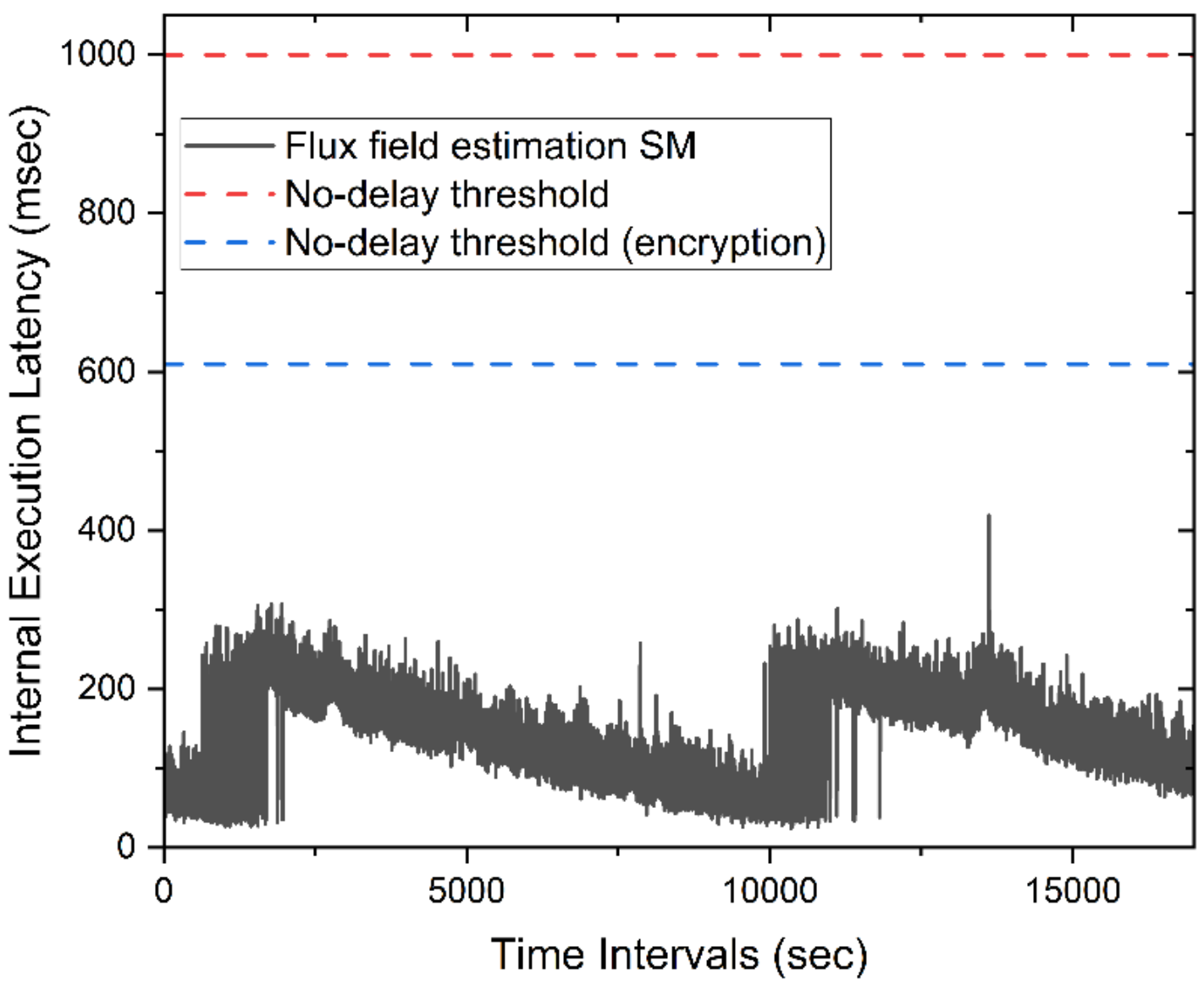

Figure 39: Execution latency of the neutron flux field SM excluding GUI rendering and presentation latency. Dashed lines indicate the synchronization threshold and the internal-latency threshold after accounting for encrypted communication latency.

# 11 Lessons Learned

The development of the PUR-1 DT showed that real-time performance must be evaluated at the system level, rather than for each module independently. Even when individual models satisfy the 1 s update interval, sequential dependencies, communication, and visualization can increase the total latency. This motivated the use of surrogate models for time-critical processes, critical-path latency analysis, and lower-frequency visualization for computationally intensive outputs.

The PUR-1 Multiphysics model development also highlighted the need to balance spatial fidelity with computational efficiency. The large dimensions of the reactor pool (approximately 5 m in depth) combined with the small thickness of the fuel plates (approximately 2 mm) substantially increase computational cost and make uniformly fine mesh computationally impractical (simulation runtimes exceeding 28 h per case). To address this challenge, a nonuniform meshing strategy was adopted, with finer discretization applied within and near the reactor core, where stronger thermal and flow gradients are expected, and progressively coarser meshes used in regions farther from the core. This approach reduced the computational burden while preserving the accuracy required to reproduce the experimentally observed thermal behavior.

Similar computational burdens were identified during the development of the SM for neutron-behavior prediction. Each OpenMC simulation initially required approximately 2 h, providing high-fidelity results but substantially increasing the total computational time required to generate the 2,500-run training dataset. To address this challenge, a parametric study was performed to optimize the number of batches and particles per batch with respect to both runtime and agreement with experimental benchmarking data. Although increasing either parameter improved simulation accuracy, the associated gains became marginal beyond a convergence threshold. The optimized configuration reduced the runtime to approximately 18 min per case while preserving the accuracy of the Monte Carlo model.

A further implementation challenge concerned the integration of CERVEROS subsystem data into the virtual system. Measurements of the AMDR position exhibited significant noise, which degraded the accuracy of position monitoring and control. In addition, the ultrasonic position sensor was susceptible to reflections from surrounding objects within the measurement area, producing intermittent outliers during operation. To mitigate these effects, a moving-average filter was applied to the acquired position data to suppress high-frequency measurement noise. The sensing configuration was also physically modified by repositioning the ultrasonic sensor and target surface to minimize interference from nearby objects.

The development of the DT also highlighted challenges associated with maintaining model validity during operation. Offline model validation alone is insufficient, as models trained on historical or synthetic data do not remain reliable when physical system conditions evolve beyond the regimes represented in the training dataset. In addition, data artifacts present in real-world systems such as sensor noise, communication disturbances, and operational variability may be insufficiently represented during model development. This necessitates data-driven models to be periodically reassessed against newly acquired operational data and, when necessary, recalibrated or retrained to maintain accurate DT performance.

# 12 Discussion and Conclusions

This paper described the design, functions and requirements, implementation, and experimental benchmarking of a real-time, cyber-physical DT for PUR-1. By coupling high-fidelity physics-based simulations with SMs and real-time data exchange between the physical and digital asset, the DT achieved full synchronization with the reactor at a 1 Hz frequency. Benchmarking against neutron flux distributions from gold foil activation and thermal-hydraulic measurements of pool temperature verified the accuracy of the Monte Carlo and Multiphysics modules, with deviations below 10% for flux mapping and approximately 2% for thermal fields.

State identification, short-term forecasting and operator decision support use case showed that the DT can extend observability beyond instrumentation limits and provide predictive insight to support enhanced control and more efficient reactor operation. Comparison between DT predictions and measurements collected from the physical system yielded an average percentage difference of 9.2%, indicating that the developed framework can maintain satisfactory predictive consistency throughout the full operational cycle. At the same time, SMs trained on extensive simulation datasets enabled real-time prediction of key reactor parameters overcoming the latency associated with high-fidelity simulations while preserving the physical interpretability of the overall framework. This hybrid approach is essential for DT deployment under reactor cycle-time constraints. Latency measurements further showed that the PKE and neutronic surrogate workflows satisfy the 1 s synchronization requirement, including the evaluated encrypted communication overhead, while high-resolution flux field visualization represents a non-critical presentation bottleneck that can be updated asynchronously.

Several limitations remain. Forecasting accuracy remains sensitive to uncertainties in control rod reactivity worth and sensor-related artifacts, particularly during reactor transients. The neutron flux SM also exhibited larger localized discrepancies in regions characterized by strong spatial gradients, while the Multiphysics model operates at a lower update frequency because of its higher computational cost. Furthermore, the reliance on continuous data exchange emphasizes the necessity of robust cybersecurity. Although anomaly detection and explainable AI methods were incorporated to mitigate false data injection, comprehensive defenses against evolving cyber threats are essential. Looking forward, the PUR-1 DT provides a reference architecture for nuclear applications, supporting potential extensions to autonomous control, predictive maintenance, operator training, and smart grid integration.

## 13 Acknowledgements

This research is part of the Consortium for Strategic Revitalization of Cyber-Physical Nuclear Infrastructure for Advanced Small Modular Reactors supported by the U.S. Department of Energy, Office of Nuclear Energy, Nuclear Energy University Program under contract DE-NE-0009525.

## 14 References

1. Nuclear Energy Institute (NEI): Nuclear Costs in Context. , Washington, DC, USA (2025).
2. Wang, Y.-J., Shirvan, K., Baglietto, E., Tsilifis, P., Amer, A., Khan, G., Wang, L.: High Fidelity Digital Twin for Critical System Assessments. (2022).
3. Wilsdon, K., Hansel, J., Kunz, M.R., Browning, J.: Autonomous control of heat pipes through digital twins: Application to fission batteries. Progress in Nuclear Energy. 163, 104813 (2023). https://doi.org/10.1016/j.pnucene.2023.104813.
4. Wang, H., Ponciroli, R., Vilim, R.B., Theos, V., Chatzidakis, S.: A Data-driven approach to Core Power distribution reconstruction in a Nuclear Reactor. Argonne National Laboratory (ANL), Argonne, IL (United States) (2024). https://doi.org/10.2172/2477295.
5. Kritzinger, W., Karner, M., Traar, G., Henjes, J., Sihn, W.: Digital Twin in manufacturing: A categorical literature review and classification. IFAC-PapersOnLine. 51, 1016–1022 (2018). https://doi.org/10.1016/j.ifacol.2018.08.474.
6. AIAA Digital Engineering Integration Committee: Digital twin: Definition & value. AIAA and AIA Position Paper. (2020).
7. Kochunas, B., Huan, X.: Digital Twin Concepts with Uncertainty for Nuclear Power Applications. Energies. 14, 4235 (2021). https://doi.org/10.3390/en14144235.
8. Bandala, M., Chard, P., Cockbain, N., Dunphy, D., Eaves, D., Hutchinson, D., Lee, D., Ma, X., Marshall, S., Murray, P., Parker, A., Stirzaker, P., Taylor, C.J., Zabalza, J., Joyce, M.J.: Digital twin challenges and opportunities for nuclear fuel manufacturing applications. Nuclear Engineering and Design. 420, 113013 (2024). https://doi.org/10.1016/j.nucengdes.2024.113013.
9. Ferrari, A., Willcox, K.: Digital twins in mechanical and aerospace engineering. Nat Comput Sci. 4, 178–183 (2024). https://doi.org/10.1038/s43588-024-00613-8.
10. Kapteyn, M.G., Pretorius, J.V.R., Willcox, K.E.: A probabilistic graphical model foundation for enabling predictive digital twins at scale. Nat Comput Sci. 1, 337–347 (2021). https://doi.org/10.1038/s43588-021-00069-0.
11. Tuegel, E.J., Ingraffea, A.R., Eason, T.G., Spottswood, S.M.: Reengineering Aircraft Structural Life Prediction Using a Digital Twin. International Journal of Aerospace Engineering. 2011, 154798 (2011). https://doi.org/10.1155/2011/154798.
12. Tao, F., Zhang, H., Liu, A., Nee, A.Y.C.: Digital Twin in Industry: State-of-the-Art. IEEE Transactions on Industrial Informatics. 15, 2405–2415 (2019). https://doi.org/10.1109/TII.2018.2873186.

13. Peldon, D., Banihashemi, S., LeNguyen, K., Derrible, S.: Navigating urban complexity: The transformative role of digital twins in smart city development. Sustainable Cities and Society. 111, 105583 (2024). https://doi.org/10.1016/j.scs.2024.105583.

14. Görtz, M., Brandl, C., Nitschke, A., Riediger, A., Stromer, D., Byczkowski, M., Heuveline, V., Weidemüller, M.: Digital twins for personalized treatment in uro-oncology in the era of artificial intelligence. Nat Rev Urol. 23, 29–39 (2026). https://doi.org/10.1038/s41585-025-01096-6.

15. Gahlot, S., Reddy, S.R.N., Kumar, D.: Review of Smart Health Monitoring Approaches With Survey Analysis and Proposed Framework. IEEE Internet of Things Journal. 6, 2116–2127 (2019). https://doi.org/10.1109/JIOT.2018.2872389.

16. Yadav, V., Zhang, H., Chwasz, C., Gribok, A., Ritter, C., Lybeck, N., Hays, R., Trask, T.C., Jain, P.K., Badalassi, V., Ramuhalli, P., Eskins, D., Gascot, R. l., Ju, D., Iyengar, R.: The State of Technology of Application of Digital Twins. United States Nuclear Regulatory Commission (U.S.NRC) (2021).

17. Kropaczek, D.J., Badalassi, V., Jain, P.K., Ramuhalli, P., Pointer, W.D.: Digital Twins for Nuclear Power Plants and Facilities. In: Crespi, N., Drobot, A.T., and Minerva, R. (eds.) The Digital Twin. pp. 971–1022. Springer International Publishing, Cham (2023). https://doi.org/10.1007/978-3-031-21343-4_31.

18. Browning, J., Hansel, J., Wilsdon, K., Houck, K., Pluth, A.: Microreactor Testbed Automation through Digital Engineering and Digital Twins. INCOSE International Symposium. 32, 375–389 (2022). https://doi.org/10.1002/iis2.12937.

19. Kothe, D.B.: CASL: The Consortium for Advanced Simulation of Light Water Reactors. , CASL: The Consortium for Advanced Simulation of Light Water Reactors (2010).

20. Cox, R.W., Kernicky, T., Khire, M., Whelan, M., Park, Y., Charkas, H., Varma, A., Vedovi, J.: Digital Twins for Monitoring Construction Quality. Presented at the ANS Winter Meeting 2022 November (2022). https://doi.org/10.13182/T127-39967.

21. Yadav, V., Sanchez, E., Gribok, A., Chwasz, C., Hays, R., Zhang, H., Lybeck, N., Gascot, R. l., Eskins, D., Ju, D., Iyengar, R.: Proceedings of the Workshop on Digital Twin Applications for Advanced Nuclear Technologies. (2021).

22. Lin, L., Athe, P., Rouxelin, P., Avramova, M., Gupta, A., Youngblood, R., Lane, J., Dinh, N.: Development and assessment of a nearly autonomous management and control system for advanced reactors. Annals of Nuclear Energy. 150, 107861 (2021). https://doi.org/10.1016/j.anucene.2020.107861.

23. Rivas, A., Delipei, G.K., Hou, J.: A System Predictive Maintenance Framework for Advanced Reactors Using a Data-Driven Digital Twin. Nuclear Science and Engineering. 199, 358–387 (2025). https://doi.org/10.1080/00295639.2024.2372515.

24. Manera, A., Duraisamy, K., Downar, T., Sun, X., Garcia, H., Vilim, R., Haugh, B., Sutter, T.: Project SAFARI Secure Automation for Advanced Reactor Innovation. (2021).

25. Stewart, R., Treviño, E., Shields, A., Heaps, K., Darrington, J., Williams, Q., Pope, C., Scott, J., Baker, B., Palmer, J., Vainqueur, B., Palmer, T.S., Palmer, C., Bays, S., Schanfein, M., Reyes, G., Ritter, C.: The AGN-201 Digital Twin: A test bed for remotely monitoring nuclear reactors. Annals of Nuclear Energy. 213, 111041 (2025). https://doi.org/10.1016/j.anucene.2024.111041.

26. Treviño, E., Shields, A., Stewart, R., Darrington, J., Scott, J., Pope, C., Ritter, C.: Autonomous anomaly detection of proliferation in the AGN-201 nuclear reactor digital twin. Annals of Nuclear Energy. 211, 110990 (2025). https://doi.org/10.1016/j.anucene.2024.110990.

27. Foundational Research Gaps and Future Directions for Digital Twins. National Academies Press, Washington, D.C. (2024). https://doi.org/10.17226/26894.

28. Theos, V., Gkouliaras, K., Miller, T., Jowers, B., Smith, R., Chatzidakis, S.: Development Of A Quantum-Based Cyber-PhysicalTestbed For Secure Communications In Nuclear Reactor Environments. Presented at the ANS Winter Meeting 2022 November (2022). https://doi.org/10.13182/T127-39659.

29. Zhang, Z., Liu, J., Zeng, W., Huang, Q., Liu, X.: Digital Twin Technology Architecture and Application for Nuclear Reactor Intelligent Operation and Maintenance. IEEE Access. 13, 91494–91504 (2025). https://doi.org/10.1109/ACCESS.2025.3570182.

30. United States Nuclear Regulatory Commission (U.S.NRC): Digital Twins Definition, https://www.nrc.gov/reactors/power/digital-twins, last accessed 2025/10/24.

31. Digital Twin Consortium: Definition of a Digital Twin, https://www.digitaltwinconsortium.org/initiatives/the-definition-of-a-digital-twin/, last accessed 2025/10/24.

32. Fuller, A., Fan, Z., Day, C., Barlow, C.: Digital Twin: Enabling Technologies, Challenges and Open Research. IEEE Access. 8, 108952–108971 (2020). https://doi.org/10.1109/ACCESS.2020.2998358.

33. Glaessgen, E., Stargel, D.: The Digital Twin Paradigm for Future NASA and U.S. Air Force Vehicles. In: 53rd AIAA/ASME/ASCE/AHS/ASC Structures, Structural Dynamics and Materials Conference. American Institute of Aeronautics and Astronautics, Honolulu, Hawaii (2012). https://doi.org/10.2514/6.2012-1818.

34. Boschert, S., Rosen, R.: Digital Twin—The Simulation Aspect. In: Hehenberger, P. and Bradley, D. (eds.) Mechatronic Futures: Challenges and Solutions for Mechatronic Systems and their Designers. pp. 59–74. Springer International Publishing, Cham (2016). https://doi.org/10.1007/978-3-319-32156-1_5.

35. Chen, Y.: Integrated and Intelligent Manufacturing: Perspectives and Enablers. Engineering. 3, 588–595 (2017). https://doi.org/10.1016/J.ENG.2017.04.009.

36. Schluse, M., Priggemeyer, M., Atorf, L., Rossmann, J.: Experimentable Digital Twins—Streamlining Simulation-Based Systems Engineering for Industry 4.0. IEEE Transactions on Industrial Informatics. 14, 1722–1731 (2018). https://doi.org/10.1109/TII.2018.2804917.

37. Liu, Z., Meyendorf, N., Mrad, N.: The role of data fusion in predictive maintenance using digital twin. Presented at the 44TH ANNUAL REVIEW OF PROGRESS IN QUANTITATIVE NONDESTRUCTIVE EVALUATION, VOLUME 37 (2018). https://doi.org/10.1063/1.5031520.

38. Mohammadi, N., Taylor, J.E.: Smart city digital twins. In: 2017 IEEE Symposium Series on Computational Intelligence (SSCI). pp. 1–5 (2017). https://doi.org/10.1109/SSCI.2017.8285439.

39. Coraddu, A., Oneto, L., Baldi, F., Cipollini, F., Atlar, M., Savio, S.: Data-driven ship digital twin for estimating the speed loss caused by the marine fouling. Ocean Engineering. 186, 106063 (2019). https://doi.org/10.1016/j.oceaneng.2019.05.045.

40. Görtz, M.: Digital twins: past, present and future. Sci Rep. 16, 10510 (2026). https://doi.org/10.1038/s41598-026-45272-z.

41. Willcox, K.E., Ghattas, O., Heimbach, P.: The imperative of physics-based modeling and inverse theory in computational science. Nat Comput Sci. 1, 166–168 (2021). https://doi.org/10.1038/s43588-021-00040-z.

42. Martin, N., Stewart, R., Bays, S.: A multiphysics model of the versatile test reactor based on the MOOSE framework. Annals of Nuclear Energy. 172, 109066 (2022). https://doi.org/10.1016/j.anucene.2022.109066.

43. Kumar, M., Shenbagaraman, V.M., Shaw, R.N., Ghosh, A.: Predictive Data Analysis for Energy Management of a Smart Factory Leading to Sustainability. In: Favorskaya, M.N., Mekhilef, S., Pandey, R.K., and Singh, N. (eds.) Innovations in Electrical and Electronic Engineering. pp. 765–773. Springer, Singapore (2021). https://doi.org/10.1007/978-981-15-4692-1_58.

44. Framatome joins with academia and industry partners to develop nuclear reactor digital twins, https://www.framatome.com/medias/framatome-joins-with-academia-and-industry-partners-to-develop-nuclear-reactor-digital-twins/, last accessed 2025/04/27.

45. Varé, C., Morilhat, P.: Digital Twins, a New Step for Long Term Operation of Nuclear Power Plants. In: Liyanage, J.P., Amadi-Echendu, J., and Mathew, J. (eds.) Engineering Assets and Public Infrastructures in the Age of Digitalization. pp. 96–103. Springer International Publishing, Cham (2020). https://doi.org/10.1007/978-3-030-48021-9_11.

46. Papadionysiou, M., Delipei, G., Avramova, M., Ferroukhi, H., Ivanov, K.: High-resolution predictions of the coolant properties for the 3D PWR core with artificial neural networks based on CTF. Nuclear Engineering and Design. 442, 114261 (2025). https://doi.org/10.1016/j.nucengdes.2025.114261.

47. Delipei, G.K., Altahhan, M., Rouxelin, P., Sen, S., George, N., Avramova, M., Ivanov, K.: Uncertainty Quantification Framework for High-Temperature Gas-Cooled Reactors Using VSOP and DAKOTA. Nuclear Science and Engineering. 1–27 (2025). https://doi.org/10.1080/00295639.2025.2486901.

48. Luciano, Ross, J.R., Seo, Hoffing, M., Gentry, C., Benjamin, C., Charlton, W., Clarno, M.: Developing a Digital Twin of the TRIGA II Research Reactor at The University of Texas at Austin. Presented at the 2025 ANS Annual June (2025). https://doi.org/10.13182/T140-48695.

49. Ross, J.R.: Development of a digital twin for online state reconstruction in TRIGA reactors. (2024).

50. ARPA-E: ARPA-E Project | Generation of Critical Irradiation Data to Enable Digital Twinning of Molten-Salt Reactors, http://arpa-e.energy.gov/technologies/projects/generation-critical-irradiation-data-enable-digital-twinning-molten-salt, last accessed 2023/06/15.

51. Zhu, E., Li, T., Xiong, J., Chai, X., Zhang, T., Liu, X.: A super-real-time three-dimension computing method of digital twins in space nuclear power. Computer Methods in Applied Mechanics and Engineering. 417, 116444 (2023). https://doi.org/10.1016/j.cma.2023.116444.

52. Rasheed, A., San, O., Kvamsdal, T.: Digital Twin: Values, Challenges and Enablers From a Modeling Perspective. IEEE Access. 8, 21980–22012 (2020). https://doi.org/10.1109/ACCESS.2020.2970143.

53. Mercier, B., Ziliang, Z., Liyi, C., Nuoya, S.: Modeling and control of xenon oscillations in thermal neutron reactors. EPJ Nuclear Sci. Technol. 6, 48 (2020). https://doi.org/10.1051/epjn/2020009.

54. Alnahdi, A.H., Alghamdi, A.A., Almarshad, A.I.: Investigation of the Fuel Shape Impact on the MTR Reactor Parameters Using the OpenMC Code. Processes. 11, 637 (2023). https://doi.org/10.3390/pr11020637.

55. Theos, V., Gkouliaras, K., Miller, T., Jowers, B., Chatzidakis, S.: Towards a Cyber-Physical Testbed for Cybersecurity Research in Nuclear Environments. In: Transactions of the American Nuclear Society. pp. 320–323. American Nuclear Society (ANS), Washington, D.C. (2023). https://doi.org/10.13182/T129-42746.

56. Aldama, D.L., Gual, M.R.: Determination of the control rod worth for research reactors. In: Research reactor utilization, safety and management. Proceedings. pp. 7–7 (2000).

57. Depriest, K.R., Kajder, K.C., Frye, J.N., Denman, M.R.: CONTROL ROD REACTIVITY CURVES FOR THE ANNULAR CORE RESEARCH REACTOR. In: Reactor Dosimetry State of the Art 2008. pp. 195–203. WORLD SCIENTIFIC, Akersloot, The Netherlands (2009). https://doi.org/10.1142/9789814271110_0028.
58. Duderstadt, J., Hamilton, L.J.: Nuclear reactor analysis. John Wiley and Sons, Inc., New York (1976).
59. U.S. Nuclear Regulatory Commission (NRC): Regulatory Guide 5.71. U.S. Nuclear Regulatory Commission (2010).
60. Gkouliaras, K., Theos, V., Chatzidakis, S.: Exploring the Feasibility of Quantum-Based Secure Communications for Nuclear Applications. Nuclear Technology. 0, 1–20 (2024). https://doi.org/10.1080/00295450.2024.2368977.
61. Garton, D.: Purdue Model Framework for Industrial Control Systems & Cybersecurity Segmentation. (2019).
62. Chatzidakis, S., Theos, V., Gkouliaras, K., Dahm, Z., Vasili, K., Miller, T., Jowers, B., Lawrence, J., Hollern, J., Eskins, D., Cottrell, K., Kim, A.: Performance Evaluation and GAP Analysis. United States Nuclear Regulatory Commission (U.S.NRC), Washington, DC 20555–0001 (2024).
63. Chatzidakis, S., Theos, V., Gkouliaras, K., Dahm, Z., Vasili, K., Miller, T., Jowers, B., Lawrence, J., Hollern, J., Eskins, D., Cottrell, K., Kim, A.: Characterizing Nuclear Cybersecurity States Using Artificial Intelligence/Machine Learning - Final Report. United States Nuclear Regulatory Commission (U.S.NRC), Washington, DC 20555–0001 (2024).
64. Gkouliaras, K., Theos, V., Dahm, Z., Richards, W., Vasili, K., Chatzidakis, S.: False Data Injection Detection in Nuclear Systems Using Dynamic Noise Analysis. IEEE Access. 12, 94936–94949 (2024). https://doi.org/10.1109/ACCESS.2024.3425270.
65. Theos, V., Gkouliaras, K., Chatzidakis, S.: Development and Analysis of a Real-time Digital Twin System in PUR-1. In: The Nuclear Plant Instrumentation and Control & Human-Machine Interface Technology (NPIC&HMIT 2025) Proceedings. pp. 464–473. , Chicago, IL (2025). https://doi.org/10.13182/NPICHMIT25-46907.
66. Reuss, P.: Neutron physics. (2008).
67. Romano, P.K., Horelik, N.E., Herman, B.R., Nelson, A.G., Forget, B., Smith, K.: OpenMC: A state-of-the-art Monte Carlo code for research and development. Annals of Nuclear Energy. 82, 90–97 (2015). https://doi.org/10.1016/j.anucene.2014.07.048.
68. Todreas, N.E., Kazimi, M.S.: Nuclear Systems Volume I: Thermal Hydraulic Fundamentals, Third Edition. CRC Press, Boca Raton (2021). https://doi.org/10.1201/9781351030502.
69. Abdou, M.A., Maynard, C.W., Wright, R.Q.: MACK: computer program to calculate neutron energy release parameters (fluence-to-kerma factors) and multigroup neutron reaction cross sections from nuclear data in ENDF Format. , Oak Ridge National Lab., Tenn. (USA) (1973).
70. Lau, J., Dahm, Z., Anwar, A., Bae, J., Takahashi, G., Akhras, A., Zhou, J., Triveri, J., Chatzidakis, S.: Enabling Online Education and Training in Nuclear Engineering Curriculum Using Virtual Labs and Augmented Reality. In: Conference on Nuclear Training and Education: A Biennial International Forum (CONTE 2023),. pp. 140–143. American Nuclear Society, Amelia Island, FL (2023). https://doi.org/10.13182/CONTE23-40498.
71. Dahm, Z., Theos, V., Vasili, K., Richards, W., Gkouliaras, K., Chatzidakis, S.: A one-class explainable AI framework for identification of non-stationary concurrent false data injections in nuclear reactor signals. Nuclear Engineering and Design. 444, 114359 (2025). https://doi.org/10.1016/j.nucengdes.2025.114359.

72. Vasili, K., Dahm, Z.T., Chatzidakis, S.: An Unsupervised Deep Explainable AI Framework for Localization of Concurrent Replay Attacks in Nuclear Reactor Signals. IEEE Access. 14, 25414–25433 (2026). https://doi.org/10.1109/ACCESS.2025.3646560.

73. Dahm, Z., Vasili, K., Theos, V., Lin, L., Oncken, J., England, R., Agarwal, V., Chatzidakis, S.: Model predictive control with state estimation and reduced-order physics for nuclear reactor operation during unanticipated communication transients. Annals of Nuclear Energy. 240, 112733 (2026). https://doi.org/10.1016/j.anucene.2026.112733.

74. Teixeira, A., Pérez, D., Sandberg, H., Johansson, K.H.: Attack models and scenarios for networked control systems. In: Proceedings of the 1st international conference on High Confidence Networked Systems. pp. 55–64. Association for Computing Machinery, New York, NY, USA (2012). https://doi.org/10.1145/2185505.2185515.

75. Mo, Y., Sinopoli, B.: Secure control against replay attacks. In: 2009 47th Annual Allerton Conference on Communication, Control, and Computing (Allerton). pp. 911–918 (2009). https://doi.org/10.1109/ALLERTON.2009.5394956.

76. Mo, Y., Weerakkody, S., Sinopoli, B.: Physical Authentication of Control Systems: Designing Watermarked Control Inputs to Detect Counterfeit Sensor Outputs. IEEE Control Systems Magazine. 35, 93–109 (2015). https://doi.org/10.1109/MCS.2014.2364724.

77. Zhao, Y., Smidts, C.: A control-theoretic approach to detecting and distinguishing replay attacks from other anomalies in nuclear power plants. Progress in Nuclear Energy. 123, 103315 (2020). https://doi.org/10.1016/j.pnucene.2020.103315.

78. Soman, R.K., Farghaly, K., Mills, G., Whyte, J.: Digital twin construction with a focus on human twin interfaces. Automation in Construction. 170, 105924 (2025). https://doi.org/10.1016/j.autcon.2024.105924.

79. Farghaly, K., Soman, R.K., Whyte, J.: Visualizing real-time information through a construction production control room. Presented at the EC3 Conference 2021 (2021). https://doi.org/10.35490/EC3.2021.169.

80. Huang, S., Wang, G., Lei, D., Yan, Y.: Toward digital validation for rapid product development based on digital twin: a framework. Int J Adv Manuf Technol. 119, 2509–2523 (2022). https://doi.org/10.1007/s00170-021-08475-4.

81. Zheng, J., Meister, M.: The unbearable slowness of being: Why do we live at 10 bits/s? Neuron. 113, 192–204 (2025). https://doi.org/10.1016/j.neuron.2024.11.008.

82. Lau, J., Miller, T., Jowers, B., Chatzidakis, S.: Streamlining PUR-1 Operation: A Case Study on Large Language Models (LLM) in Reactor Informational Systems. In: Proceedings of the Nuclear Plant Instrumentation and Control & Human-Machine Interface Technology (NPIC&HMIT 2025). American Nuclear Society, Chicago (2025). https://doi.org/10.13182/NPICHMIT25-46965.

83. Yao, S., Zhao, J., Yu, D., Du, N., Shafran, I., Narasimhan, K., Cao, Y.: ReAct: Synergizing Reasoning and Acting in Language Models. International Conference on Learning Representations (ICLR). (2023).

84. Darrington, J.W.: The DeepLynx Data Warehouse. Idaho National Laboratory (INL), Idaho Falls, ID (United States) (2022).

85. Gkouliaras, K., Theos, V., Miller, T., Jowers, B., Kennedy, G., Grant, A., Cronin, T., Evans, P.G., Chatzidakis, S.: A framework for quantum-secure communications in cyber-physical control systems with experimental demonstration in a nuclear reactor. Sci Rep. 16, (2026). https://doi.org/10.1038/s41598-026-49514-y.

86. International Atomic Energy Agency: Calibration of Radiation Protection Monitoring Instruments. International Atomic Energy Agency (2000).

87. Ponciroli, R., Wang, H., Theos, V., Lau, J., Chatzidakis, S., Vilim, R.B.: Ex-Core Neutron-Flux Reconstruction Using Machine-Learned Green's Kernels and Flux-Gradient Measurements. Nuclear Science and Engineering. 0, 1–27 (2026). https://doi.org/10.1080/00295639.2026.2690565.

88. Madden, R., Pike, H., Kabir, N.B., Miller, T., Jowers, B., Chatzidakis, S.: Design and Characterization of the Modified Purdue Subcritical Pile for Nuclear Research Applications, https://www.preprints.org/manuscript/202502.0812/v1, (2025). https://doi.org/10.20944/preprints202502.0812.v1.

89. ASTM: Test Method for Determining Thermal Neutron Reaction Rates and Thermal Neutron Fluence Rates by Radioactivation Techniques, http://www.astm.org/cgi-bin/resolver.cgi?E262-17. https://doi.org/10.1520/E0262-17.

90. ASTM: Test Methods for Detector Calibration and Analysis of Radionuclides, http://www.astm.org/cgi-bin/resolver.cgi?E181-17. https://doi.org/10.1520/e0181-17.

91. ASTM: Practice for Characterizing Neutron Fluence Spectra in Terms of an Equivalent Monoenergetic Neutron Fluence for Radiation-Hardness Testing of Electronics, http://www.astm.org/cgi-bin/resolver.cgi?E722-19. https://doi.org/10.1520/E0722-19.

92. ASTM: Practice for Determining Neutron Fluence, Fluence Rate, and Spectra by Radioactivation Techniques, http://www.astm.org/cgi-bin/resolver.cgi?E261-16R21. https://doi.org/10.1520/e0261-16r21.

93. International Atomic Energy Agency: Neutron fluence measurements. (1970).

94. Westcott, C.H., Chalk River, Atomic Energy of Canada Limited: Effective cross section values for well-moderated thermal reactor spectra. Atomic Energy of Canada Limited, Chalk River, Ontario (Canada) (1970).

95. 8. Specifying Tallies — OpenMC Documentation, https://docs.openmc.org/en/latest/usersguide/tallies.html, last accessed 2024/12/13.

96. Miller, T., Kong, R., Xu, Y., Storz, D., Smith, R., Palmer, J., Raman, T., Anwar, A., Kim, S., Chatzidakis, S.: Development of a Temperature Compensated Correlation Between Core Power and Coolant Temperature Change Rate for PUR-1. In: Transactions of the American Nuclear Society. pp. 573–576. American Nuclear Society (ANS) (2022). https://doi.org/10.13182/T126-38382.

97. Rasmussen, C.E., Williams, C.K.I.: Gaussian Processes for Machine Learning. MIT Press, Cambridge, MA, USA (2005). https://doi.org/10.7551/mitpress/3206.003.0019.

98. West, A., Tsitsimpelis, I., Licata, M., Jazbec, A., Snoj, L., Joyce, M.J., Lennox, B.: Use of Gaussian process regression for radiation mapping of a nuclear reactor with a mobile robot. Sci Rep. 11, 13975 (2021). https://doi.org/10.1038/s41598-021-93474-4.

99. Williams, C., Rasmussen, C.: Gaussian Processes for Regression. In: Advances in Neural Information Processing Systems. MIT Press (1995).